\pdfoutput=1
\documentclass[fleqn,usenatbib,usedcolumn]{mnras}
\usepackage[british]{babel}

\usepackage{txfonts}
\usepackage[T1]{fontenc}
\usepackage{graphicx,url}

\usepackage{microtype}
\begin{document}

\label{firstpage}

\title[Interacting Large-Scale Magnetic Fields and Ionised Gas in the W50/SS433 System]{Interacting Large-Scale Magnetic Fields and Ionised Gas in the W50/SS433 System}

\author[J.~S.\ Farnes, et al.]{J.~S.\ Farnes$^{1,2}$\thanks{email:\texttt{j.farnes@astro.ru.nl}}, B.~M.\ Gaensler$^{3,2}$, C.\ Purcell$^{2}$, X.~H.\ Sun$^{2}$, M.~Haverkorn$^{1}$, \newauthor E.\ Lenc$^{4,2}$, S.~P.\ O'Sullivan$^{5,2}$, T.\ Akahori$^{6,2}$\\
  $^{1}$Department of Astrophysics/IMAPP, Radboud University, PO Box 9010, NL-6500 GL Nijmegen, the Netherlands.\\
  $^{2}$Sydney Institute for Astronomy, School of Physics, The University of Sydney, NSW 2006, Australia.\\
  $^{3}$Dunlap Institute for Astronomy and Astrophysics, The University of Toronto, 50 St. George Street, Toronto, ON M5S 3H4, Canada.\\ 
  $^{4}$ARC Centre of Excellence for All-sky Astrophysics (CAASTRO).\\
  $^{5}$Instituto de Astronom\'{i}a, Universidad Nacional Aut\'{o}noma
de M\'{e}xico (UNAM), A.P. 70-264, 04510 M\'{e}xico, D.F., Mexico.\\
  $^{6}$Graduate School of Science and Engineering, Kagoshima University, Kagoshima 890--0065, Japan.\\  }

\date{Accepted ---}

\pagerange{\pageref{firstpage}--\pageref{lastpage}}

\pubyear{2016}

\maketitle

\begin{abstract}
The W50/SS433 system is an unusual Galactic outflow-driven object of debatable origin. We have used the Australia Telescope Compact Array (ATCA) to observe a new 198 pointing mosaic, covering $3^\circ \times 2^\circ$, and present the highest-sensitivity full-Stokes data of W50 to date using wide-field, wide-band imaging over a 2~GHz bandwidth centred at 2.1~GHz. We also present a complementary H$\alpha$ mosaic created using the Isaac Newton Telescope Photometric H$\alpha$ Survey of the Northern Galactic Plane (IPHAS). The magnetic structure of W50 is found to be consistent with the prevailing hypothesis that the nebula is a reanimated shell-like supernova remnant (SNR), that has been re-energised by the jets from SS433. We observe strong depolarization effects that correlate with diffuse H$\alpha$ emission, likely due to spatially-varying Faraday rotation measure (RM) fluctuations of $\ge48$ to 61~rad~m$^{-2}$ on scales $\le4.5$ to 6~pc. We also report the discovery of numerous, faint, H$\alpha$ filaments that are unambiguously associated with the central region of W50. These thin filaments are suggestive of a SNR's shock emission, and almost all have a radio counterpart. Furthermore, an RM-gradient is detected across the central region of W50, which we interpret as a loop magnetic field with a symmetry axis offset by $\approx$90$^{\circ}$ to the east--west jet-alignment axis, and implying that the evolutionary processes of both the jets and the SNR must be coupled. A separate RM-gradient is associated with the termination shock in the Eastern ear, which we interpret as a ring-like field located where the shock of the jet interacts with the circumstellar medium. Future optical observations will be able to use the new H$\alpha$ filaments to probe the kinematics of the shell of W50, potentially allowing for a definitive experiment on W50's formation history.
\end{abstract}


\section{Introduction}\label{intro}
The large radio nebula W50, also referred to as G39.7--2.0 and the ``Manatee nebula'', is both unusual and of an undetermined origin. The nebula appears to be interacting with the central compact source and microquasar SS433 -- the first known stellar source of relativistic jets \citep{1979MNRAS.187P..13F,1979ApJ...230L..41M}. The W50/SS433 system was first identified at radio wavelengths by \citet{1958BAN....14..215W}. Initial suggestions for the system's origin were made by, e.g.\ \citet{1978Natur.276..571R}, which identified SS433 as a new class of stellar remnant (the first-known Galactic microquasar) and proposed that the surrounding radio nebula of W50 was the remnant from a supernova explosion. 

The nebula is widely catalogued and considered as a supernova remnant \citep[SNR;][]{2014BASI...42...47G}, however this origin is still uncertain and debated. While the circular shell of W50 is reminiscent of shell-type SNRs, the additional elongated morphology is completely anomalous. This elongation is at least partly due to the interaction between SS433 and W50, with ``blobs'' of material being ejected at $0.26c$ from the central engine of SS433 by precessing jets that are aligned along the east--west axis of elongation \citep[e.g.][]{1995yera.conf...12J}. The elongated morphology of the radio nebula could therefore have been formed by (i) \emph{a Wind+Jets model}: with the SS433 jets expanding into a swept-up interstellar medium (ISM) -- a stellar-wind bubble -- that was itself created either entirely by the jets of SS433 or even by a continuous wind from the SS433 binary \citep{1980ApJ...238..722B,1980Natur.287..806S,1983MNRAS.205..471K}, or alternatively by (ii) \emph{a SNR+Jets model}: with an encounter between the jets from SS433 and an already formed SNR, with the ram pressure from the jets `punching' through the shell to form the ears \citep{1980MNRAS.192..731Z,1986MNRAS.218..393D,1996PASJ...48..819M,2000A&A...362..780V}. What is certain is that the W50/SS433 system -- whether a wind-blown bubble or a SNR -- is an anomalous object. It could be a one-of-a-kind, and could also form part of a rare community of other peculiar Galactic radio nebulae \citep{1987MNRAS.225..221G,1998AJ....116.1842D,1998MNRAS.299..812G,1987AuJPh..40..855K,2001ApSSS.276..139D}.

To further complicate matters, the morphology of the system is also affected by interactions between the swept-up media and the surrounding environment -- which consists of a dense surrounding interstellar medium (ISM) and an ambient interstellar magnetic field \citep[e.g.][]{1987A&A...183..118K,2008MNRAS.387..839Z}. These environmental factors are important for W50, as the west of the nebula appears in projection as dipping into the Galactic plane -- likely resulting in a density gradient across the source and expansion into an inhomogeneous ISM. In addition, previous attempts at modelling W50 have largely neglected the impact of magnetic fields on the evolution of the nebula. The large-scale magnetic fields have likely played a crucial role in the development of W50 -- independent of the chief mechanism that drives the object's formation -- as the interaction of magnetic fields is known to influence the evolution of wind-blown bubbles \citep[e.g.][]{2015ApJ...804...22P}, SNRs \citep[e.g.][]{2015HiA....16..395K}, and other radio nebulae \citep[e.g.][]{2015ApJ...799..198R}. Moreover, previous observations have revealed regular magnetic fields, including a strong field in the rim to the north \citep[e.g.][]{1981A&A...103..277D,1986MNRAS.218..393D}. The polarization angle of this region is indicative of the ordered field strength being dominated by standard compression of an interstellar field that is oriented approximately east--west, with the radio ears that are generally believed to be punched out by ram pressure from the jets of SS433 also approximately aligned along this same axis. This further complicates understanding the evolution of W50's morphology since, for example, a SNR evolving in a strong ambient field is known to become elongated along the field direction \citep{1991MNRAS.252...82I,1995MNRAS.274.1157R}, while in a wind-driven nebula a stellar wind with frozen-in toroidal fields can evolve into collimated outflows \citep{1996ApJ...469L.127R}. The presence of magnetic fields may also influence the collimation of the jets of SS433 \citep[e.g.][]{2015MNRAS.453.3213S}, which itself has likely had a strong influence on the nebula's evolution. Furthermore, more complex models essentially consisting of either \emph{SNR+Wind+Jets} or \emph{Wind+SNR+Jets} could also influence the shape of W50 \citep[e.g.][]{1991MNRAS.251..318T}. Models have recently been proposed for generating asymmetric supernova remnants \citep[e.g.][]{2015MNRAS.450.3080M}, and even off-centred or asymmetric supernova explosions within a pre-existing wind-driven bubble could give rise to a morphology that resembles that of W50 \citep{1989A&A...215..347C,1993MNRAS.261..674R}.

Attempting to separate the magnitude of each of these multiple effects on the elongation of W50 ultimately requires a multi-wavelength study. For example, diffuse H$\alpha$ emission is a reliable tracer of the ionised hydrogen content/thermal electrons in the surrounding environment, which can depolarize synchrotron radio emission emitted at a similar or greater distance along the line-of-sight and can thereby conceal the presence of ordered magnetic fields in the nebula \citep[e.g.][]{2014MNRAS.437.2936S}. Furthermore, the presence of optical filamentary emission associated with the central region of the nebula would most likely originate from shock-excited material in a SNR with the surrounding ISM being heated by the outgoing shock wave, and producing a set of filamentary shell-like structures that occur due to cooling behind the shock front. This shock-excited optical emission results in both filamentary and diffuse faint H$\alpha$ emission, while the presence of diffuse optical emission can also be associated with a wind-blown bubble \citep{1977ApJ...218..377W,2007MNRAS.381..308B,2008IAUS..250..341C}. A filamentary shell-like structure in H$\alpha$ is therefore one way to distinguish a SNR from a wind-driven bubble. However, neither of these scenarios are mutually exclusive. Nevertheless, the discovery of such optical emission would allow for the optical spectra to be obtained -- thereby providing a direct test of shock physics in the object, and potentially allowing for a definitive experiment to determine the origin of W50 \citep[e.g.][]{1977MNRAS.181..541L}. In addition, if new optical filaments can be found, they would help us to understand the connections between the filamentary emission and the charged particles and ordered magnetic fields within W50. If shocked optical filaments exist in W50, do any of them have coincident synchrotron emission? Are the magnetic fields more strongly ordered in such filaments? To what extent does the creation of optical filaments influence the orientation of the magnetic fields? And conversely, to what extent do the magnetic fields influence the creation of the optical filaments? 

In this paper, we set out to understand the origin and formation of W50, by performing a needed multi-wavelength study -- obtaining new broadband 1 to 3~GHz radio data and complementary 1~arcsec spatial resolution H$\alpha$ images that allow us to investigate both the large-scale magnetic fields and the ionised hydrogen content. The paper is structured as follows: in Section~\ref{backgroundw50} we provide extensive background information and review of the much-studied W50/SS433 system, in Section~\ref{datareduction} we discuss the technical details of the data processing, in Section~\ref{w50} we present and analyse our new multi-wavelength results, in Section~\ref{discussion} we provide a discussion of our findings, and in Section~\ref{conclusions} we provide our conclusions. Throughout, we assume a distance to SS433 (and by association W50) of $5.5\pm0.2$~kpc \citep{2004ApJ...616L.159B,2007MNRAS.381..881L}. This distance is still a source of some debate \citep[e.g.][]{2010arXiv1001.5097P}. The total intensity spectral index, $\alpha$, is defined in the sense $S_{\nu} \propto \nu^{\alpha}$. When referring to `polarization', we refer to linear polarization unless otherwise specified. 

\section{Background \& Review}
\label{backgroundw50}
\subsection{Overview of the W50/SS433 System}
The radio nebula of W50 (G39.7--2.0) has been studied thoroughly since its identification in the radio catalogue of \citet{1958BAN....14..215W}, with a comprehensive literature dedicated to the object. W50 is a non-thermal Galactic source of large angular extent ($2^\circ \times 1^\circ$). The shape of the object has been compared to that of a conch shell \citep[e.g.][]{2011ApJ...735L...7B} and of an interstellar manatee. At radio wavelengths, the morphology of the nebula of W50 is seen in projection as an approximately-circular shell that is $\sim1^{\circ}$ in diameter with the bright compact point source SS433 located close to the geometric centre. There are two bright radio ears (or `ansae') to the east and west of the nebula, and the entire radio structure appears elongated and aligned along this east--west axis -- with the entire nebula covering $2^\circ \times 1^\circ$ on the sky \citep[e.g.][]{1980A&A....84..237G}. The western edge of the nebula dips into the Galactic plane, as the system is at a Galactic latitude of $\approx-2^\circ$. The surrounding field of view contains several bright extragalactic radio sources, and a nearby H{\sc II} region S74 to the NW. W50 is one of the largest Galactic supernova remnants. Such objects are poorly studied at high angular resolution due to their overall angular size and the considerable number of pointings required for full imaging at radio wavelengths. 

It has been shown that the nebula can be reproduced to a good approximation by the combined effects of the evolution of a SNR shell and its interaction with the precessing jets from SS433 \citep{2011ApJ...735L...7B,2011MNRAS.414.2828G,2011MNRAS.414.2838G}. While an accumulating body of evidence tends to be in favour of a SNR origin \citep[see e.g.][and references therein]{2004ASPRv..12....1F}, this is still a source of debate with some studies suggesting it is more likely a stellar-wind/interstellar bubble (see Section~\ref{intro}). The object is certainly in a unique class of outflow-driven objects \citep[e.g.][]{1998MNRAS.299..812G}. The progenitor that formed SS433 is typically presumed to have exploded $10^4$--$10^5$~years ago \citep{1980ApJ...238..722B,2007MNRAS.381..881L,2011MNRAS.414.2838G}. The morphology of W50 hints at an approximately circular explosion, with the ``ears'' having been ``punched'' through the rim by the relativistic jets moving outwards from SS~433 at 0.26$c$ \citep{1984ARA&A..22..507M,1987AJ.....94.1633E,2011MNRAS.414.2828G,2011MNRAS.414.2838G}. Importantly, the contributions from an initial SN explosion, that probably preceded formation of the compact object SS433, have not been convincingly distinguished from the impact of the jet and wind activity of the central compact system \citep{2010AN....331..412A}. Recent studies in the optical and X-ray wavebands \citep{2007MNRAS.381..308B,2007A&A...463..611B,2011ApJ...735L...7B} possibly suggest at least two distinctly different states of jet activity \citep{2011MNRAS.414.2828G}. 

Optical emission was first associated with W50 in the 1980's \citep{1980ApJ...236L..23V,1980MNRAS.192..731Z}, with filaments located $\sim30$~arcmin to the E and W of SS433, although none were located within the central region of the nebula itself. Diffuse X-ray emission is also coincident with the optical filaments. The extended X-ray lobes or jets extend to the E and W along the precession axis of the radio jets, with the X-ray emission peaking in the region of the optical filaments \citep{1980Natur.287..806S,1983ApJ...273..688W}. There has also been additional soft X-ray emission discovered in the ear to the E, coincident with the radio ear and associated with the terminal shock of the SS433 jets \citep{1997ApJ...483..868S,1999ApJ...512..784S}. It has been found that the jet terminates in a ring-like termination shock \citep{2007A&A...463..611B}. No X-ray emission is found in the ear to the W, suggesting an inhomogeneous surrounding medium. Both the X-ray emission discovered in the lobes, and the presence of the optical filaments, strongly implies a density enhancement to the E and W at a radius of $\sim30$~arcmin from SS433. This is consistent with the traced out shell that is visible at radio wavelengths to the N and S, and together with the optical filaments being oriented perpendicularly to the radio jets, implies an interaction between the jets with a preexisting shell.

\citet{2007MNRAS.381..308B} reported the first faint optical emission associated with the central region of the nebula itself, albeit in only a small rectangular region of $\approx4$~arcmin width and height. The identification of more extensive optical filaments, especially if there was good correlation between the radio structure and the filaments, would lend increasing support to a \emph{SNR+Jets} hypothesis. Such emission is challenging to observe, as it is expected to be faint, and possibly not visible at all due to dust extinction. To observe faint H$\alpha$ emission also requires a continuum-correction. New sensitive and high-resolution surveys such as the Isaac Newton Telescope Photometric H$\alpha$ Survey of the Northern Galactic Plane (IPHAS) may feasibly allow for their detection \citep{2005MNRAS.362..753D}.

\subsection{Radio Polarimetry of W50}
\label{radiopolbackground}
As mentioned in Section~\ref{intro}, the large-scale magnetic fields in W50 are likely playing an influential role in the development of the object. The chief method with which to study cosmic magnetic fields is via radio polarimetry. Observations of the linear polarization in radio nebulae can provide information on the physical processes driving the emission and propagation of synchrotron radiation. In the case of W50, such observations have traditionally been complicated by the object's large angular size, although contemporary analyses have now been applied to numerous angularly-large astrophysical systems \citep[see e.g.][and references therein]{2013ApJ...777...55H,2013ApJ...764..162O,2015ApJ...804...22P}. As polarization is a direct tracer of magnetic field structure, its detection can enable investigations of the order and orientation of fields. This is particularly important in shocked regions -- which compress and order magnetic fields. In particular, the interaction between SNRs and magnetic fields is a crucial one, as the shocked material from a SN explosion compresses fields that are frozen-in to the plasma. Furthermore, magnetohydrodynamic turbulence can give rise to Rayleigh--Taylor instabilities in the SN shell that allows for a distinction between younger and older objects \citep{1976AuJPh..29..435D}. In the case of W50, the magnetic fields may also be influencing the jets of SS~433.

Measurements of the Faraday rotation, in the form of rotation measures (RM; see \citet{2005A&A...441.1217B} for a full review), are strong probes of the magnetic field structure in any foreground medium along the line-of-sight and across the face of a radio nebula, allowing us to distinguish between the presence of coherent and random fields. Furthermore, detection of a RM gradient across the source and any resulting depolarization can allow us to infer the geometry and also the anisotropy of the field -- which could be caused by either turbulent or systematically varying regular fields \citep{1966MNRAS.133...67B,1975A&A....41..307V,1991MNRAS.250..726T,2014ApJS..212...15F}. Such scenarios are presently unconstrained for W50. While previous studies at radio wavelengths have shown that there is an inhomogeneous distribution of polarization from W50 \citep{1981A&A...103..277D,1986MNRAS.218..393D}, there is no understanding of whether this inhomogeneity could be the result of a very turbulent and randomly-distributed magnetic field internally to the nebula itself, depolarization in the surrounding environment of W50, or due to unassociated magnetoionic material that is located in the foreground. Furthermore, the previous RM measurements of \citet{1986MNRAS.218..393D} used only two data points at different frequencies and were also subject to the well-known $n\pi$-ambiguity. In addition, the limited sensitivity of these observations resulted in gaps where the polarization was too low to calculate a reliable RM across the entire Eastern part of the central region, and several additional large gaps in the Northern, Western, and Southern parts of the shell, and also the Eastern ear. No spectropolarimetry of W50 using new less-ambiguous techniques, such as RM Synthesis \citep{2005A&A...441.1217B}, has ever been performed -- despite considerable technical advances in the handling of such data. The RMs can be determined using the high sensitivity, broadband, and improved resolution observations now available with contemporary instruments such as the Australia Telescope Compact Array (ATCA) -- allowing us to understand the magnetic field geometry within the object, and how this varies across different parts of the nebula. Such a magnetic field study is a useful tool for separating different contributions to the evolution of the object, as distinct patterns are expected to be imprinted on the observed RM in both the case of a \emph{Wind+Jets} \citep{2013ApJ...765...19I,2014ASTRP...1....1I} or a \emph{SNR+Jets} origin \citep{2009IAUS..259...75K,2010ApJ...712.1157H}.

There have been numerous previous studies of the W50 radio nebula at radio wavelengths. \citet{1981A&A...103..277D} and \citet{1986MNRAS.218..393D} observed the total intensity and the polarized emission of W50 at 1.7~GHz, 2.7~GHz, and 5~GHz with the 100-m Effelsberg radio telescope. The brightest polarization was found to be associated with the Northern and Eastern edges at 1.7~GHz, with fractional linear polarizations up to $\sim10$\%. At 2.7~GHz, these areas are $\sim30$\% polarized, with the brightest polarization emanating within the roughly circular region of size $1$~degree centred on SS433, and from the extreme Eastern end of the source. At 5~GHz, the Northern edge is $\sim50$\% polarized, with a magnetic field direction approximately tangential to the shell -- typical of old SNRs \citep{1987AuJPh..40..771M}. This bright Northern edge therefore appears to be a typical example of compression of the large-scale ambient interstellar magnetic field. The region that connects the central region and the Eastern ear, and several other areas within the object were all found to be completely unpolarized. The entire Western ear was also found to be completely unpolarized -- even at a resolution of 30~arcsec \citep{1987AJ.....94.1633E}. What creates this inhomogeneous distribution of polarization across the radio nebula is still unknown, and it could be the result of a very turbulent and randomly-distributed magnetic field internally to the nebula itself, depolarization in the surrounding environment of W50, or due to unassociated magnetoionic material that is located in the foreground. 

Using the Urumqi survey at 5~GHz, \citet{2011A&A...529A.159G} also found strong polarized emission along the northeastern shell of the central part of W50. They derived an integrated spectral index of $\alpha = -0.48\pm0.03$, consistent with that found by \citet{1998AJ....116.1842D}. They find that measurements between 6~cm and 11~cm gave $\alpha = -0.41 \pm 0.01$ for the central part, and $\alpha = -0.52 \pm 0.02$ for the eastern ear. The eastern ear therefore appears to have a slightly steeper spectrum than the central region, consistent with the finding of \citet{1986MNRAS.218..393D}. Meanwhile, interferometer-based studies such as that by \citet{1998AJ....116.1842D} tend to find that the spectra flattens towards the western ear with $\alpha\sim-0.4$ and steepens towards the eastern ear with $\alpha\sim-0.8$. These spectral index variations are likely due to stronger shocks and higher compression to the West where the shock is encountering a denser ISM, and consistent with the location of the Galactic plane. This interpretation is also consistent with the depolarization, which is less in the Eastern part of the source and greater across the rest of the object \citep{1981A&A...103..277D} -- possibly implying variations on smaller angular scales due to increased structure in the magnetic field towards the West. The flattest spectra of all appear to be located in the Southwestern rim, possibly as low as $\alpha\sim-0.3$. In addition, \citet{1998AJ....116.1842D} also find other small features such as a ``chimney'' radiating away from the eastern ear of W50. Many of these features are also still not fully understood.

The spectral index and RM distributions of the very brightest emission were studied in detail by \citet{1986MNRAS.218..393D}, and two-frequency estimates were made of the depolarization and Faraday rotation (albeit subject to $n\pi$-ambiguities, with low sensitivity, and with an angular resolution of 3~arcmin). RMs were found to vary from 0~rad~m$^{-2}$ to as high as 240~rad~m$^{-2}$. The RM in the Eastern part of the source is either $60+376n$~rad~m$^{-2}$ or ($140$--$160$)$+174m$~rad~m$^{-2}$. In addition to $n\pi$-ambiguities, the limited sensitivity of these observations resulted in gaps where the polarization was too low to calculate a reliable RM across the entire Eastern part of the central region, and several additional large gaps in the Northern, Western, and Southern parts of the shell, and the Eastern ear. The RMs of these regions can be determined using the high sensitivity, broadband, and improved resolution observations now available with contemporary instruments such as the ATCA \citep[e.g.][]{2016arXiv160401403A} -- allowing us to understand the magnetic field configuration within the object, and how this varies across different parts of the nebula. Sensitive polarimetry has the potential to separate the extent to which the morphology of W50 is dominated by extrinsic effects or coupling between the jet/remnant. If the ear-like structures to the E \& W do originate due to the jets of SS433, then there should in principle be consistent variations in the spectral index, fractional polarization, and RM in both sides of the source -- allowing us to distinguish between competing scenarios for the remnant's formation such as the \emph{Wind+Jets} and the \emph{SNR+Jets} models, as described in Section~\ref{intro}.

\subsection{Optical Filaments in W50}
As discussed in Section~\ref{intro}, filamentary H$\alpha$ emission associated with large-scale radio nebulae can allow distinction between competing models for an object's formation. An optical filamentary nebulosity has already been identified in the W50/SS433 system \citep{1980ApJ...236L..23V,1980MNRAS.192..731Z}, that is visible in both H$\alpha$ and SII. These bright filaments are located in the eastern and western ears, and are aligned perpendicular to the axis of the jets from SS433. As the relativistic jet from SS433 appears to have broken through the edge of the central region of W50, it is of interest that this bright optical emission lies at these breakout regions and is therefore likely a consequence of the jet activity. These breakout regions have been discussed further by \citet{2007MNRAS.381..308B}. In an ideal case, the breakouts are rings of emission that occur where the expanding jet envelope shocks a dense shell of swept-up and compressed material at the boundary of the circular region of W50. Optical emission should then not be observed throughout the circular region, which is too hot and rarefied, and with the ambient medium being of too lower density in front of the jet. As the jet axis of SS433 is tilted with the eastern edge pointing towards us \citep{1979Natur.279..701A,1979IAUC.3358....1M}, this orientation modifies the rings of optical emission into ellipses. It has been argued that the observed arc morphology is due to heavy, patchy foreground interstellar absorption, so that only the component of the ring nearer to us along the line-of-sight remains visible \citep{2007MNRAS.381..308B}. However, such a mechanism would likely require absorption within the nebula, as purely foreground material should affect both the front and back of the ring. While this overall picture is consistent with the observed arc morphology, it does not account for the known `corkscrew' structure of the jet close to SS433, which is also markedly imprinted onto the radio features of the eastern ear \citep[e.g.][]{2004ASPRv..12....1F}.

Until recently, there was no known optical emission associated with the central region of W50. However, a small rectangular region of $\approx4$~arcmin width and height was observed by \citet{2007MNRAS.381..308B}, who found faint optical emission associated with this central region of W50. The observed region was not large enough to make definitive statements about the formation of the object, or to connect this optical emission with the radio morphology. These results suggest that a deep optical mosaic across the entirety of W50 would help to locate additional optical filaments. Filamentary H$\alpha$ emission tends to be produced by shocks (see Section~\ref{intro}), and has been identified in several SNRs. The \emph{SNR+Jets} model would suggest that there would be faint filaments that cover all of the circular portion of the object, and which due to different pressures would be expected to be considerably fainter than the previously identified, jet-driven, optical emission. Identifying a network of such optical filaments would lend significant support to the \emph{SNR+Jets} model, and future studies would be able to use any new filaments to obtain in-depth information on the kinematics throughout the shell -- thereby allowing an experiment to determine the formation history of W50. 

Such observations are complicated in that they require a large optical mosaic that deals with continuum-contamination, and are also hindered by possible interstellar dust extinction towards the Galactic plane. Nevertheless, many of the observational challenges have already been overcome (see Section~\ref{halpha}). From the H$\alpha$/H$\beta$ ratios, it has also been inferred that there is likely patchy absorption of optical emission across W50 \citep{2007MNRAS.381..308B} -- consistent with the distance and Galactic latitude. While the presence of foreground dust that heavily obscures the optical emission is potentially problematic, both the patchiness and the small piece of faint emission in the interior of W50 suggest there may be new fainter filaments waiting to be discovered.

\section{Data Reduction}
\label{datareduction}
\subsection{Radio Data}
\label{radiodata}
\subsubsection{Observations}
We obtained full spectropolarimetric observations of the W50/SS433 system using the ATCA. We collected six days of data, using three different hybrid array configurations (H75, H168, and H214) which are all compact and well-suited for observing at low declinations, with baseline lengths distributed between 31 to 230~m. We used the CABB system \citep{2011MNRAS.416..832W}, which provided 2~GHz of instantaneous bandwidth centred at 2.1 GHz in 2048 spectral channels. As W50 is above the elevation limit of ATCA for $\sim$9h~49m per day, we used six full-track observations consisting of 66~hours in total which included overheads for standard calibration, full parallactic angle coverage of a leakage calibrator, and antenna slew times (i.e.\ $2\times11$~hours in each hybrid configuration). Our $3^\circ \times 2^\circ$ mosaic consisted of 198 pointings. In order to obtain an approximately uniform sensitivity pattern across W50, and to limit instrumental effects, the pointings were arranged in a hexagonal pattern such that they Nyquist-sample the sky at the high-frequency end of the band. Details of the observations are provided in Table~\ref{table1}.

   \begin{table}
      \caption[]{Details of the ATCA CABB Observations.}
         \label{table1}
     $$ 
         \begin{array}{p{0.5\linewidth}l}
            \hline
            \noalign{\smallskip}
           Parameter      &  \textrm{Value} \\
            \noalign{\smallskip}
            \hline
            \noalign{\smallskip}
            Central Frequecy                & 2.1~\textrm{GHz}     \\
            Bandwidth                         & 2.0~\textrm{GHz}            \\
            No. Channels                         & 2048            \\
            Primary beam FWHM          & 15.1^\prime \\
            Array Configurations           & \textrm{H75, H168, H214}      \\
            \noalign{\smallskip}
            \hline
         \end{array}
     $$ 
    \small{The observations for each configuration were taken on -- H75: 2013 July 05 and 2013 July 06, H168: 2013 August 19 and 2013 August 20, and H214: 2013 September 15 and 2013 September 17. The stated primary beam FWHM is that at the central frequency of 2.1~GHz.}
   \end{table}

In order to measure the large-scale magnetic fields, we only require the detection of diffuse extended linearly polarized emission. This requires only low-angular resolution data that have sensitivity to large angular scales, i.e.\ short baselines. Note that observations of a large Galactic object such as W50 can be well-suited to a radio interferometer, rather than a single dish, as the former naturally filters out the very large-scale Galactic emission that would otherwise require subtraction and would thereby affect our interpretation. The observing strategy here is appropriate for our scientific questions. The hybrid configurations at the ATCA are ideal for this, as they are nominally sensitive to a largest angular scale of $\approx36$~arcmin in mosaiced total intensity, while simulataneously providing a synthesised beam of $59.7$~arcsec to $103.6$~arcsec across the band. Mosaicing is also a standard procedure with the ATCA.

\subsubsection{Flagging \& Calibration}

The excision of narrow band radio frequency interference (RFI) and calibration were carried out using standard techniques in the \textsc{miriad} package. Data were loaded into \textsc{miriad} and flagged for known bad channels and edge effects using the automated flagging routine \textsc{pgflag}. The default settings were found to discard significant quantities of good data, and were adjusted to provide several iterative loops of flagging and calibration using customised parameters -- allowing for lighter flagging in each loop. Due to the compact configurations used, flags were also applied to remove shadowed antennas. During the data reduction, the fixed antenna CA06 was also flagged throughout the observations -- and was thereby not included during the imaging stage -- in order to avoid a very significant gap in the $uv$-coverage.

A standard initial calibration was performed to correct for the effects of the bandpass and the complex antenna gains using the tasks \textsc{mfcal} $\rightarrow$ \textsc{gpcal} $\rightarrow$ \textsc{gpcopy} to transfer solutions from the flux calibrator (PKS1934-638) to the phase calibrator (J1859+129). The frequency-dependent instrumental leakages were then calculated in 16 bins across the full bandwidth using \textsc{gpcal} and the phase calibrator -- which was observed over a large range in parallactic angle. The absolute flux scale was bootstrapped using \textsc{gpboot}, and the final solutions then linearly interpolated across the 198 pointings of the target source (W50). All of the sources (including calibrator scans and target pointings) were then gently flagged (being careful to avoid over-flagging) before iteratively performing a new calibration loop (which provided improved solutions due to the removal of bad data). Once flagging was complete, a final calibration loop was then performed to provide the complete dataset. No self-calibration was performed. Channels at the edges of the bandpass were discarded. Due to the RFI environment, which at ATCA is selectively worse at lower-frequencies, it was found that the usable region of the band extended from 1.4~GHz to 3.1~GHz. Channels outside of this range tended to have very limited $uv$-coverage, particularly on shorter baselines.

\subsubsection{Imaging and Deconvolution}
\label{imaging}
High-quality mosaic imaging of this field of view is complicated, as it contains diffuse extended emission from W50, compact sources embedded in diffuse emission, and large-scale diffuse emission from the Galaxy. The field of view is also at a low declination ($\approx+5^\circ$), has imperfect and limited $uv$-coverage (as shown in Fig.~\ref{figuvcoverage}), and was observed with a broad 2~GHz bandwidth. Two techniques were therefore used for separate wide-field, wide-band imaging of both the total intensity (Stokes $I$) and the linearly polarized intensity (Stokes $Q$ and $U$) emission.

To image Stokes $I$, the Common Astronomy Software Applications (\textsc{casa~v4.2.0}) package was used to process the wide-field, wide-band data, as the package implements the full-mosaic, multi-frequency synthesis algorithms that are now available. The mosaic imaging was not able to simultaneously solve for the spectral index in the current implementation of \textsc{casa}, as mosaics can currently only be processed with \textsc{nterms=1}. However, the incorporation of the large 2~GHz bandwidth substantially improved the $uv$-coverage available for imaging the complex field of view -- albeit with some frequency-dependent artefacts appearing around very bright sources such as SS433. As the mosaic imaging process employed in \textsc{casa} combines the individual pointings in $uv$-space, we are essentially imaging a single $\approx3^{\circ}\times2^{\circ}$ field-of-view; we therefore also used the $w$-projection algorithm with 256 terms. Subtraction during deconvolution using the clean algorithm in \textsc{casa} also takes place in the $uv$-plane, allowing for a more accurate clean, and the Hogbom-type deconvolution was performed slowly with a very-low loop gain, 0.005, in order to avoid confusing emission and sidelobes when extended emission was present. A high cyclefactor of 5 was also used in order to complete major clean-cycles more often. The final image has a resolution of $2.5\times1.9$~arcmin with a position angle of $75.9^{\circ}$.

To create Stokes $Q$ and $U$ image datacubes, the \textsc{miriad} package was used, which is specifically designed to handle data with linear feeds (i.e.\ $XX$, $YY$, $XY$, and $YX$) such as that from the ATCA. Deconvolution of the images from the synthesised dirty beam took place in the image-plane. The data were imaged using multi-frequency synthesis of both 10~MHz and 100~MHz channel-subsets in order to balance maintaining good $uv$-coverage while also minimising bandwidth depolarization. Due to the large observational frequency range, the bandwidth depolarization varies across the band. For the 10~MHz channel dataset, significant bandwidth depolarization\footnote{We define significant bandwidth depolarization as a depolarization factor of 0.85 or less, i.e.\ a source that is polarized at $\le0.85\times \sqrt{Q_{0}^2+U_{0}^2}/I$, where $Q_{0}$/$U_{0}$ are the intrinsic polarization.} is expected for RMs $\ge$1,367~rad~m$^{-2}$ at 1.35~GHz and for RMs $\ge$19,965~rad~m$^{-2}$ at 3.3~GHz. For the 100~MHz channel dataset, significant bandwidth depolarization is expected for RMs $\ge$137~rad~m$^{-2}$ at 1.35~GHz and for RMs $\ge$1,997~rad~m$^{-2}$ at 3.3~GHz. Note that the 100~MHz dataset has improved $uv$-coverage relative to the 10~MHz dataset, whereas the 10~MHz dataset is more sensitive to high RMs. The data were imaged with a robust setting of $-1.0$, giving moderately uniform weights and avoiding excessive weight being given to visibilities in relatively sparsely-filled regions of the $uv$-plane. An image of size 168 pixels was made for each pointing before the mosaic operation was performed, and each image was over-sampled, with a cell-size of 15~arcsec. The images were deconvolved from the synthesised dirty beam using \textsc{mossdi}, and a gain of 0.01. A constant restoring beam was used for all images across the band, using the best angular resolution at the low-frequency end of the band of $3.8\times2.8$~arcmin with a position angle of $72.1^{\circ}$.

\subsubsection{Primary Beam Effects}
\label{PB}
The effects of the primary beam were included in both \textsc{casa} and \textsc{miriad}. The mosaicing process in \textsc{casa} includes the effects of the frequency-dependent total intensity primary beam. In \textsc{miriad}, a frequency-dependent primary beam model (the `ATCA16' model) for the CABB feeds was used. In addition, the Stokes $Q$ and $U$ images are also affected by the off-axis instrumental polarization beam. With CABB, this instrumental leakage has been estimated to be as large as $\sim13$\% at the half-power point \citep{ATmemo,2015ApJ...815...49A}. We here make the critical assumption that this instrumental polarization is oriented radially outwards from the phase-centre so that the altitude--azimuth mounts will cancel out instrumental effects via vector averaging over the large parallactic angle range. The mosaicing procedure will further reduce the wide-field polarimetric aberrations by a factor of at least $\approx2$, and will be aided further as we have closely-packed (better than Nyquist-sampled) mosaic spacings. Wide-field polarization effects are therefore expected to be very small within the central portion of the mosaic, while strong linear polarization at the edges of the mosaic is likely spurious.

\subsubsection{Rotation Measure Synthesis}
\label{RMSynth}
\begin{figure}
   \centering
   \includegraphics[trim=0.5cm 0.3cm 0cm 0cm,clip=true,angle=0,origin=c,width=\hsize]{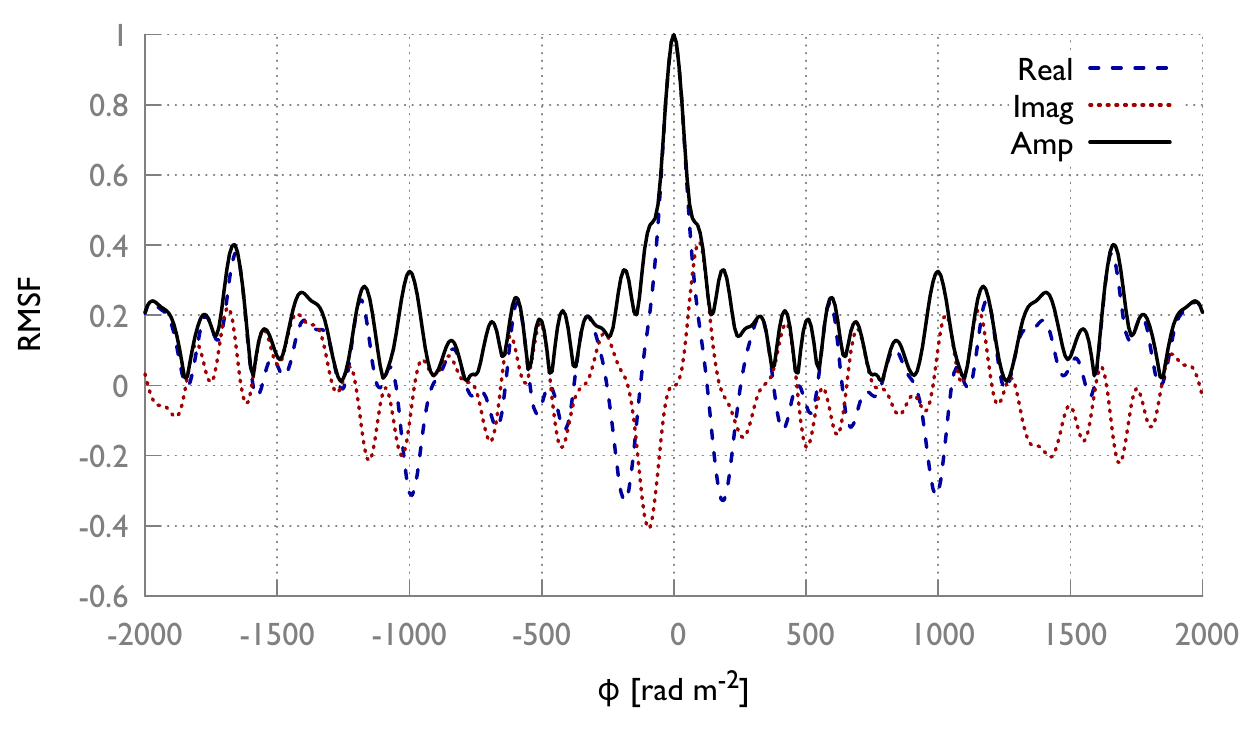}
   \caption{The Rotation Measure Spread Function (RMSF), when using the 100~MHz channel images for our observations of W50. Due to the large bandwidth, and few gaps in the frequency-coverage, there is a well-defined peak with a FWHM of 107.2~rad~m$^{-2}$, and minimal sidelobes within the range $-1000$ to $1000$~rad~m$^{-2}$. Note that the axes display the range $-2000$ to $2000$~rad~m$^{-2}$. The amplitude of the RMSF ($\sqrt{Q^2+U^2}$) is shown as a solid black line, the real component (Stokes $Q$) is shown as a dashed blue line, and the imaginary component (Stokes $U$) is shown as a dotted red line.}
              \label{figRMSF}%
    \end{figure}

For the linear polarization data, we do not need to image the entire band simultaneously into one image -- we only need to produce a Stokes $Q$ and $U$ datacube of narrow-band channels as a function of frequency, so that we can Fourier transform the data and measure the peak Faraday depth using the technique of Rotation Measure Synthesis \citep[see e.g.][for further details]{2005A&A...441.1217B}. The observations allow for RM Synthesis across a bandwidth that to our knowledge is unprecedented for any large radio nebulae. The large bandwidth provides a RM spread function (RMSF; the point spread function in Faraday space) of FWHM 107.2~rad~m$^{-2}$, as shown in Fig.~\ref{figRMSF}. No weights were applied during RM Synthesis. The few gaps in the frequency-coverage also provide low sidelobe levels within the RM range in which we are primarily interested.

We do not deconvolve the data from the RMSF using the technique of RM-clean \citep{2009IAUS..259..591H}. As we are sensitive to different angular scales in our $I$ and $Q$/$U$ images respectively, we cannot calculate accurate polarized fractions or correct for the effects of spectral index (see Section~\ref{limitations}). This prevents accurate RM-deconvolution, but is only an amplitude effect \citep{2005A&A...441.1217B}, i.e.\ our estimates of RMs are unaffected. In addition, the technique of RM-deconvolution has been found to have several limitations and makes many simplifying assumptions \citep{2015AJ....149...60S} -- we therefore avoid all issues associated with deconvolution techniques. 

The full Faraday structure of the remnant can likely only be obtained with $QU$-fitting, to include the effects of depolarization across the band -- although this has not yet been applied to extended objects \citep[e.g.][]{2012MNRAS.421.3300O,2014PASJ...66...65A}. Full $QU$-fitting is beyond the scope of this current work, and would likely be required to fully separate the magnetic field contributions that arise both within the local environment of W50 and along the line-of-sight in any intervening Faraday rotating media. Nevertheless, other properties such as the depolarization and correlations with other structural features of the radio nebula can also allow elucidation of these properties.

Using our data, we are in principle also able to detect Faraday thick structures along the line-of-sight (i.e.\ as $\lambda_{\textrm{min}}^2 < \Delta\lambda^2$). We define a Faraday thick source as those with $\lambda^2\Delta\phi \gg 1$, where the extent of the source in Faraday depth, $\phi$, is given by $\Delta\phi$. Our ATCA data are therefore sensitive to thicknesses $\Delta\phi\ge 106$~rad~m$^{-2}$ at 3.1~GHz, and to $\Delta\phi\ge 21$~rad~m$^{-2}$ at 1.4~GHz. A line-of-sight can therefore appear to be subtly extended in Faraday space, relative to the FWHM of the RMSF$=107.2$~rad~m$^{-2}$, however we cannot \emph{resolve} Faraday thick structures. Furthermore, with a noise level in our Faraday cubes of 0.125~mJy~beam$^{-1}$~rmsf$^{-1}$, this is the most sensitive mosaic of W50 that has been created in linear polarization.

\subsubsection{Data Limitations}
\label{limitations}

\begin{figure}
   \centering
   \includegraphics[trim=6.5cm 2.5cm 9cm 4cm,clip=true,angle=0,width=\hsize]{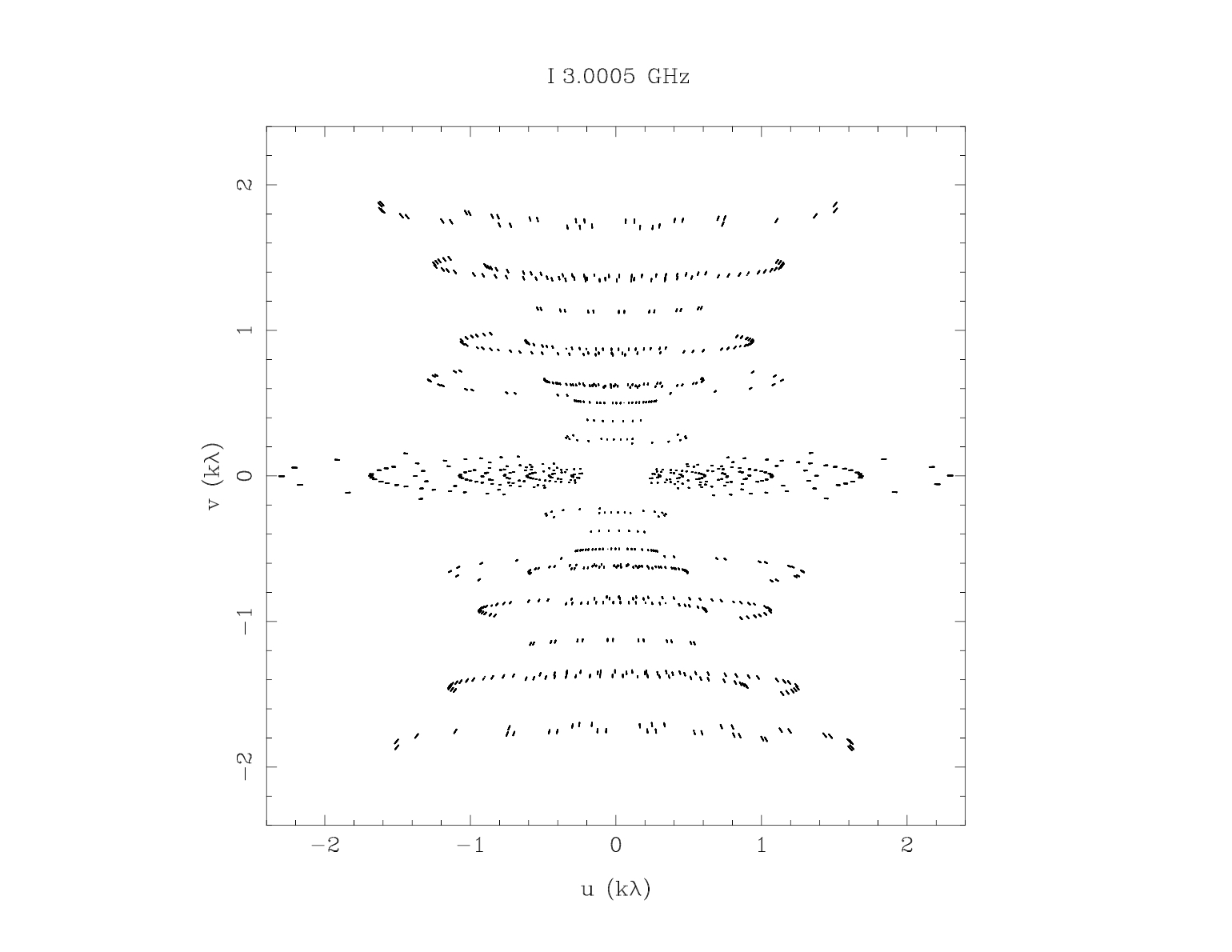}
   \includegraphics[trim=6.5cm 2.5cm 9cm 4cm,clip=true,angle=0,width=\hsize]{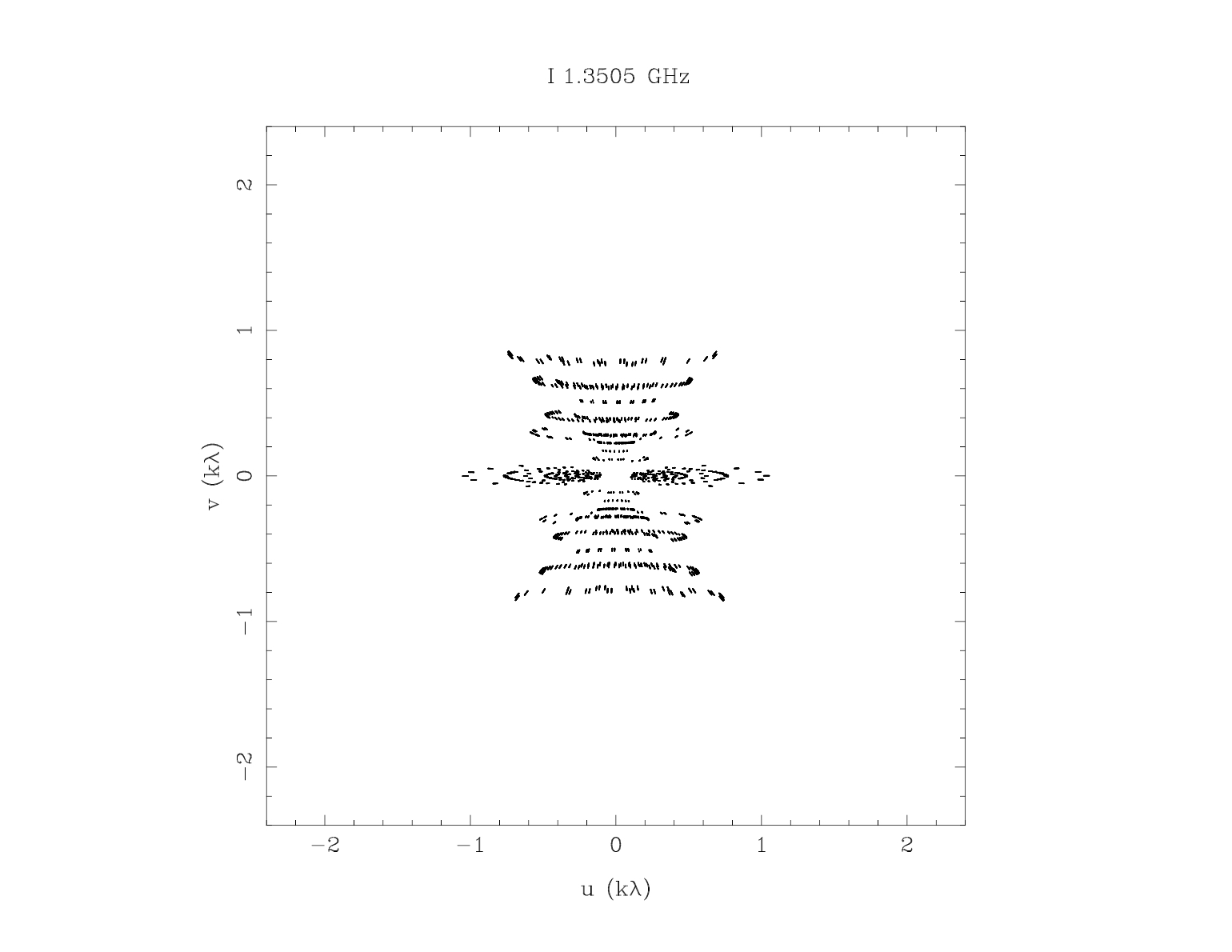}
   \caption{The $uv$-coverage, in k$\lambda$, of a single-pointing from the observations centred at a frequency of 3.00~GHz (top) and 1.35~GHz (bottom), when imaged using 100~MHz channels. The $uv$-coverage varies substantially across the band, providing more sensitivity to large-scale extended emission and less missing flux density at the low-frequency end of the band.}
              \label{figuvcoverage}%
    \end{figure}

Due to the large observational bandwidth, there is substantial variation in the $uv$-coverage across the band, as shown in Fig.~\ref{figuvcoverage}. This is problematic for consistent imaging of the very extended emission in the total intensity images. As shorter $uv$-spacings are better sampled at lower observational frequencies, the extended emission is both better reconstructed and has less missing flux at these low frequencies. This tends to provide overly steep spectral indices that vary with position in the image depending on the scale of the emission. This overestimation of a steep spectral index due to frequency-dependent $uv$-coverage is an inherent property of ultra broadband observations that is potentially generally overlooked. Due to the limited number of baselines, attempts at imaging with a constant $uv$-range across the band produced artefacts in the final images. In addition, the capability to perform both mosaic imaging and to simultaneously obtain a spectral index is not yet available (see Section~\ref{imaging}), and our data cannot therefore be easily used to recover the spectral index. The recovery of spectral indices is particularly difficult for this field of view as the western edge of the nebula dips into the Galactic plane, with increased diffuse background emission. This background emission changes across the band, along with DC offsets in the images, and also causes low-level frequency-dependent changes in the antenna system temperature. In addition, inaccuracies in the primary beam model and its frequency dependence can also affect the spectral index -- although the closely-packed mosaic and the applied primary beam model should minimise such an effect. Furthermore, there is also increased RFI at the low-frequency end of the CABB band -- which selectively tends to have a larger effect on shorter baselines. The flagging of these RFI affected data can therefore tend to remove larger scale structures at the low-frequency end of the band. The ability to use techniques such as $T$--$T$ plots \citep[e.g.][]{1962MNRAS.124..297T,1990AJ....100.1927G,1993ApJ...408..514A} and Spectral Tomography \citep{2000ApJ...529..453K} is limited as we have multiple narrow channels across our band. Attempts to reimage our data as two separate channels and to apply these techniques did not provide consistent results as the $uv$-coverage still varies significantly. Additional observations to more fully sample shorter spacings in the $uv$-plane would assist in recovering all of the extended total intensity emission across the band -- however another facility is likely required, as with the ATCA hybrid arrays we are essentially sampling the shortest spacings that the interferometer can provide (although RFI preferentially affects the very shortest baselines).

These issues also prevent us from calculating reliable polarization fractions, $\Pi = P/I = \sqrt{Q^2+U^2}/I$, since different scales, negative bowls, and background emission are retrieved in both the $P$ and $I$ images. Fortunately, these effects are much less significant for our polarization images, as the emission is only ever a fraction of the total intensity emission, and is also broken up into much smaller scale structures -- with the polarized emission being separated into individual $Q$ and $U$ components, and also being broken up in the plane of the sky due to polarization angle variations (and Faraday rotation). Since we use RM Synthesis to recover the peak in the Faraday spectra, any such residual amplitude effect will not affect the derived RMs and will only affect the measured polarized intensity at a given RM, i.e.\ it is only an amplitude effect in Faraday space. In cases of multiple Faraday components, the highest amplitude component could feasibly be misidentified, although we have not identified such multiple components in our data. We can therefore obtain reliable RMs despite variations in $uv$-coverage across the band and without having to assume a spectral index during RM Synthesis.

\subsection{IPHAS H$\alpha$ Data}
\label{halpha}
\subsubsection{The Observations}
We created a large continuum-corrected H$\alpha$ mosaic covering the full $2.5^{\circ} \times 1.5^{\circ}$ region surrounding W50/SS433, using data release 2 (DR2) observations from IPHAS \citep{2005MNRAS.362..753D,2014MNRAS.444.3230B}.

IPHAS is a 1800~deg$^2$ CCD survey of the entire northern Galactic plane in the latitude range $-5^{\circ}$ < $b$ < $+5^{\circ}$. The data obtained are Wide Field Camera images in the H$\alpha$ narrow-band, and Sloan $r^{\prime}$ and $i^{\prime}$ broad-band filters. The survey is particularly well-suited to studies of spatially resolved nebulae, which indicate ionisation of circumstellar gas, as demonstrated by previous studies of a continuum-subtracted mosaic image of Shajn~147 (S147), a $3^\circ$ diameter supernova remnant \citep{2005MNRAS.362..753D}. The observations were carried out between 2003 and 2012 with a median seeing of 1.1~arcsec (sampled at 0.33~arcsec~pixel$^{-1}$) and to a mean $5\sigma$ depth of 21.2 ($r$), 20.0 ($i$), and 20.3 (H$\alpha$) in the Vega magnitude system \citep{2014MNRAS.444.3230B}. 

Imaging of extended nebulae requires a well-behaved background on the CCDs, and the limitations of such a technique are largely due to background variations. The sky generally subtracts fairly well in an H$\alpha - r^\prime$ image, albeit with a few imaging artefacts that can occur on occasion \citep{2005MNRAS.362..753D}.

\subsubsection{Mosaicing}
To create the mosaic, CCD images were initially downloaded from the IPHAS data repository, alongside the corresponding confidence maps of each field. The confidence maps were used to create an image mask, and to filter out confidence values below 0.9. This tended to remove bright corner regions that were present in CCD3, bad pixels streaks across the images, and also removed some additional bad data around the CCD edges.

The images were all then individually regridded into the `SIN' world-coordinate system projection corresponding to that used for the ATCA images. Large mosaics of H$\alpha$ and $r^\prime$ were then made using \textsc{montage} \footnote{\url{http://montage.ipac.caltech.edu/}}. Using this software, the initial mosaic images were corrected for background differences between the individual images in order to produce a final mosaic of the sky.

We are interested in imaging both the fine filamentary $H\alpha$ features, as well as the diffuse emission. One challenge associated with this is the uneven illumination of the CCDs, which is partially due to moonlight scattering off the inside of the telescope dome onto the CCDs (N.~Wright, priv. comm.). This is only at the level of a few counts in each image, and so is not visible in single-pointing mosaics, but can dominate the emission at larger scales. It was found that by limiting the size of the mosaic to $4999 \times 4999$ pixels, that the features in the images were still clearly visible with no obvious large-scale background variation. Attempts to increase the mosaic size beyond this would likely require an alternative approach to remove any gradient, but this was not necessary to cover the angular extent of W50.

As we are only interested in the morphology of optical emission associated with W50, rather than attempting to obtain any quantitative flux measurements, we were able to provide a continuum-correction by subtracting the median image value individually from the H$\alpha$ and $r^\prime$ mosaics, and then creating a final image of H$\alpha/r^\prime$. In combination with the lack of gradient removal, this method has the advantage that it also highlights the diffuse emission -- rather than subtracting it. The result is a continuum-corrected mosaic covering the entire W50/SS433 system. We are missing part of the shell from DR2 of IPHAS, due to a few missing pointings, but these will likely be covered by future data releases.

\section{Results}
\label{w50}

\subsection{Radio Continuum}

\begin{figure*}
   \centering
   \includegraphics[totalheight=0.55\textwidth]{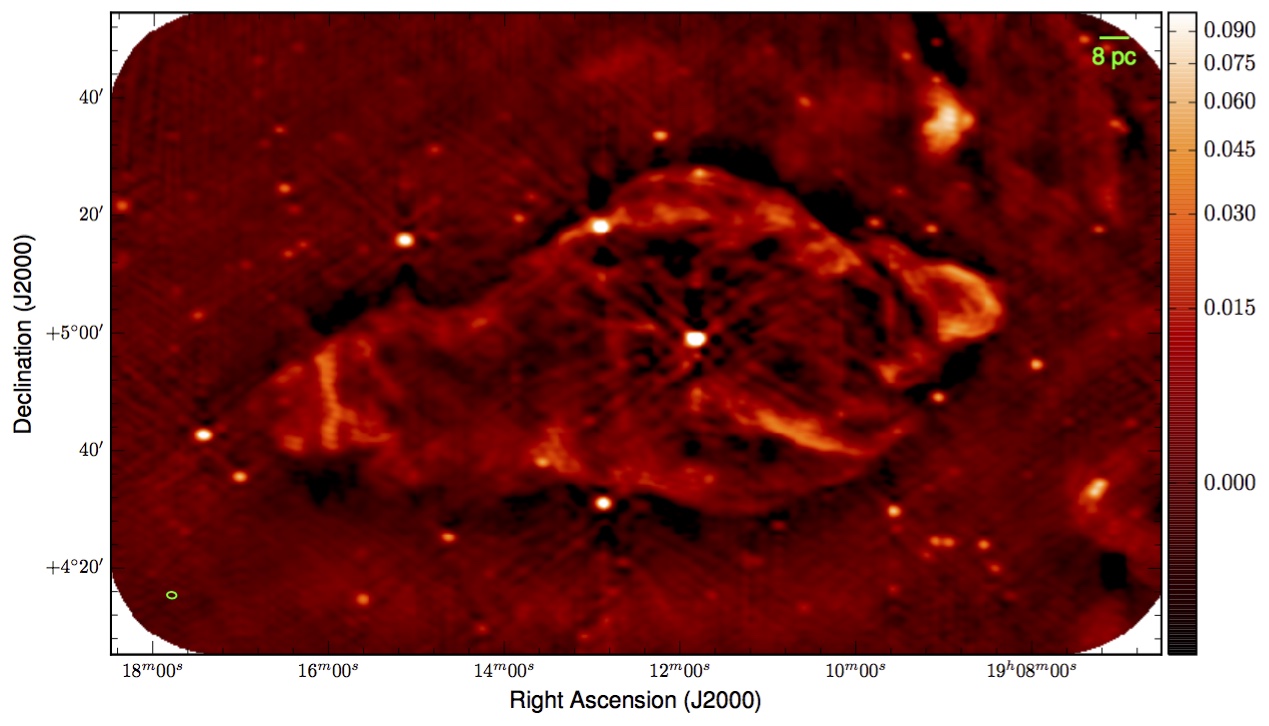}
   \caption{A wide-field, wide-band, mosaiced, multi-frequency synthesis image of the total intensity emission (Stokes $I$) from W50/SS433. The image combines data over the frequency range from 1.4~GHz to 3.1~GHz, with an effective bandwidth of 1.7~GHz. There are artefacts around bright point sources due to the assumption of zero spectral index during simultaneous imaging. The surrounding field of view contains several bright extragalactic radio sources, the nearby HII region S74 to the NW, and one half of the shell of the Galactic supernova remnant candidate G38.7--1.4 to the South West. The bright compact source and microquasar, SS433, is clearly visible at the approximate geometric centre of the nebula. The pseudo-colour scale is in units of jansky per beam. The synthesised beam is shown in green to the lower left. A physical scale of the remnant in parsec is shown in the upper right, assuming a distance to SS433 of 5.5~kpc.}
              \label{fig1}%
    \end{figure*}

A wide-field, wide-band, multi-frequency synthesis mosaiced image of the total intensity emission (Stokes $I$) of the W50/SS433 system is shown in Fig.~\ref{fig1} as observed using the ATCA. The image covers the frequency range from 1.4~GHz to 3.1~GHz, with an effective bandwidth of 1.7~GHz. There are artefacts around bright point sources due to the assumption of zero spectral index during simultaneous imaging (see Section~\ref{imaging}). 

The surrounding field of view contains several bright extragalactic radio sources, the nearby HII region S74 to the northwest, and one half of the shell of the Galactic supernova remnant candidate G38.7--1.4 to the southwest. The bright compact source and microquasar, SS433, is clearly visible at the approximate geometric centre of the nebula. There is emission throughout the field-of-view, including diffuse extended emission from the nebula itself, and also background emission from the Galaxy which is brighter towards the west. As seen in projection, there are some compact sources embedded in the extended emission associated with W50, although these are believed to be extragalactic and therefore in the background of the object itself.

The extended emission associated with W50 forms an approximately circular ring that is equidistant from SS433, giving an appearance consistent with that of a shell-type SNR. To the east and west are two `ears' that appear to have been `punched' through the shell by pressure from the jets of SS433 itself (see Section~\ref{intro}). There are several small-scale filamentary structures that appear to be associated with these ears, including two filaments that are oriented approximately along north--south in either ear (roughly perpendicular to the presumed impact angle of the jets). The eastern ear appears to extend much further from SS433 than the western ear, likely due to a denser environment to the west. At the very faintest levels of the image, an additional feature that corresponds to the ``chimney'' seen in previous studies \citep{1998AJ....116.1842D} is visible as originating from the north--south oriented filament in the eastern ear and extends $\approx10$~arcmin towards the north. This has previously been interpreted as a Rayleigh--Taylor instability originating from the material associated with the ears accelerating into a denser ambient medium. 

A potentially new second chimney is seen $\sim 20$~arcmin to the west of the first chimney, also located in the eastern ear, and also extending towards the north -- in the direction of a bright compact background source. When using the ATCA to observe at a declination of $+5^{\circ}$, the $uv$-coverage is sparse in the north--south direction even when using the hybrid arrays. This creates an asymmetric dirty beam and can cause low-level cleaning artefacts that occur around bright compact sources -- the feature is therefore likely spurious.

Although limited by the imaging artefacts, a significant number of radio filaments are visible. One bright filament extends from near to SS433 and curves down towards the southwest of the shell, terminating below the western ear. Other faint filaments are present to the east of SS433, with one filament extending away from SS433 towards the east, at a slight angle ($\approx5^{\circ}$) to the east--west axis. Another filament extends inwards from the northeast of the shell, towards the central northern region of the shell. Numerous other radio filaments are also present in the eastern and western ears, all of which have been identified in previous studies \citep{1998AJ....116.1842D}. 

\subsection{Linear Radio Polarization}
\label{linpol}
\begin{figure*}
   \centering
   \includegraphics[trim=0cm 0cm 0.5cm 5cm,clip=true,angle=90,origin=c,totalheight=0.95\textheight]{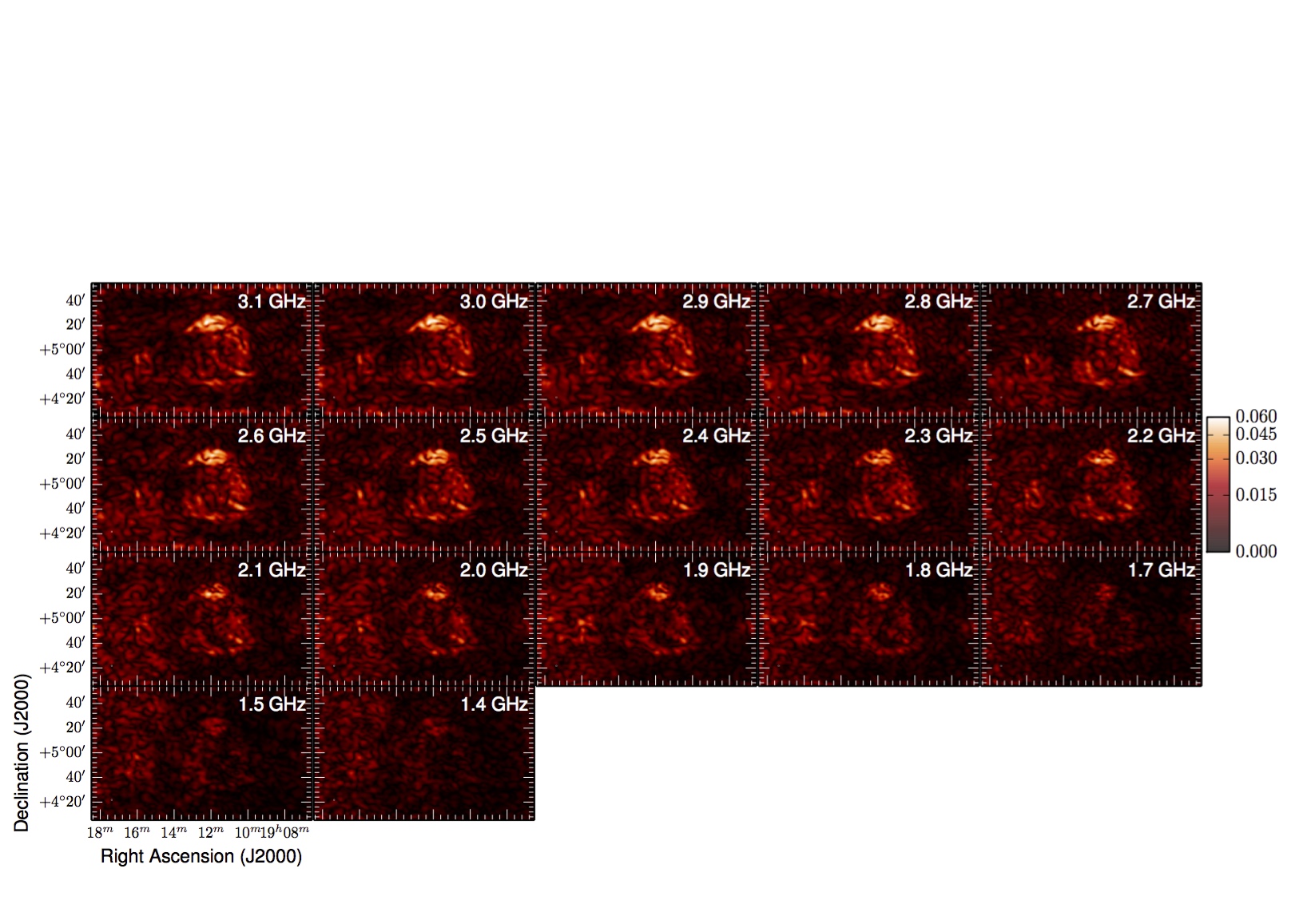}
   \caption{ATCA images of the polarized intensity in W50 as a function of frequency. Images are shown at frequencies from 3.1~GHz to 1.4~GHz, with each image having a 100~MHz bandwidth. The 1.6~GHz image was of low-quality due to RFI, and is not shown. The pseudo-colour scale is in units of jansky per beam, and is held fixed for all images. There is strong linear polarization associated with the central region of W50 and the eastern ear. The emission is clearly depolarizing at lower radio frequencies.}
              \label{fig2}%
    \end{figure*}

The linearly polarized intensity images of the mosaic surrounding W50 are shown in Fig.~\ref{fig2} at frequencies from 1.4~GHz to 3.1~GHz, with each image having a 100~MHz bandwidth. There is clear structure in the polarization towards the interior of the mosaic, which is spatially uncorrelated with total intensity. This indicates that it is real, rather than instrumental polarization. There is also apparent polarization towards the very outer edges of the mosaic, which is consistent with our assumptions about the off-axis instrumental polarization in Section~\ref{imaging}.

Using the sensitivity of our ATCA observations, we are able to see a `ring' of linearly polarized emission surrounding SS433 and corresponding to the shell of the central region of W50. The brightest polarized emission in the north and east of the central ring has been identified previously \citep{1981A&A...103..277D,1986MNRAS.218..393D}. Strong polarization is also associated with a radio filament that extends from SS433 down towards the south-west of the central region of W50.

In general, there appears to be a divide between the polarization in the east and the west of the image. There is clear linear polarization associated with the Eastern ear of W50, while there is no polarization associated with the Western ear. Polarized emission that is presumably Galactic in origin, and unrelated to W50 itself, is also visible across the entire eastern half of the image, while there is conversely almost no polarization in the western half. There is a polarized feature at the most western part of the image that appears as a faint semi-circle of polarized emission of $\approx10$~arcmin radius with an additional extension to the north (with any additional morphology located outside of the FOV). Even this faint polarized emission is at the $20\sigma$ level. All of this feature is located further to the west than the ear, and again appears to be Galactic polarization. No linear polarization was detected from either the nearby HII region S74 or the supernova remnant candidate G38.7--1.4.

All of the visible polarized emission exhibits Faraday depolarization, with the polarization having decreased significantly at 1.4~GHz compared to at 3.1~GHz.

\subsection{Depolarization and Diffuse H$\alpha$ Emission}
\label{depol}
In Section~\ref{linpol}, we were able to see a `ring' of linearly polarized emission surrounding SS433 and corresponding to the shell of the central region of W50. However, this does not take depolarization effects into account (see Section~\ref{depol}). It is not obvious whether the rest of the object should be linearly polarized at these observational frequencies, or whether Faraday effects could be expected to have depolarized the emission elsewhere. It is known that Faraday rotation is related to the presence of thermal electrons which can be traced by H$\alpha$ emission.

\begin{figure*}
   \centering
   \includegraphics[trim=3.3cm 5cm 2cm 5.75cm,clip=true,angle=0,origin=c,totalheight=0.55\textwidth]{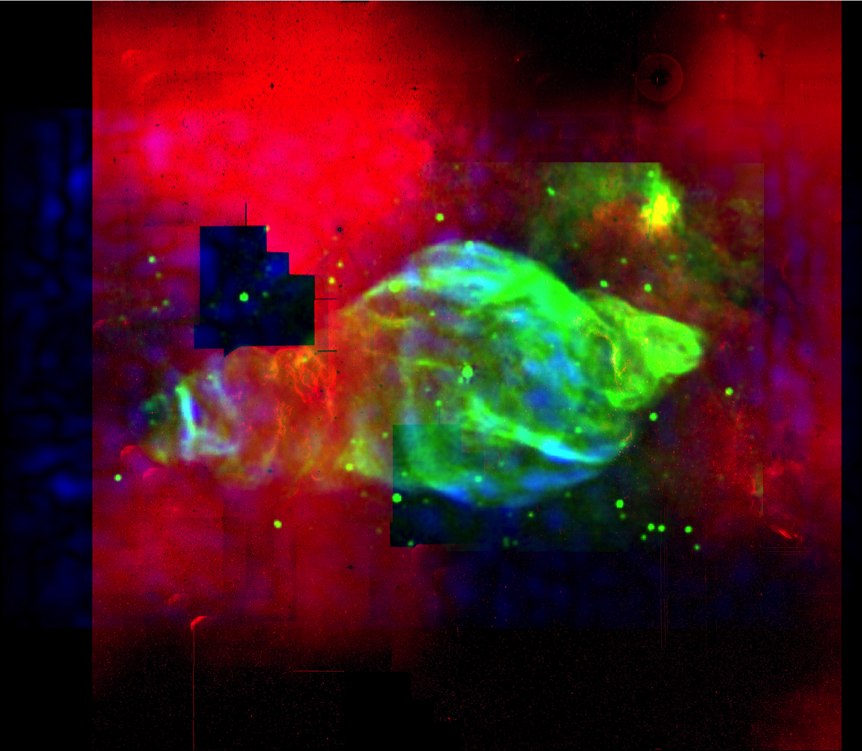}
   \includegraphics[trim=3.3cm 5cm 2cm 5.75cm,clip=true,angle=0,origin=c,totalheight=0.55\textwidth]{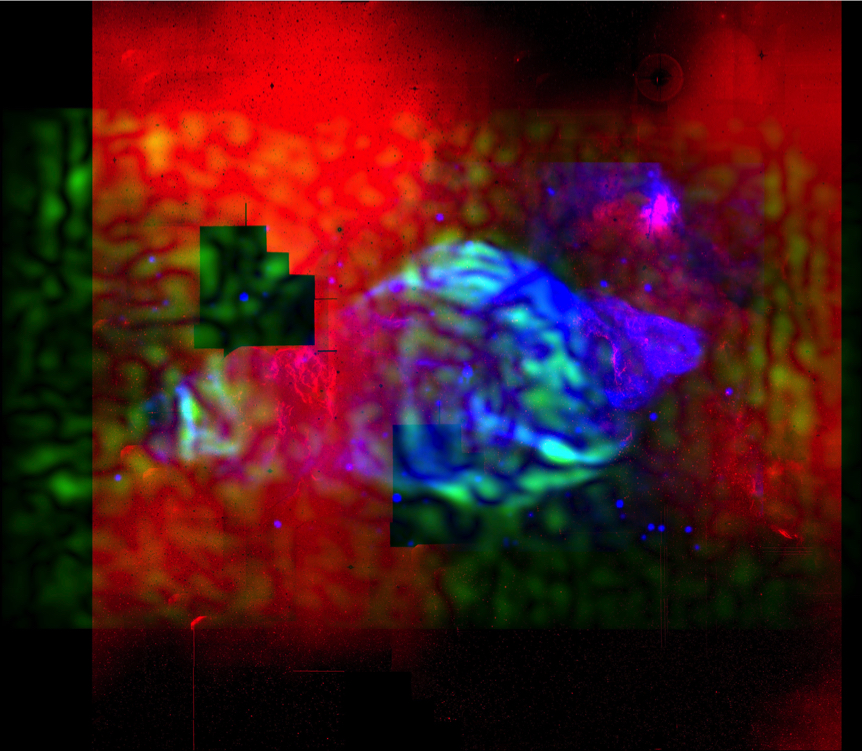}
   \caption{Top: the IPHAS H$\alpha$ image (red), the Very Large Array (VLA) \citet{1998AJ....116.1842D} radio continuum image at 1.4~GHz (green), and the ATCA polarized intensity at a low Faraday depth of 0~rad~m$^{-2}$ (blue). Bottom: a similar image to the above, with the green and blue images swapped in order to highlight the polarized intensity (green). Note the very strong anti-correlation between the polarized intensity and the diffuse H$\alpha$ emission. An additional view is shown in Fig.~\ref{figrgb2}.}
              \label{figrgb}%
    \end{figure*}

\begin{figure*}
   \centering
   \includegraphics[trim=0cm 0cm 0cm 0cm,clip=false,angle=0,origin=c,totalheight=0.62\textwidth]{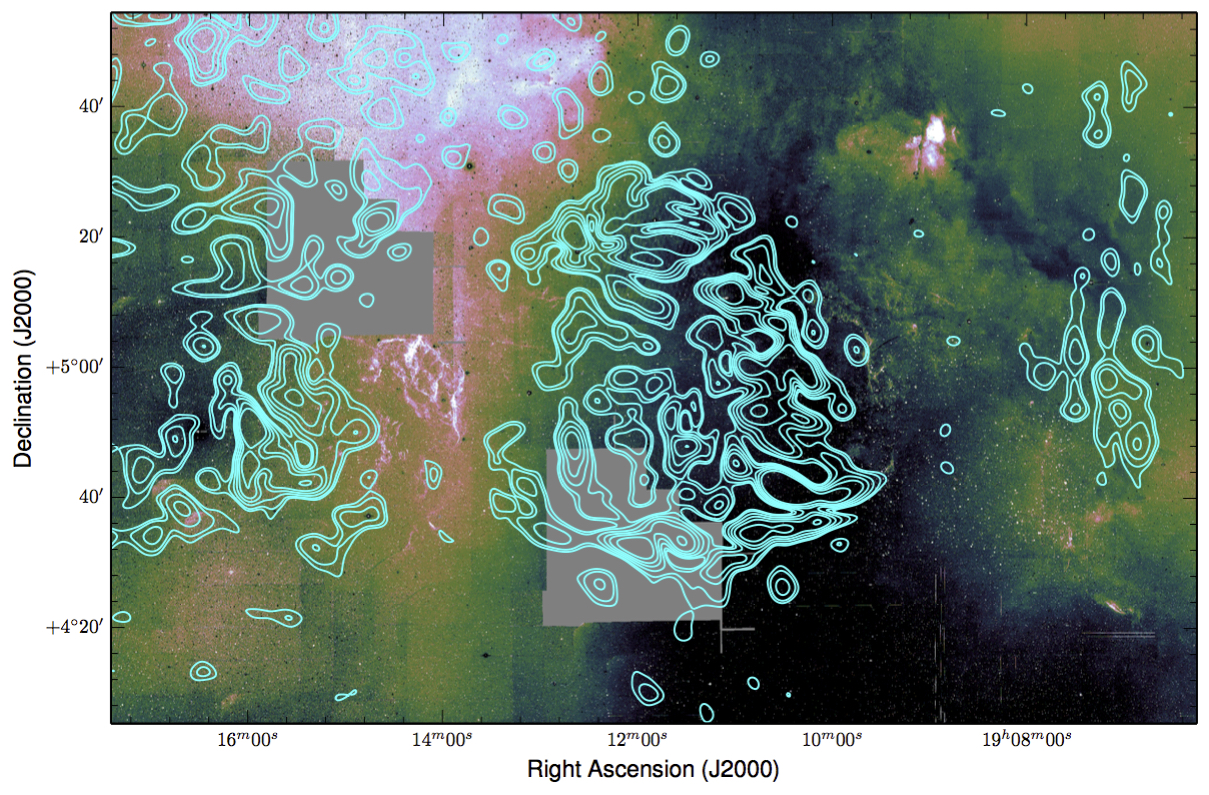}
   \caption{An additional perspective on the anti-correlation between the polarized intensity and the diffuse H$\alpha$ emission, as shown in Fig.~\ref{figrgb}. The IPHAS H$\alpha$ image (background pseudocolour image), and the ATCA polarized intensity at a low Faraday depth of 0~rad~m$^{-2}$ (cyan contours) are shown. Two large H$\alpha$ filaments can be seen running approximately north--south across the face of W50, and correspond to areas with no polarization from W50.}
              \label{figrgb2}%
    \end{figure*}

Combined red--green--blue images of W50, showing the radio total intensity, polarized intensity at a low Faraday depth (see Section~\ref{faradayrotation}), and the H$\alpha$ emission are shown in Fig.~\ref{figrgb}. The Figure shows a strong correlation between bright diffuse H$\alpha$ and complete depolarization of the region surrounding W50, or conversely there is an anticorrelation between bright H$\alpha$ and linearly polarized intensity. This anticorrelation is highlighted further in Fig.~\ref{figrgb2}, which overlays polarized intensity contours at a low Faraday depth over the H$\alpha$ image. In the presence of the diffuse H$\alpha$, the polarized emission from W50 itself, and also some of the Galactic polarized emission, both exhibit complete depolarization. 

\begin{figure}
   \centering
   \includegraphics[width=\hsize]{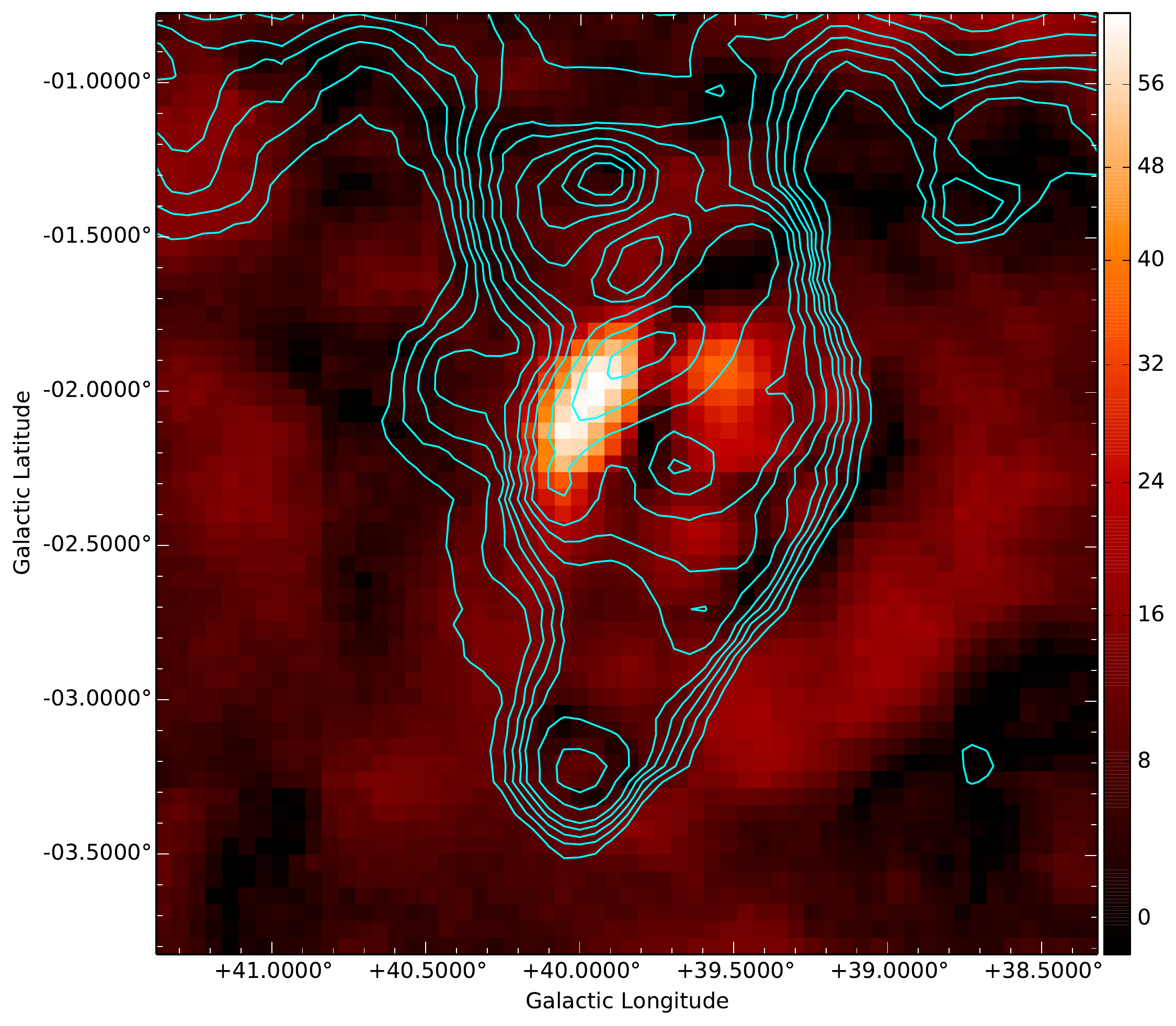}
   \caption{The single dish Urumqi image of the W50/SS433 system at 5~GHz, as observed by \citet{2011A&A...529A.159G}. The image is in Galactic coordinates to highlight the proximity to the Galactic plane, and is in units of kelvin. The pseudocolour image shows the polarized intensity, overlaid with cyan total intensity contours.}
              \label{fig7}%
    \end{figure}

The type of depolarization can be influenced if (i) the synchrotron-emitting and thermal medium are mixed, or (ii) the thermal medium is in the foreground of the nebula. In the former case, we expect broad Faraday thick features in our Faraday spectra. In the latter case, we expect narrow Faraday thin peaks. Our Faraday spectra are almost completely dominated by a single Faraday thin peak, with the sole exception of a slight indication of two Faraday components from the southern part of the polarized emission associated with the Eastern ear (see Section~\ref{discussion}). A measurement of the FWHM of the strongest Faraday components, provides values between $=107$--$109$~rad~m$^{-2}$, consistent with our RMSF$=107.2$~rad~m$^{-2}$ (see Section~\ref{RMSynth}). An extra constraint on the depolarization can be obtained by comparing our data with the 5~GHz observations of \citet{2011A&A...529A.159G}. These data are shown in Fig.~\ref{fig7}. Even at higher radio frequencies of 5~GHz, there is clearly polarization associated with the central region of W50, and complete depolarization in the regions in which we have identified diffuse H$\alpha$ emission.

\subsection{Filamentary H$\alpha$ Emission}
\begin{figure*}
   \centering
   \includegraphics[trim=0cm 0cm 0cm 0cm,clip=true,angle=0,origin=c,totalheight=0.65\textwidth]{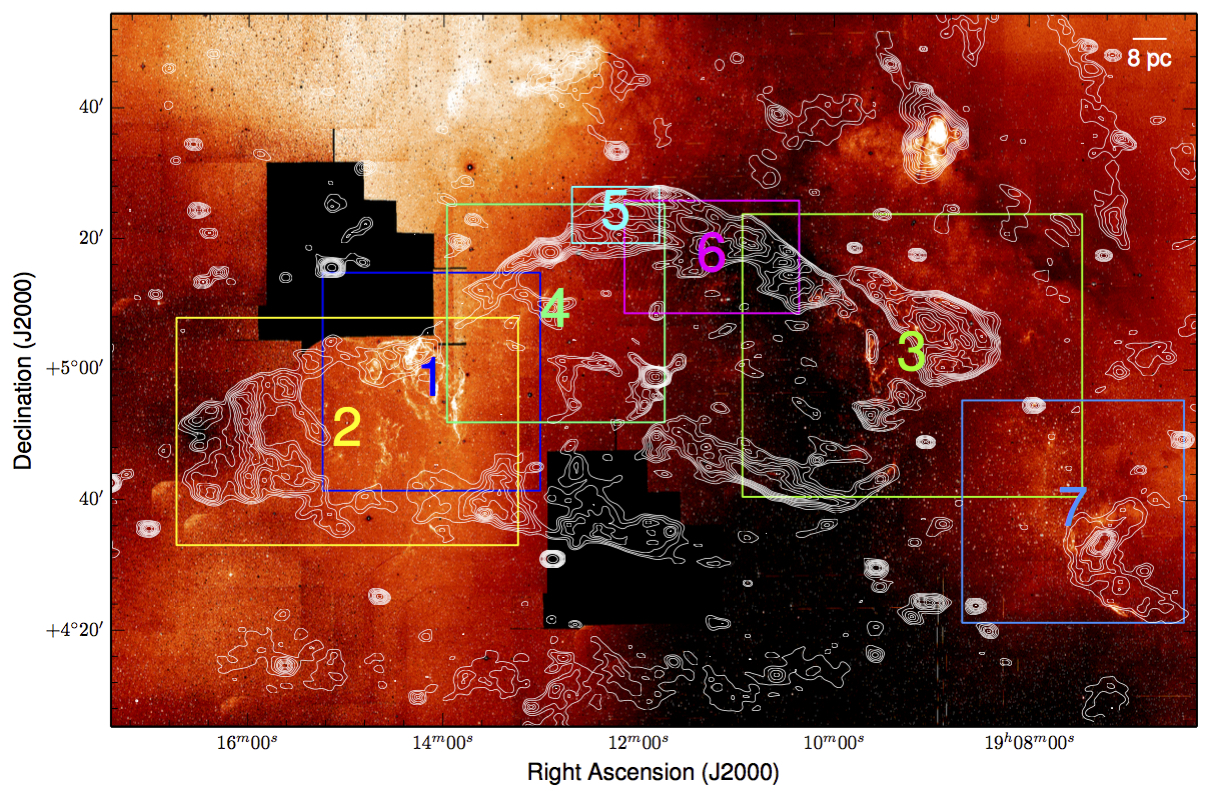}
   \caption{An image of the IPHAS continuum-corrected H$\alpha$, overlaid with ATCA radio continuum contours shown in white. The multi-coloured inset regions indicate the areas shown in Figs.~\ref{figreg1}--\ref{figreg7}.}
              \label{figregions}%
    \end{figure*}

We use the large continuum-corrected H$\alpha$ mosaic from IPHAS, as described in Section~\ref{halpha}. We use this to investigate the connection between the optical emission and both the total intensity and linearly polarized radio counterparts, as described in Section~\ref{radiodata}. The complete IPHAS H$\alpha$ mosaic, overlaid with ATCA total intensity contours, is shown in Fig.~\ref{figregions}. The discrete rectangular regions used are shown alongside their respective numerical labels from 1 to 7. To allow the reader to skip the qualitative descriptions, we provide additional detail of the morphology and features of each of these regions within Appendices~\ref{1stregion} to \ref{7thregion}. We identify new previously-undiscovered filamentary structures that are associated with the W50/SS433 system, and which are far more extensive than has been previously reported. Most of these have radio continuum counterparts, and in a few cases there is a possible correspondence to the polarized emission.

\begin{figure*}
   \centering
   \includegraphics[trim=0cm 0cm 0cm 0cm,clip=true,angle=0,origin=c,totalheight=0.65\textwidth]{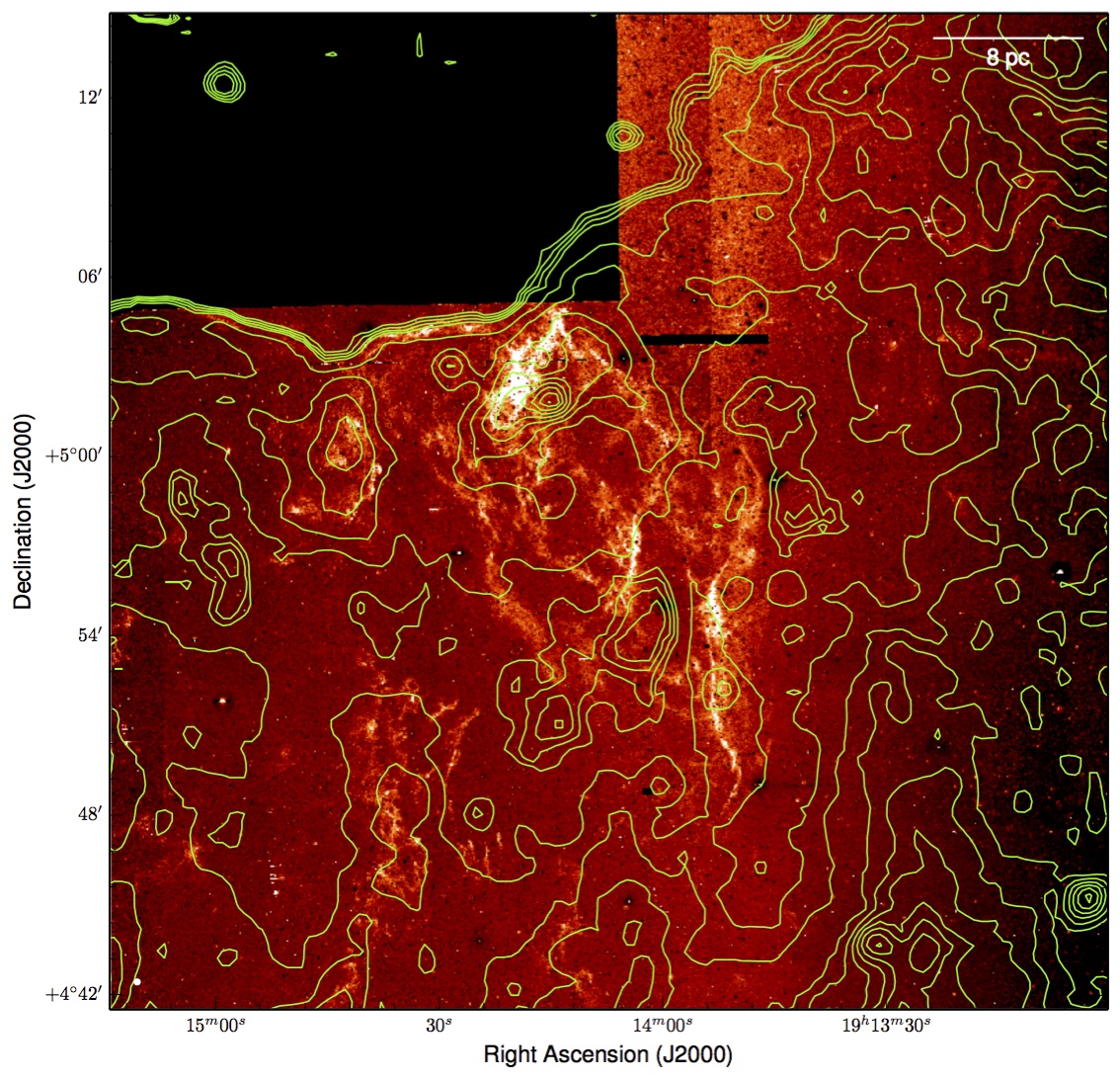}
   \includegraphics[trim=0cm 0cm 0cm 0cm,clip=true,angle=0,origin=c,totalheight=0.65\textwidth]{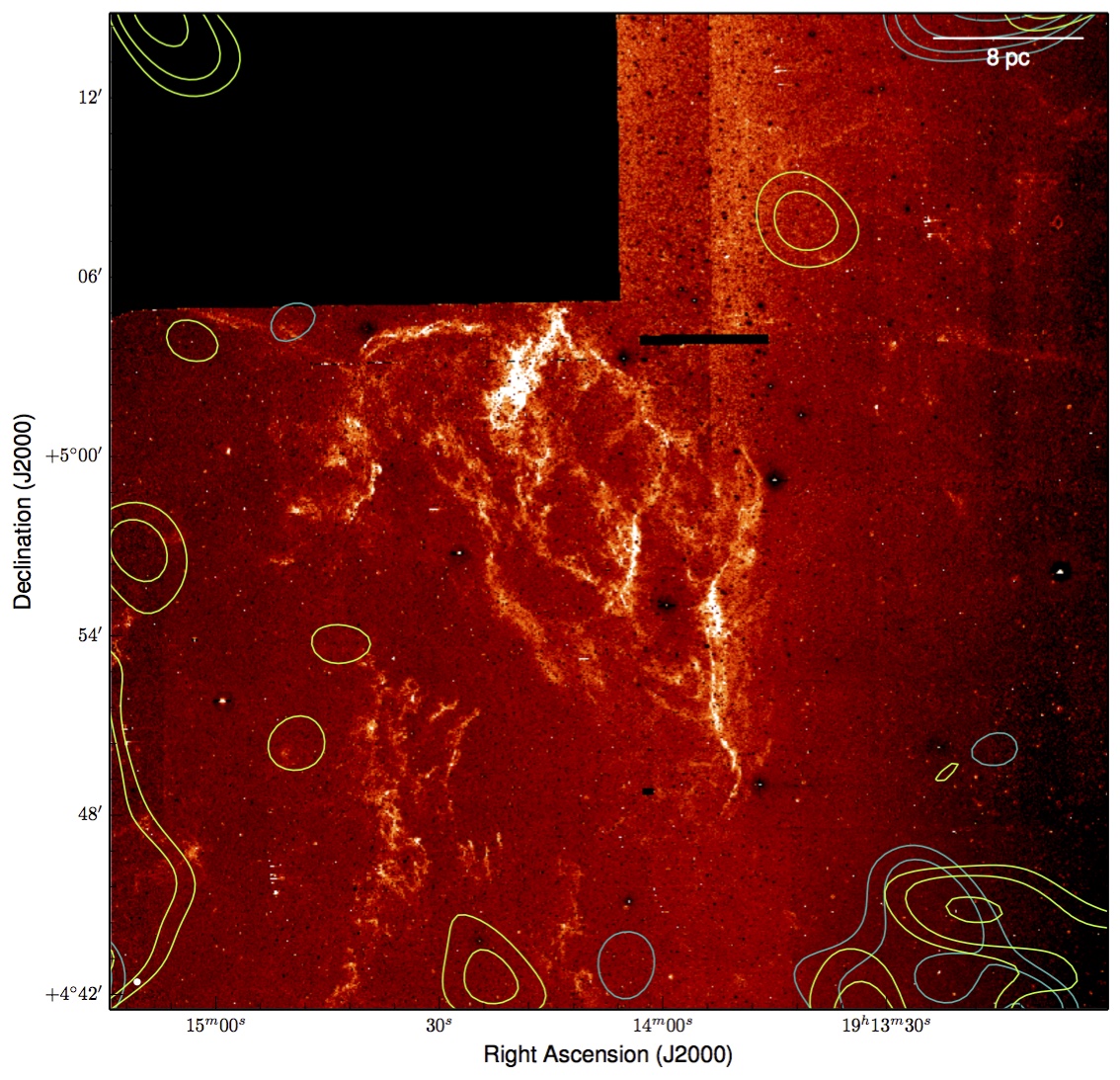}
   \caption{Top: The VLA contours (bright green; \citet{1998AJ....116.1842D}) overlaid on IPHAS continuum-corrected H$\alpha$ for Region 1. Bottom: Similar to top, but instead showing ATCA contours of the 3.1~GHz (dark green) and 2.2~GHz (light green) polarized intensity within a 100~MHz band.}
              \label{figreg1}%
    \end{figure*}

\begin{figure*}
   \centering
   \includegraphics[trim=0cm 0cm 0cm 0cm,clip=true,angle=0,origin=c,totalheight=0.65\textwidth]{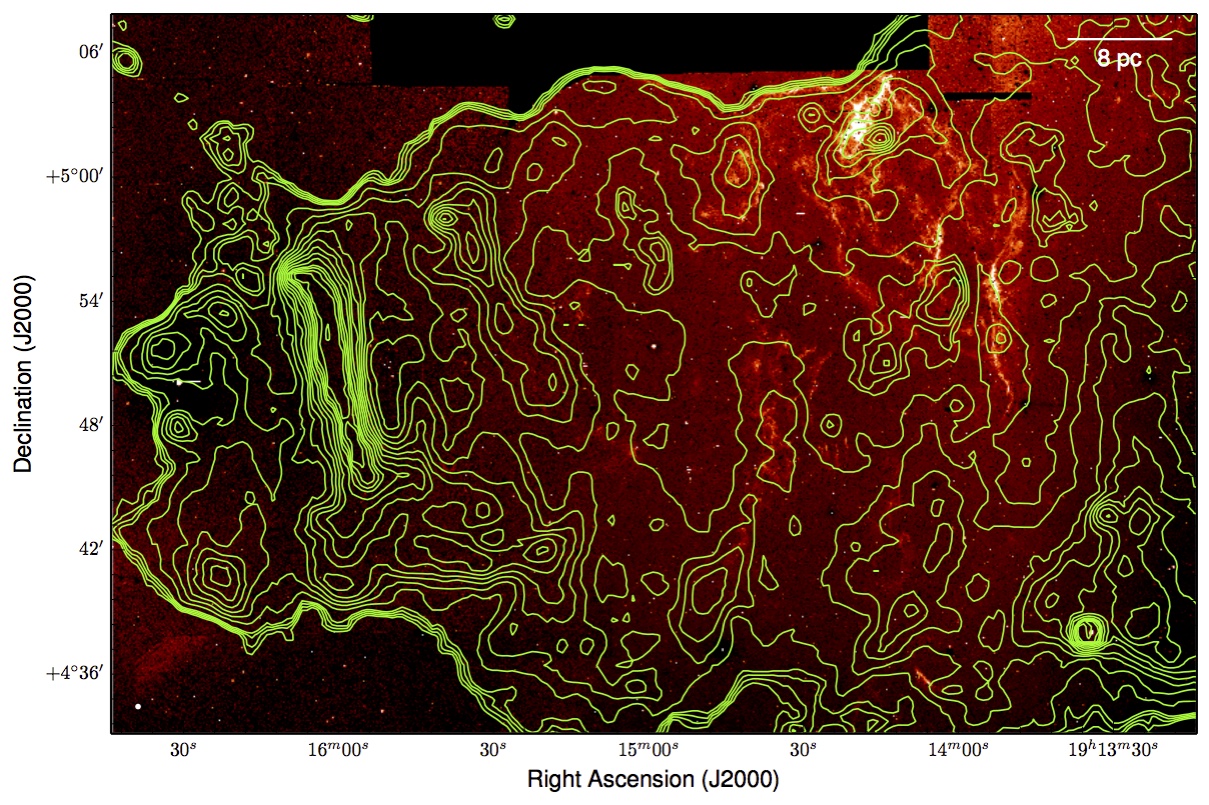}
   \includegraphics[trim=0cm 0cm 0cm 0cm,clip=true,angle=0,origin=c,totalheight=0.65\textwidth]{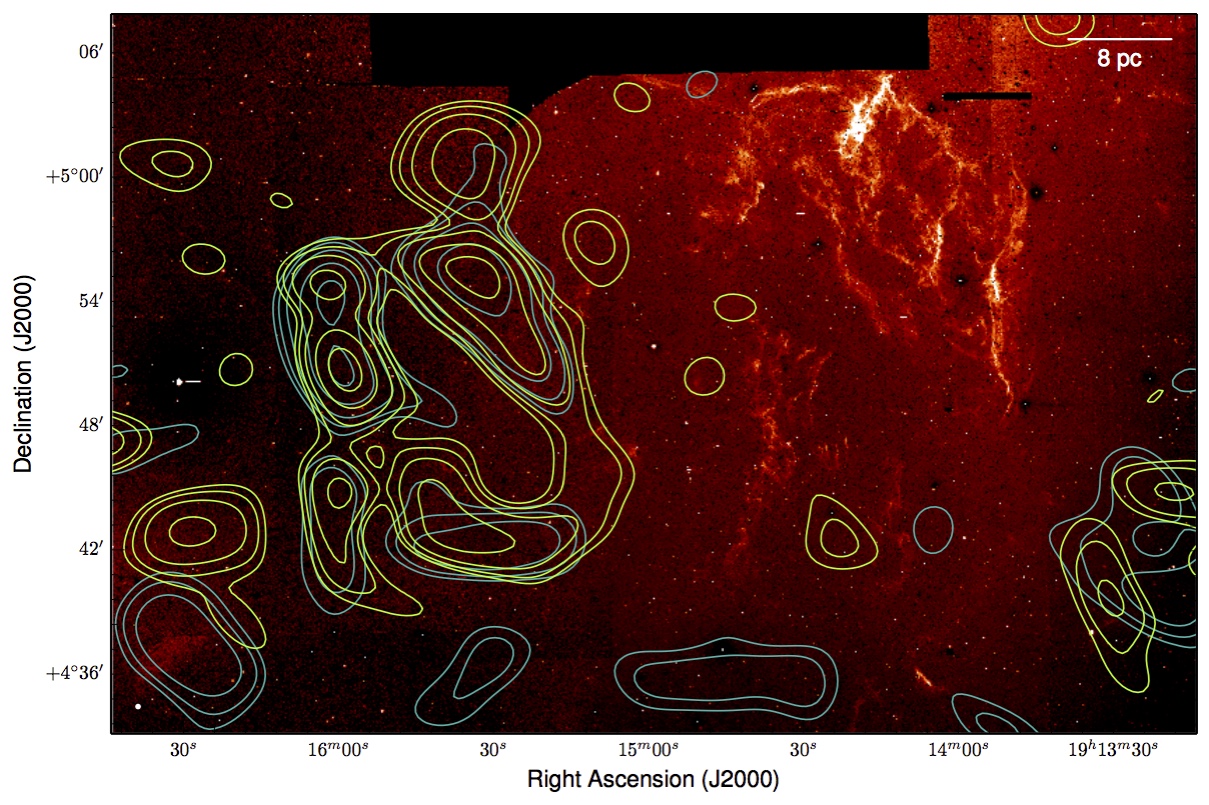}
   \caption{Similar to Fig.~\ref{figreg1}, for Region 2.}
              \label{figreg2}%
    \end{figure*}

\begin{figure*}
   \centering
   \includegraphics[trim=0cm 0cm 0cm 0cm,clip=true,angle=0,origin=c,totalheight=0.65\textwidth]{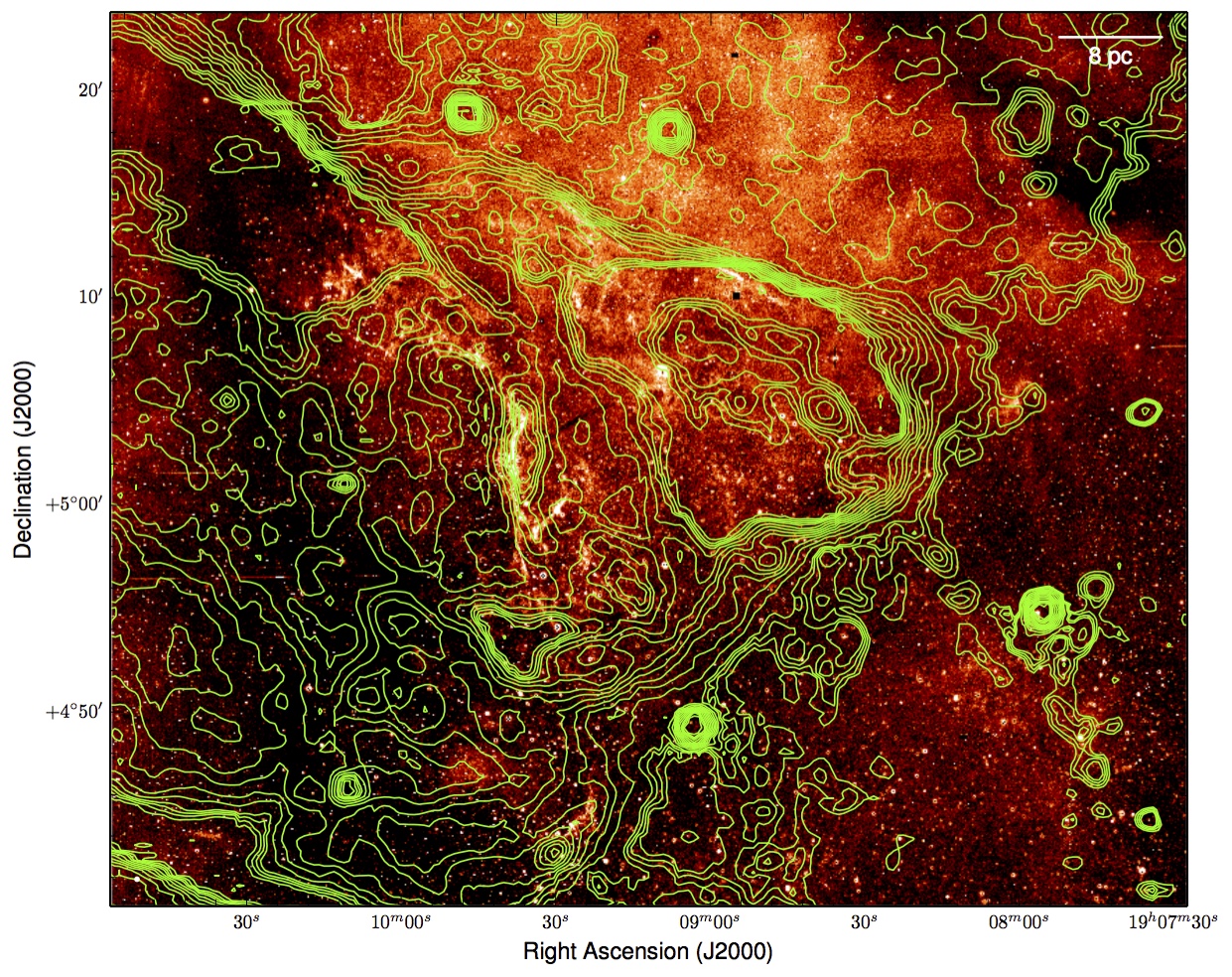}
   \includegraphics[trim=0cm 0cm 0cm 0cm,clip=true,angle=0,origin=c,totalheight=0.65\textwidth]{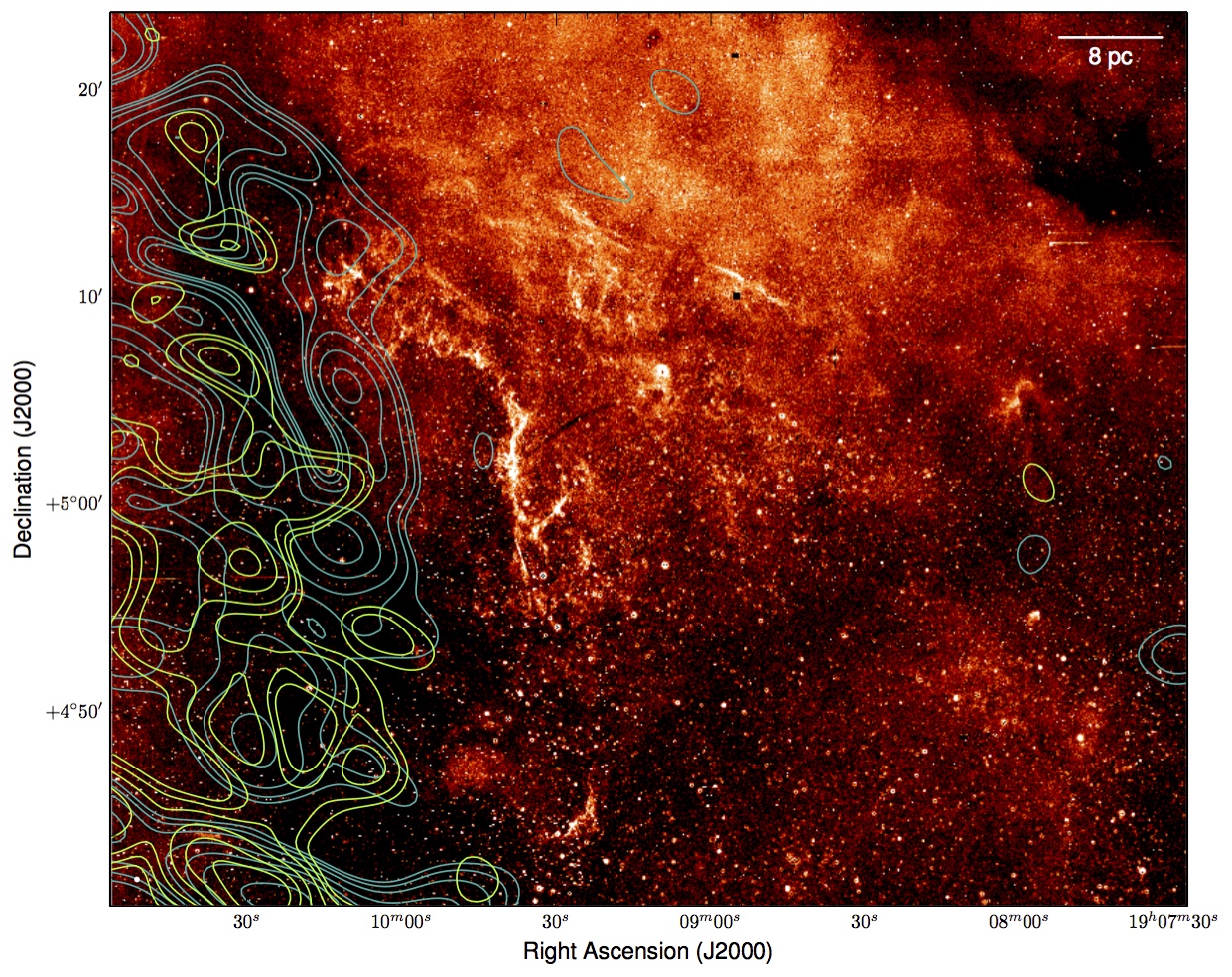}
   \caption{Similar to Fig.~\ref{figreg1}, for Region 3.}
              \label{figreg3}%
    \end{figure*}

\begin{figure*}
   \centering
   \includegraphics[trim=0cm 0cm 0cm 0cm,clip=true,angle=0,origin=c,totalheight=0.65\textwidth]{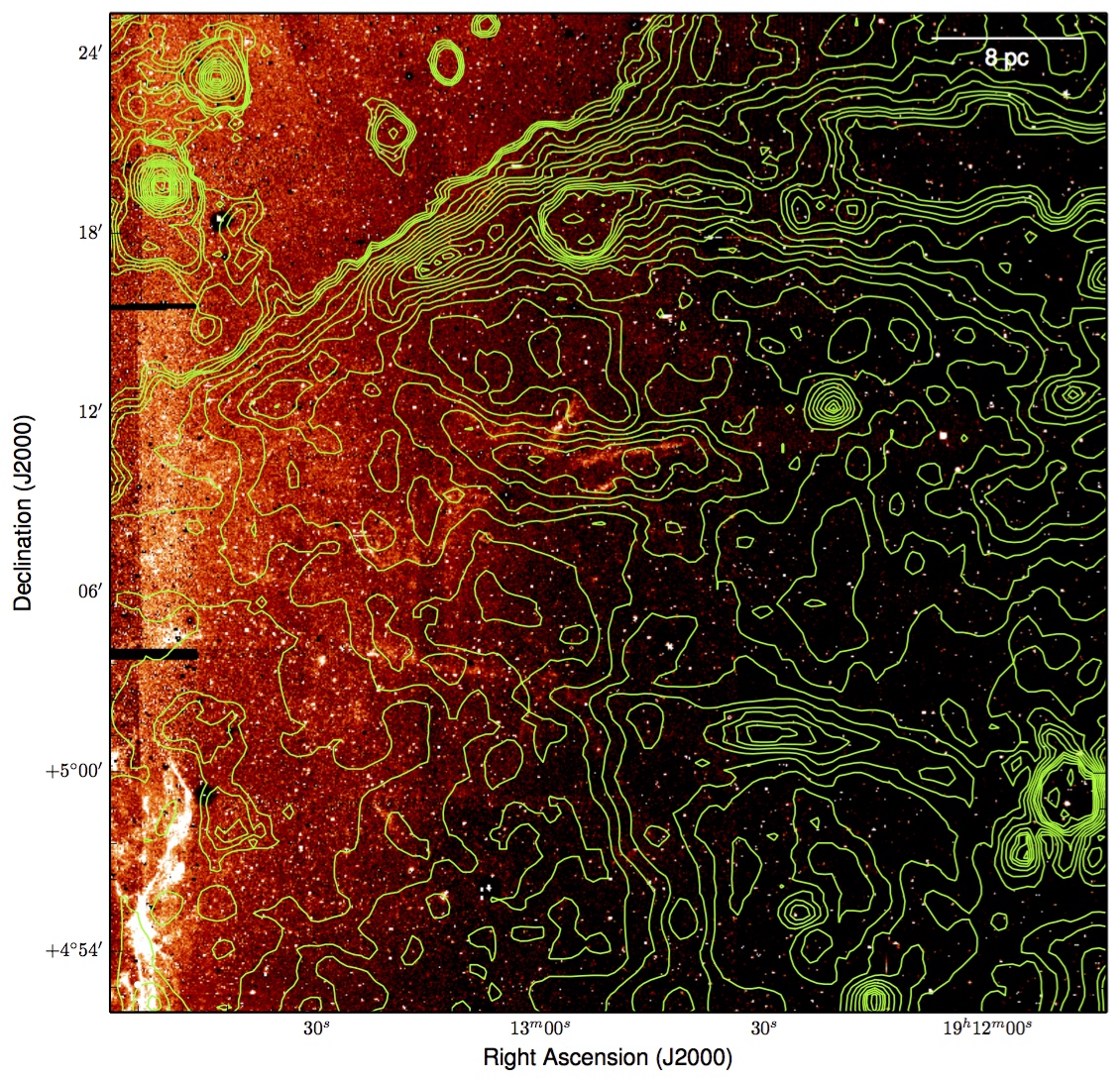}
   \includegraphics[trim=0cm 0cm 0cm 0cm,clip=true,angle=0,origin=c,totalheight=0.65\textwidth]{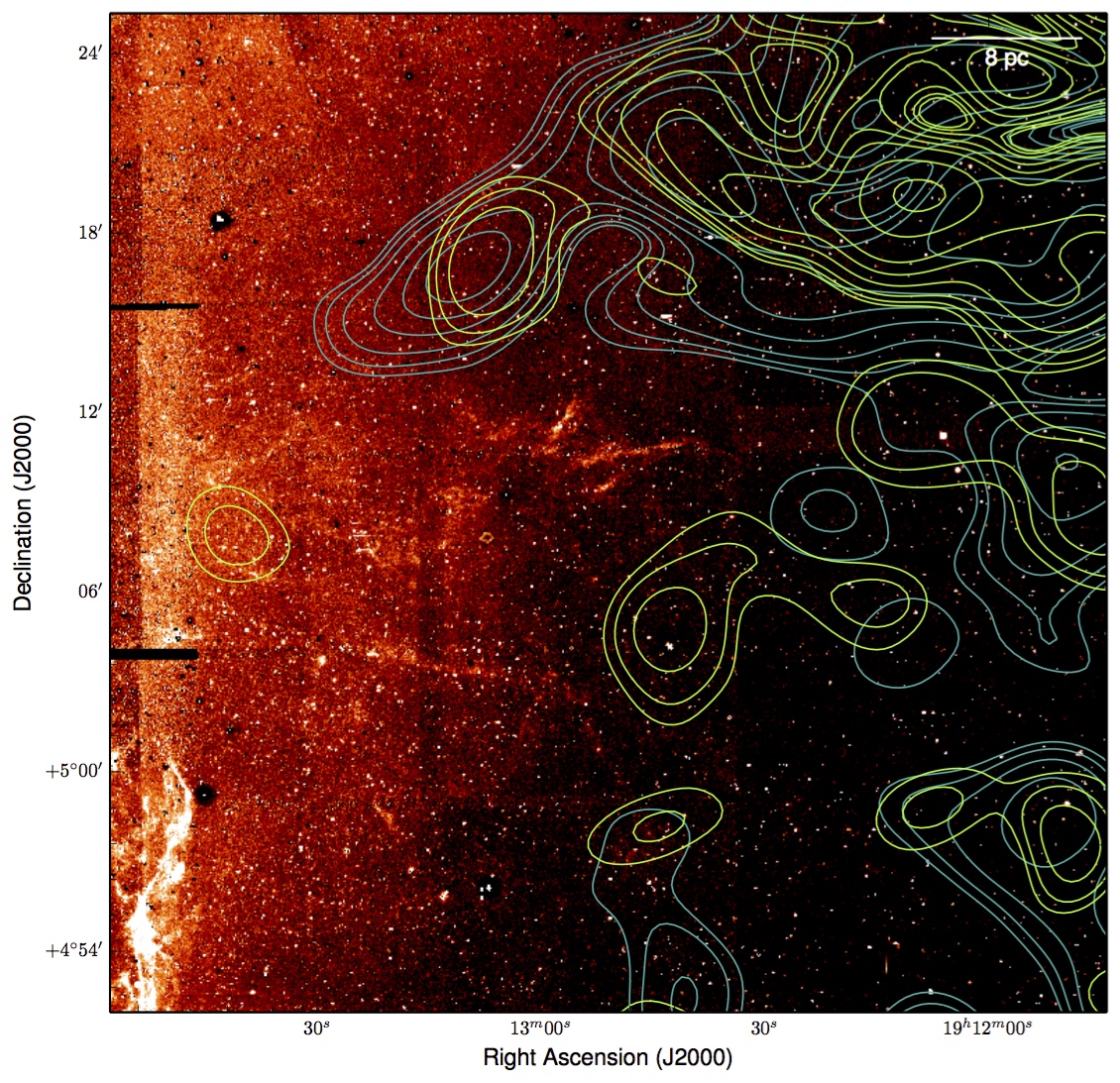}
   \caption{Similar to Fig.~\ref{figreg1}, for Region 4.}
              \label{figreg4}%
    \end{figure*}

\begin{figure*}
   \centering
   \includegraphics[trim=0cm 0cm 0cm 0cm,clip=true,angle=0,origin=c,totalheight=0.65\textwidth]{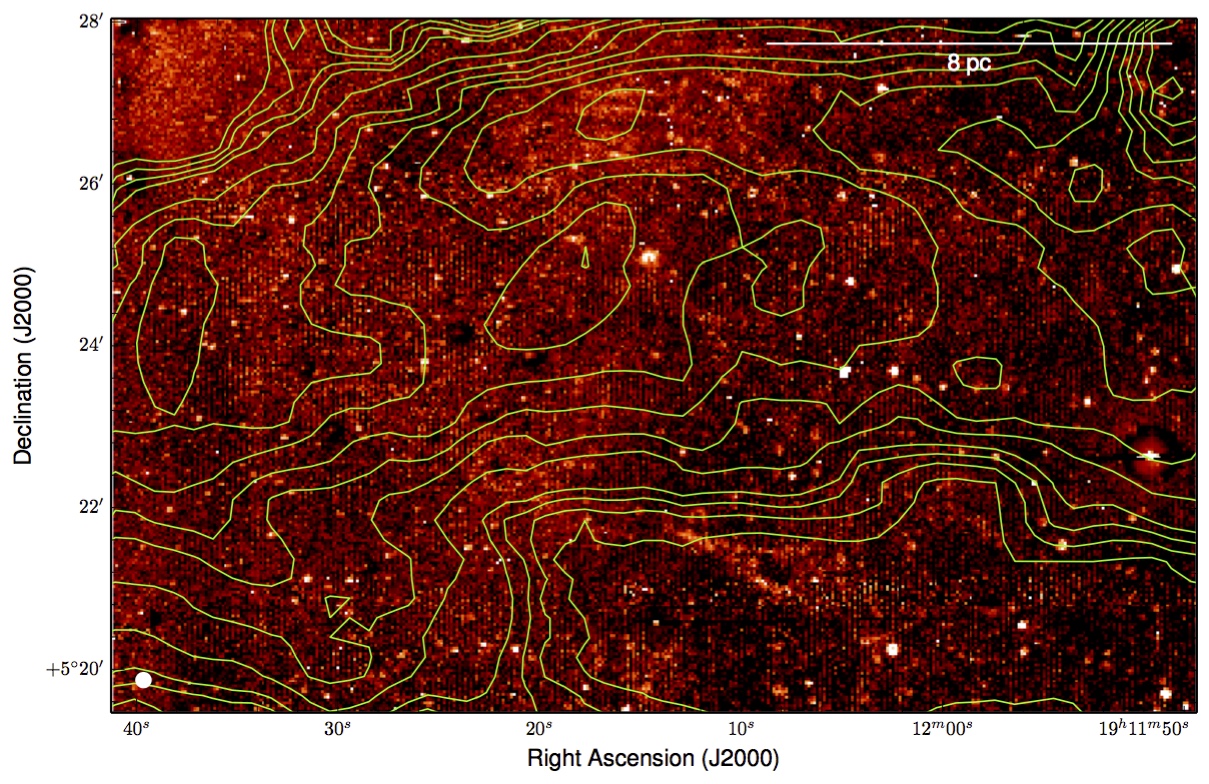}
   \includegraphics[trim=0cm 0cm 0cm 0cm,clip=true,angle=0,origin=c,totalheight=0.65\textwidth]{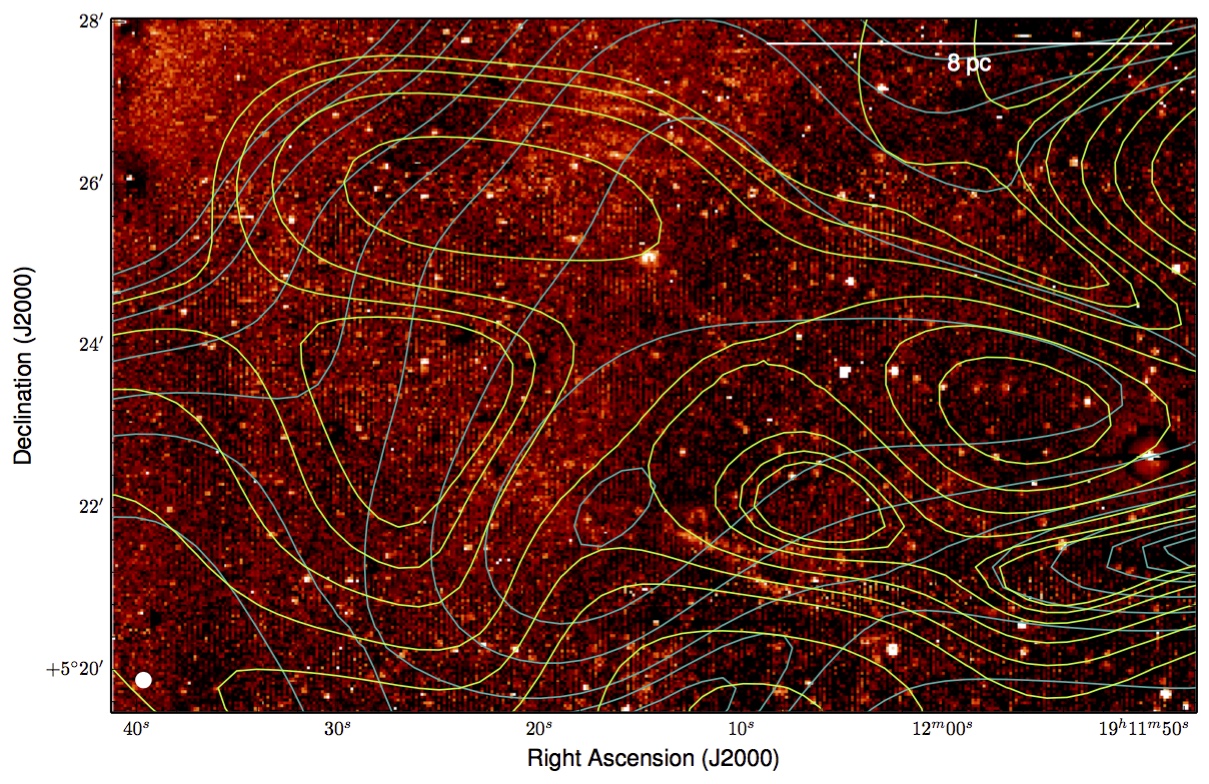}
   \caption{Similar to Fig.~\ref{figreg1}, for Region 5.}
              \label{figreg5}%
    \end{figure*}

\begin{figure*}
   \centering
   \includegraphics[trim=0cm 0cm 0cm 0cm,clip=true,angle=0,origin=c,totalheight=0.65\textwidth]{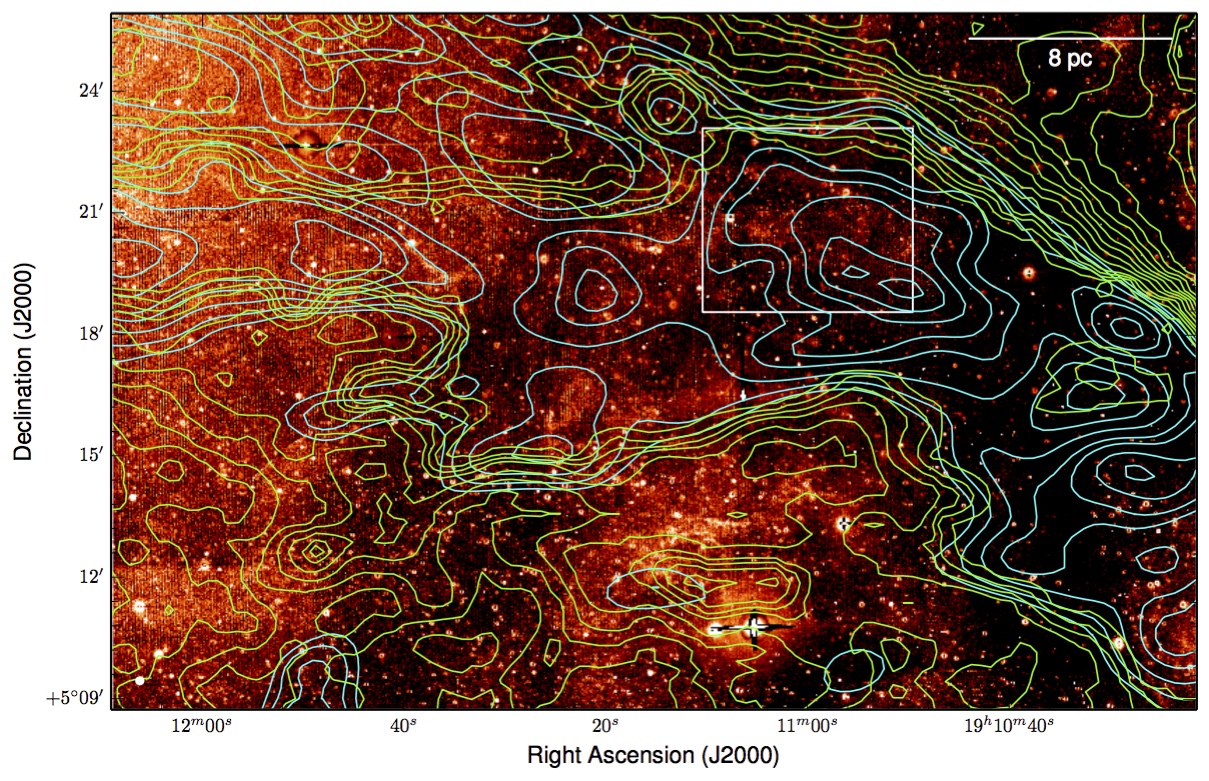}
   \includegraphics[trim=0cm 0cm 0cm 0cm,clip=true,angle=0,origin=c,totalheight=0.65\textwidth]{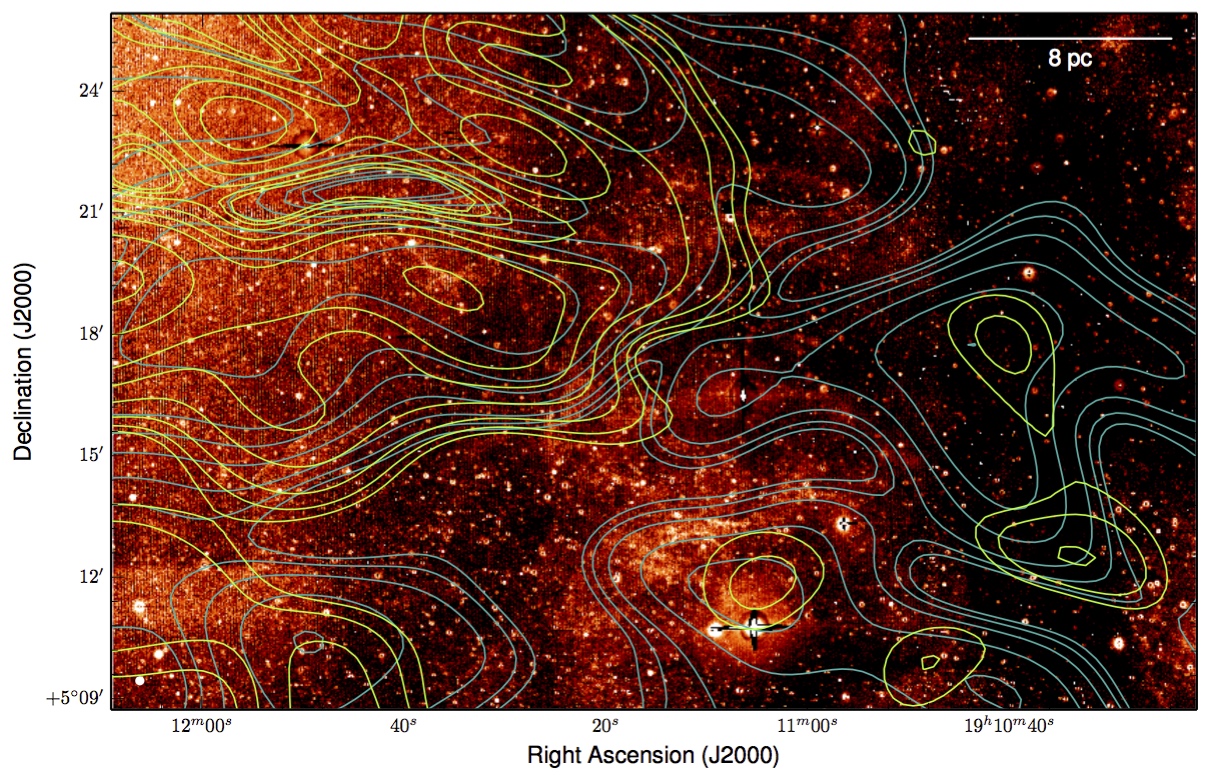}
   \caption{Similar to Fig.~\ref{figreg1}, for Region 6. The top panel also shows ATCA total intensity contours in cyan. The region indicated by the white box shows the approximate area in which H$\alpha$+NII emission has been previously reported, albeit without continuum-subtraction \citep{2007MNRAS.381..308B}.}
              \label{figreg6}%
    \end{figure*}

\begin{figure*}
   \centering
   \includegraphics[trim=0cm 0cm 0cm 0cm,clip=true,angle=0,origin=c,totalheight=0.65\textwidth]{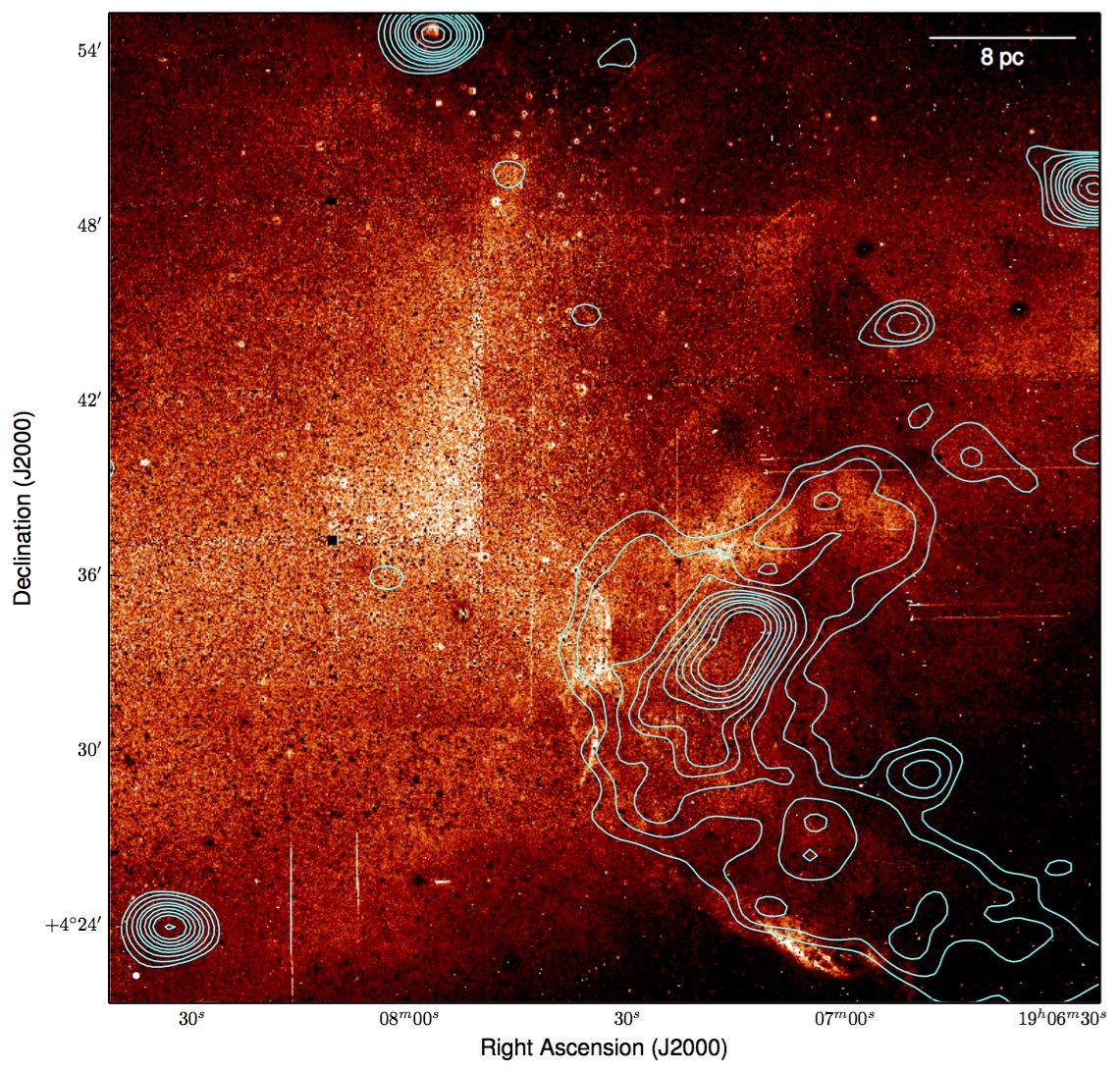}
   \includegraphics[trim=0cm 0cm 0cm 0cm,clip=true,angle=0,origin=c,totalheight=0.65\textwidth]{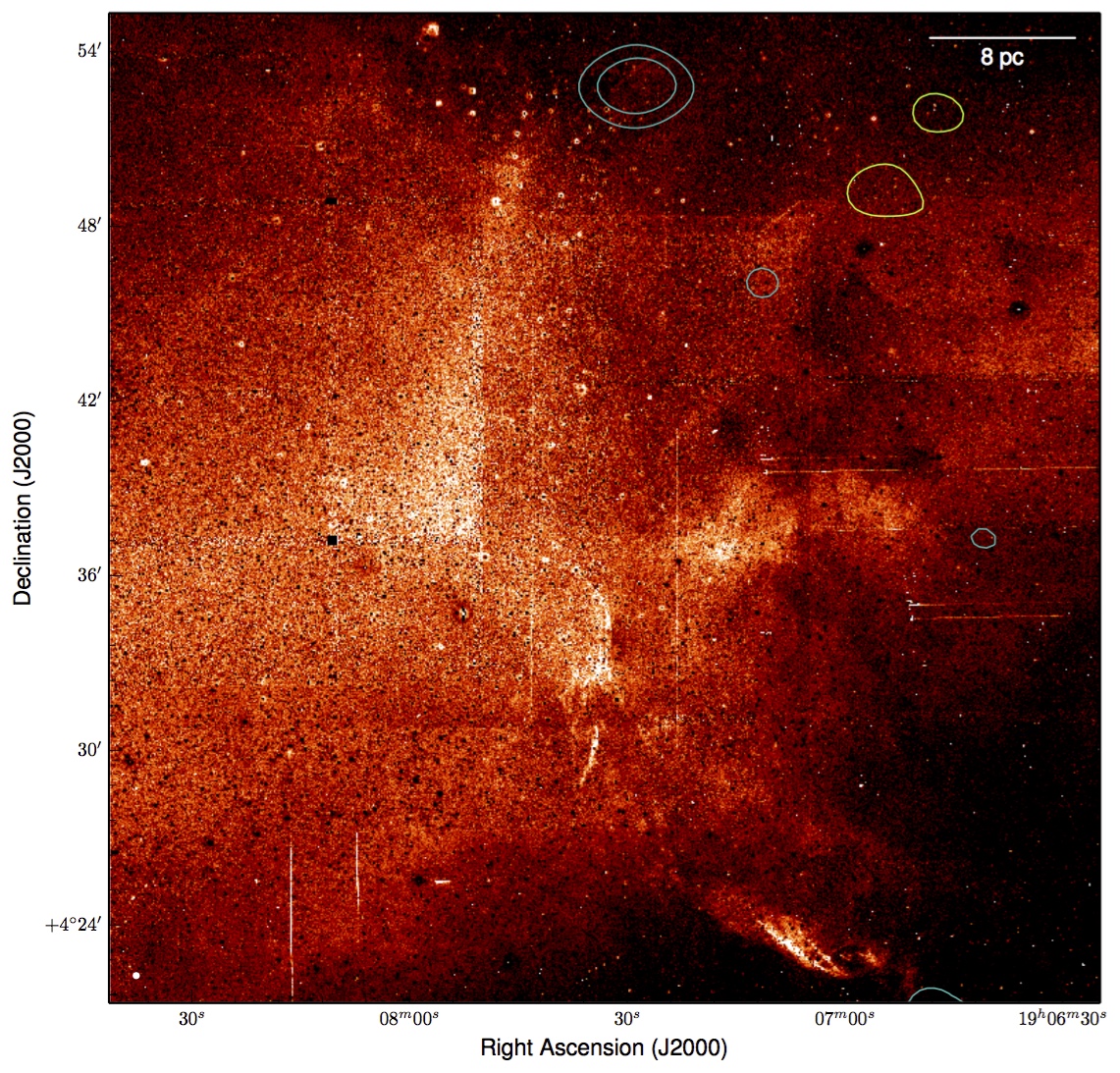}
   \caption{Similar to Fig.~\ref{figreg1}, for Region 7. The top panel shows ATCA total intensity contours in cyan, and does not show VLA \citet{1998AJ....116.1842D} contours which do not extend out to this region.}
              \label{figreg7}%
    \end{figure*}

The region showing the eastern breakout region, where the eastern ear punches through the eastern boundary of the circular region of W50, is shown in Fig.~\ref{figreg1}. The images show the pseudo-colour IPHAS continuum-corrected H$\alpha$ image, overlaid with the total intensity radio continuum contours from the VLA \citet{1998AJ....116.1842D} image at 1.4~GHz, and with the linearly polarized intensity contours from the ATCA images at both 3.1~GHz and 2.2~GHz. The region showing the eastern ear, and where the radio emission from the eastern ear terminates, is shown in Fig.~\ref{figreg2}. The region showing the western ear, the western breakout region where the jet punches through the western boundary of the circular region of W50, and where the radio emission from the western ear terminates, is shown in Fig.~\ref{figreg3}. The region showing the northeastern interior of W50, including SS433 within the field of view, is shown in Fig.~\ref{figreg4}. The region showing a small section from the northern rim of W50 is shown in Fig.~\ref{figreg5}. The region showing a small section from the northwestern rim of W50 is shown in Fig.~\ref{figreg6}. The region surrounding the SNR candidate G38.7--1.4 is shown in Fig.~\ref{figreg7}. For all images from Fig.~\ref{figreg2} to Fig.~\ref{figreg7}, the representation of various emission by different colours and contours is the same as in Fig.~\ref{figreg1}. The exact position of each region relative to the full extent of the W50/SS433 system is shown in Fig.~\ref{figregions}.

\subsection{Faraday Rotation}
\label{faradayrotation}
\subsubsection{Line-of-Sight Rotation Measure}
\label{atcaRMs}
The Stokes $Q$ and $U$ datacubes as a function of frequency were converted into Faraday-space using the technique of RM Synthesis, as per Section~\ref{RMSynth}. This was done pixel-by-pixel in order to reconstruct a full image of the Faraday rotation across the radio nebula. The RM corresponding to the peak in Faraday space for each pixel was recorded, and pixels with a signal--to--noise (s/n) $\le12\sigma$ were blanked -- making our final image very reliable. This procedure implicitly assumes that a single Faraday component is responsible for the Faraday rotation \citep[e.g.][]{2015AJ....149...60S} - although the situation can be more complicated, $QU$-fitting was not possible using these data (see Section~\ref{limitations}). Nevertheless, RM Synthesis will still retrieve an image showing spatial variations of the mean RM -- which can inform us of the average magnetic field structure, order, and geometry.

\begin{figure*}
   \centering
   \includegraphics[trim=0cm 0cm 0cm 1.5cm,clip=true,angle=90,origin=c,totalheight=0.98\textheight]{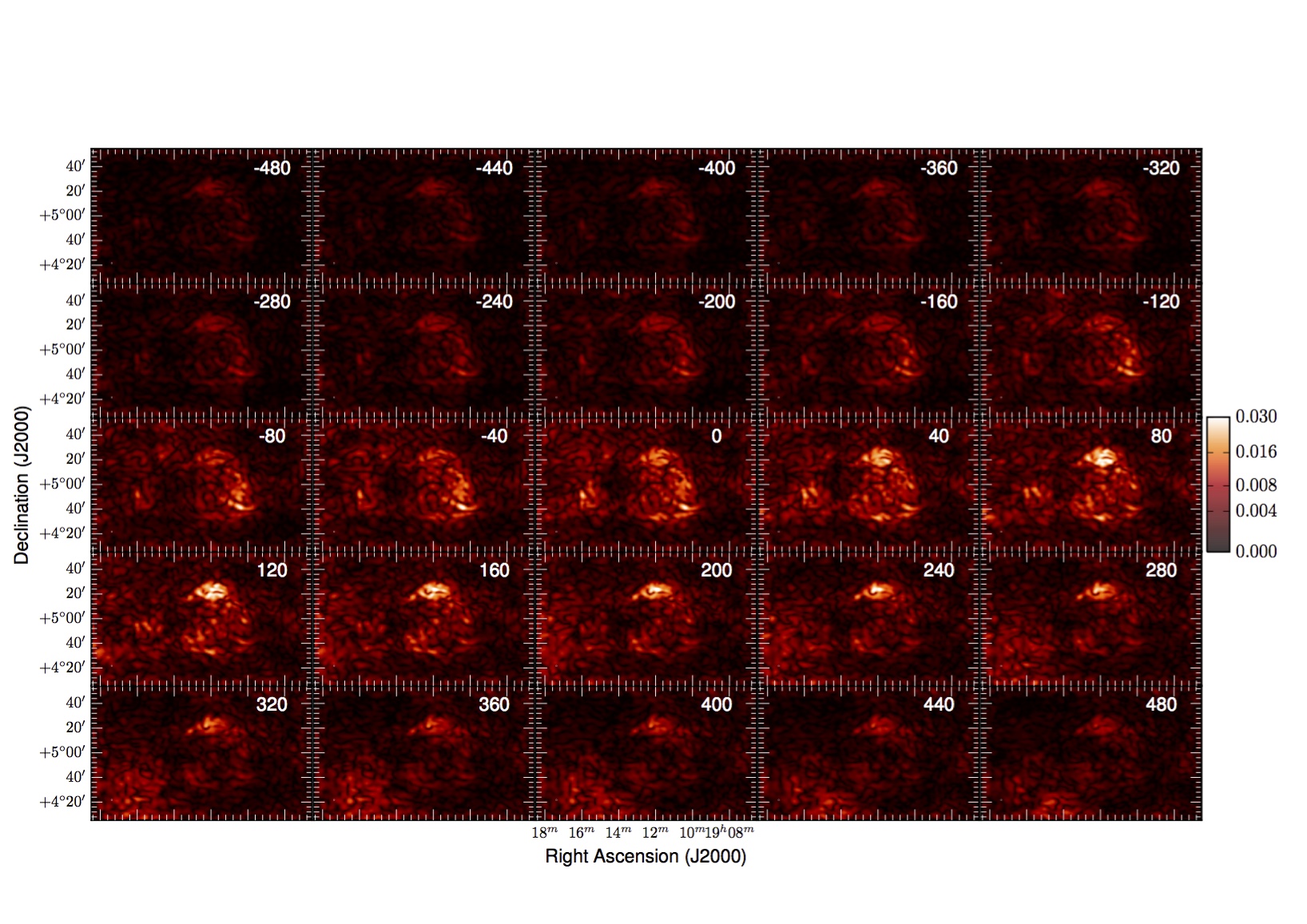}
   \caption{Slices from the Faraday depth cube of W50. The Faraday depth is shown in the upper-right of each slice in units of rad~m$^{-2}$. The pseudocolour scale is shown to the right of the slices, in units of Jy~beam$^{-1}$~rmsf$^{-1}$.}
              \label{fig3}%
    \end{figure*}

\begin{figure*}
   \centering
   \includegraphics[totalheight=0.57\textwidth]{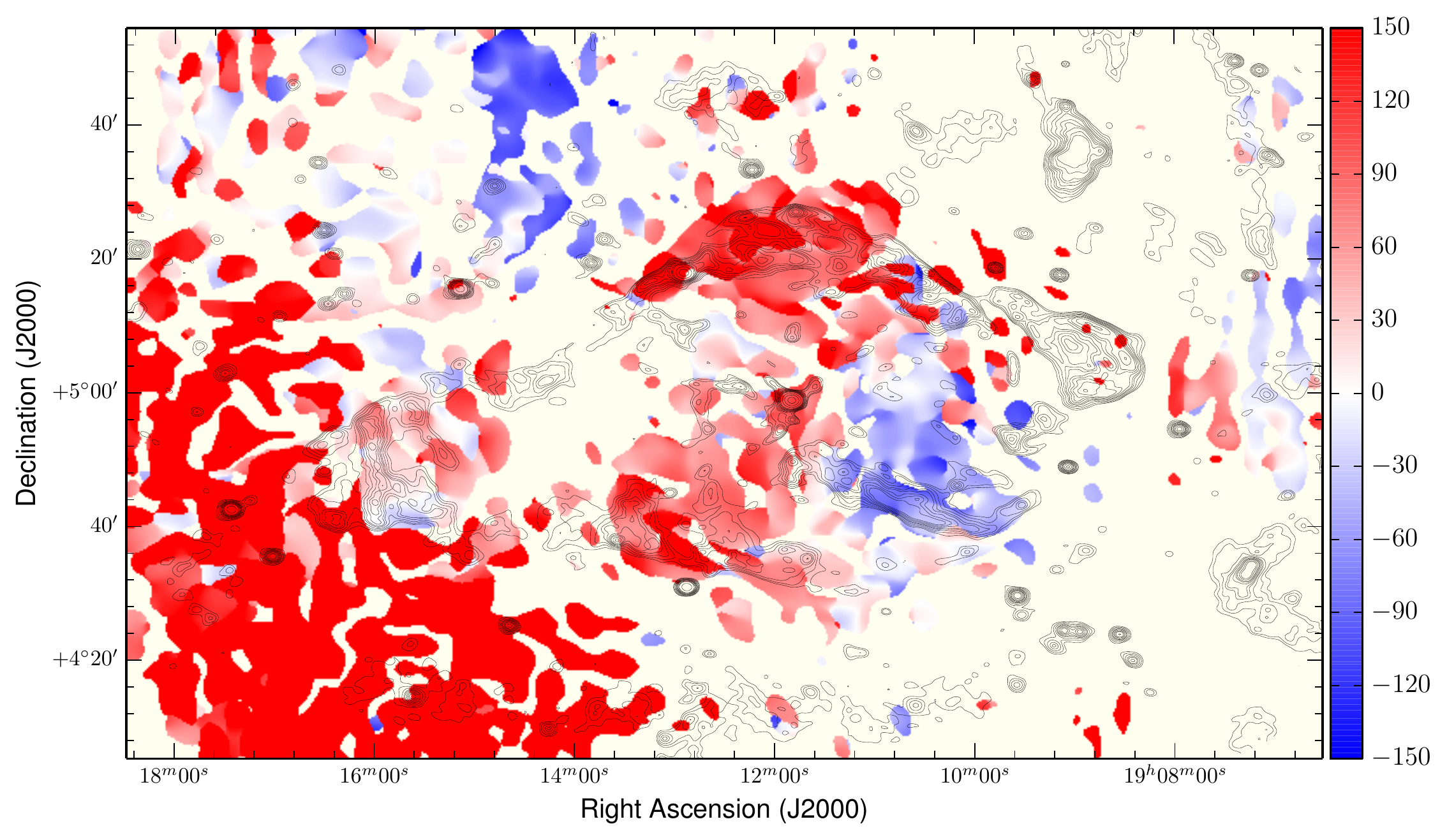}
   \caption{A peak RM image of W50. Pixels with RM detections below $12\sigma$ in the Faraday spectrum are blanked and shown in ivory. Overlaid are the ATCA Stokes $I$ contours in black to show the main features of the W50/SS433 system. The pseudocolour scale ranges from -150 to +150~rad~m$^{-2}$.}
              \label{fig4}%
    \end{figure*}

\begin{figure*}
   \centering
   \includegraphics[totalheight=0.43\textwidth]{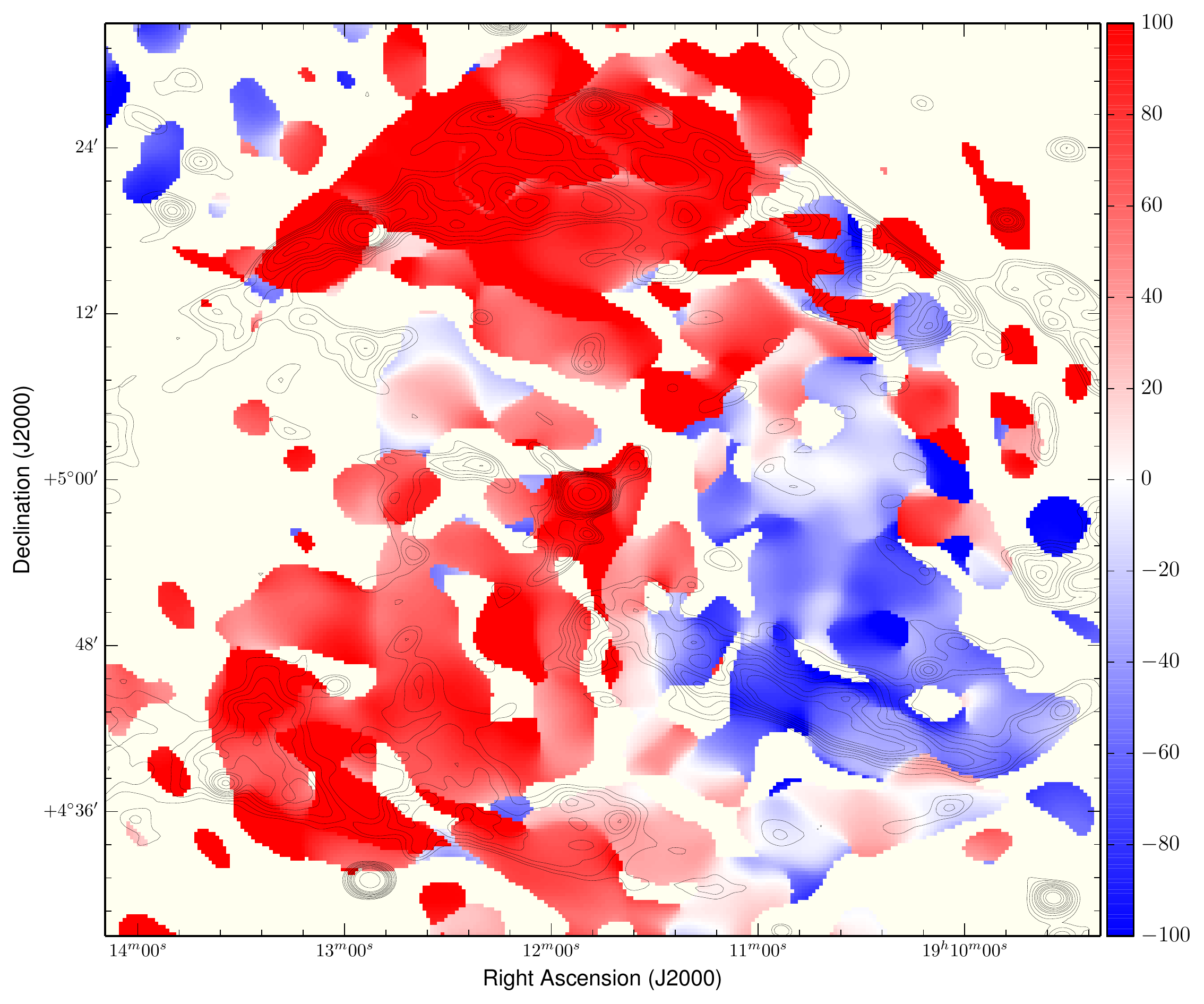}
   \includegraphics[totalheight=0.43\textwidth]{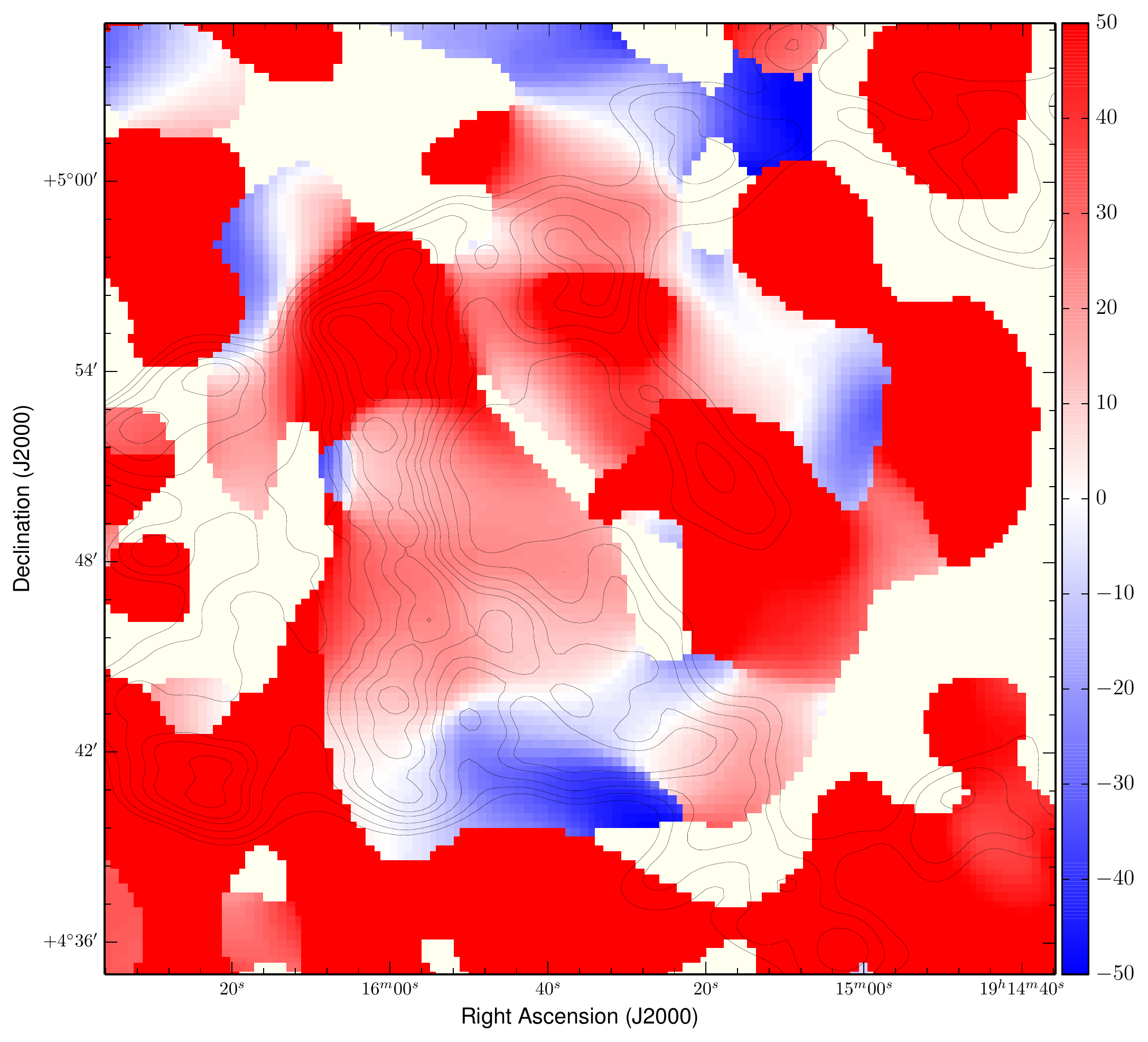}
   \caption{A peak RM image of two zoomed-in regions of W50, with Left: the central region of W50, and Right: the termination shock in the eastern ear. RMs below $12\sigma$ are blanked and shown in ivory. Overlaid are the ATCA Stokes $I$ contours in black to show the main features of the W50/SS433 system. The images are similar to Fig.~\ref{fig4}, except the pseudocolour scale ranges from -100 to +100~rad~m$^{-2}$ in the left panel, and from -50 to +50~rad~m$^{-2}$ in the right panel. In the left panel, while the northern region has exclusively positive RMs, this central region otherwise shows a strong asymmetry in RM with positive RMs to the left and negative RMs to the right. In the right panel, there is a similar asymmetry across the terminal shock -- which is bright in radio continuum. At the top of the shock are positive RMs, and at the bottom of the shock are negative RMs.}
              \label{fig4a}%
    \end{figure*}

Slices from the Faraday depth cube of W50 are shown in Fig.~\ref{fig3}. The response from $-480$~rad~m$^{-2}$ to $+480$~rad~m$^{-2}$ is clearly shown. There is clearly an RM gradient across W50, with both positive and negative RMs, and diffuse linearly polarized emission that fills the entire field of view. Note that the Faraday spectra could not be cleaned, as explained in Section~\ref{RMSynth}. An image of the peak RM at each pixel is shown in Fig.~\ref{fig4}, with the ATCA Stokes $I$ contours overlaid to guide the eye. Further zoomed-in regions of the central region of W50, and of the eastern ear, are shown in Fig.~\ref{fig4a}. The image is broadly consistent with the results of \citet{1986MNRAS.218..393D} albeit with sensitivity to much fainter polarized emission, larger spatial scales, and without $n\pi$-ambiguities. Within the circular central region of W50, there is clearly a strong asymmetry in the RM, with positive RMs to the north and south east, and negative RMs to the west. The highest positive RM is in the northern rim with $+280$~rad~m$^{-2}$, while the highest negative RM to the west is $-125$~rad~m$^{-2}$. Both the western ear and the funnel leading to the eastern ear are completely depolarized as described in Section~\ref{depol}. The polarized emission from the eastern ear also displays a strong asymmetry, with positive RMs ($+40$ to $+60$~rad~m$^{-2}$) to the north and negative RMs ($-20$ to $-40$~rad~m$^{-2}$) along the southern rim. Note that the negative region to the north-west of the ear does not originate from the bright polarization in the ear, but is rather from the surrounding diffuse Galactic emission. The RM foreground estimate further strengthens the suggestion that there is an RM asymmetry in the ear (see Section~\ref{RMforeground}), and an RM asymmetry is also consistent with the expected physics of the region (see Section~\ref{earfield}). There is a sharp transition between the RM of the eastern ear in contrast to the surrounding Galactic emission, which mostly has an RM of $+230$~rad~m$^{-2}$ in this area, and shows how the environment associated with W50 is modifying the line-of-sight magnetic field.

Away from W50, the diffuse polarized Galactic emission also fills the majority of the field of view, and separately also displays a strong RM gradient that encircles the radio nebula. The strongest emission is visible along the east of W50, with negative RMs up to $-100$~rad~m$^{-2}$ in the north which gradually increase to become positive in the south with positive RMs up to $+670$~rad~m$^{-2}$. Additionally, to the west of the field-of-view, the semi-circular gap in the diffuse H$\alpha$ emission that also has polarized emission (see Section~\ref{depol}) also displays an RM gradient, with a positive RM of $+120$~rad~m$^{-2}$ nearest to W50 and changing to $-80$~rad~m$^{-2}$ further from W50 in the north (and at the edge of the field).

\subsubsection{Foreground Rotation Measure}
\label{RMforeground}
Interpreting the magnetic field structure from the RMs requires an understanding of the foreground Galactic Faraday screen. This is a particularly complicated field of view, due to the large angular extent of W50, and due to the source's location -- which dips into the Galactic plane at a latitude of $2^{\circ}$. The best model of the Galactic foreground currently available is \citet{2015A&A...575A.118O}. Such foreground models are not fully applicable to Galactic objects, as the model is derived using extragalactic RMs, whereas W50 is immersed within the Galaxy. Consequently, the line-of-sight magnetic field towards W50 does not sample the full magnetoionic content between us and the periphery of the Milky Way. The integrated magnetic field between us and W50 could have been modified (and could even have changed sign) several times before exiting the Galaxy, particularly due to magnetic field reversals in the spiral arms of the Milky Way \citep{2006ApJ...637L..33H,2006AN....327..483H,2007ApJ...663..258B,2008ApJ...680..362H}. 

Another technique is therefore required in order to estimate the foreground Galactic contribution to the Faraday Rotation. The use of pulsars as Galactic magnetic field probes is viable as the Faraday rotation from a pulsar magnetosphere is expected to be negligible and to not exhibit significant variations \citep{2011MNRAS.417.1183W}. Pulsars therefore provide a very clean RM signal, which measures the Galactic magnetic field, and which is not contaminated by strong intrinsic variations. Following \citet{2015ApJ...804...22P}, we use v1.50 of the ATNF Pulsar Catalogue\footnote{\url{http://www.atnf.csiro.au/research/pulsar/psrcat/}} \citep{2005AJ....129.1993M} which collates the properties of more than 2300 rotation-powered pulsars and is continually revised as new discoveries are made. The ATNF Pulsar Catalogue contains RMs, dispersion measures (DMs), and pulsar distance estimates. Within a $5^{\circ}$ radius of SS433, we find 13 pulsars with measured RMs, with most located to the East of W50 within the Galactic plane. It is difficult to obtain accurate distances to pulsars, with only a few precise measurements determined using annual parallaxes for relatively nearby pulsars. All of the pulsars used here have a distance determined from the DM. The DM-derived distance relies on a model of the Galactic free-electron distribution \citep[e.g.][]{1993ApJ...411..674T}, which may be unreliable. We here use the NE2001 model for our thermal-electron density estimates \citep{2002astro.ph..7156C}. In order to contrast the pulsar RM measurements in the foreground of W50 with the entire integrated Galactic RM contribution, we also use the extragalactic RM catalogue of \citet{2011ApJ...728...97V}. Within a $5^{\circ}$ radius of SS433, we find 12 extragalactic sources with measured RMs.

\begin{figure}
   \centering
   \includegraphics[trim=0.5cm 0.5cm 0.5cm 1.0cm,clip=true,width=\hsize]{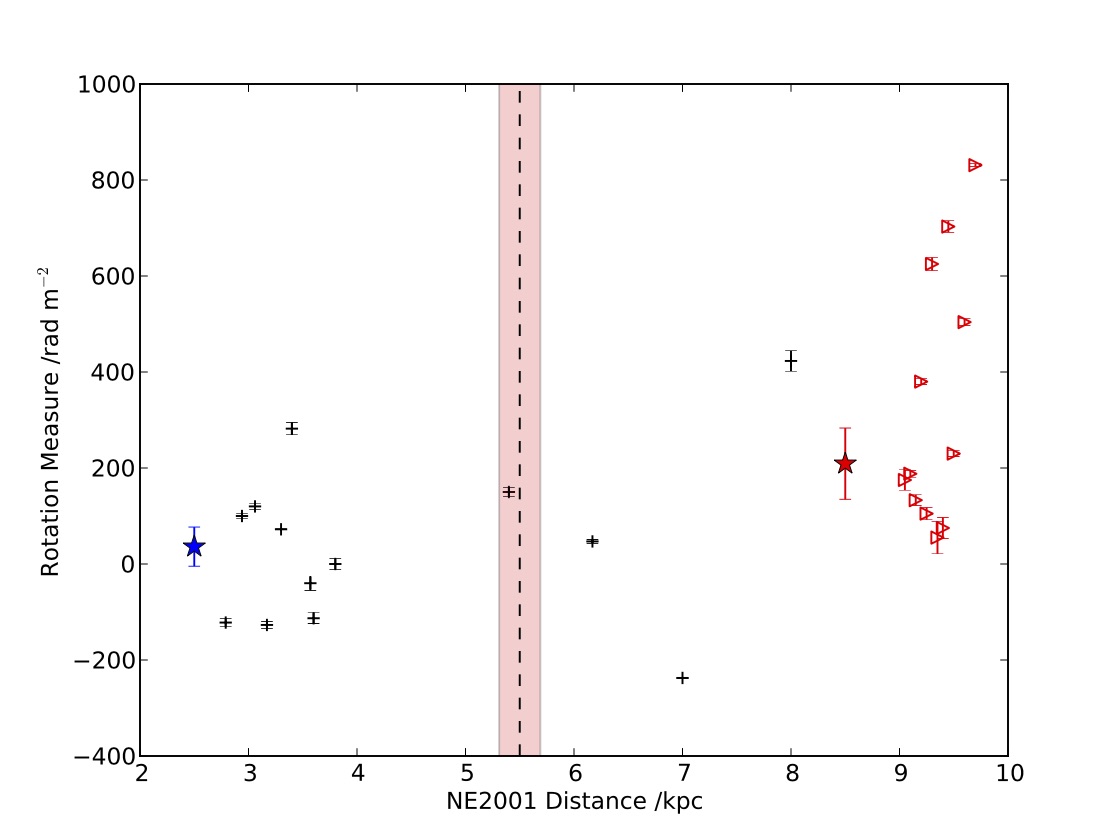}
   \includegraphics[trim=0.5cm 0.5cm 0.5cm 1.0cm,clip=true,width=\hsize]{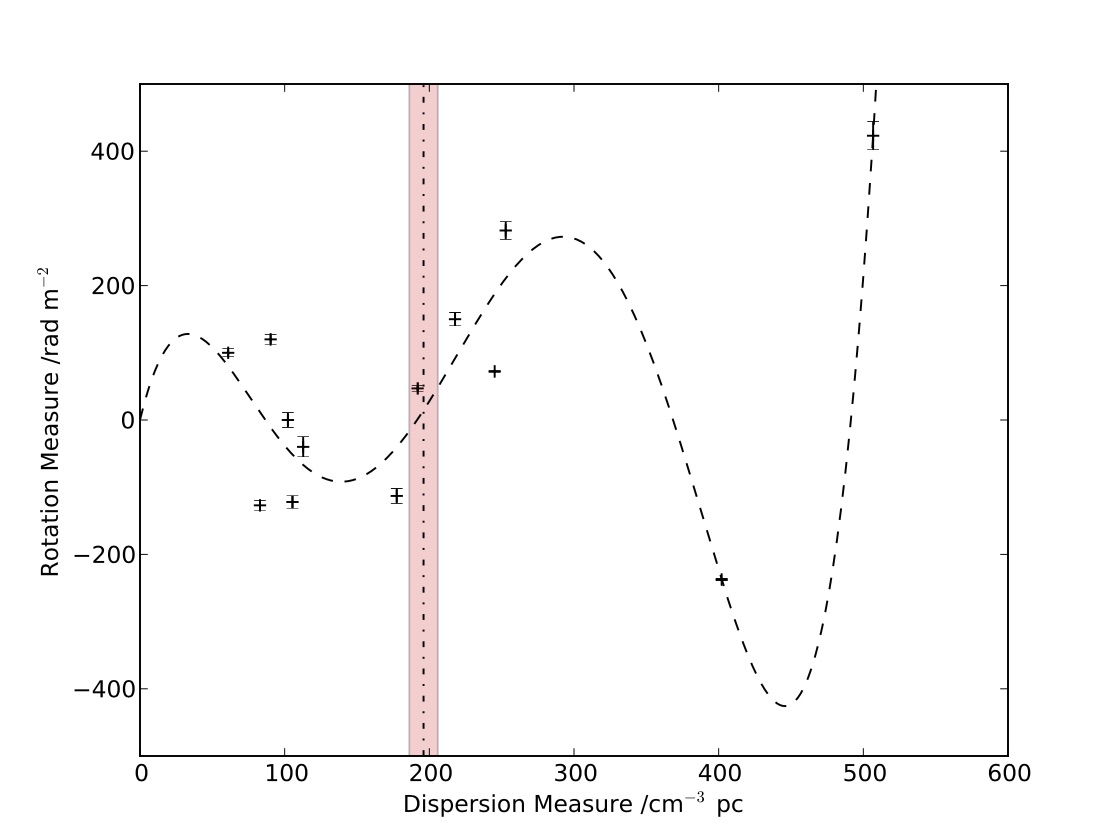}
   \caption{Plots showing our pulsar-based estimation of the Faraday rotation associated with the Galactic foreground in front of W50. Top panel: The RM versus NE2001 estimated distance. Pulsar measurements are shown in black, while extragalactic measurements are shown in red. The median pulsar RM in the foreground of W50 is shown as a blue star, while the median extragalactic RM is shown as a red star. The distance to SS433 is indicated by the shaded, red, vertical bar. Errors in distance are not shown, and are typically on the order of 10--20\%. Bottom panel: The RM versus DM for all of the pulsar measurements, with a black-dashed interpolated line added to guide the eye. The DM-distance to SS433, as estimated using NE2001, is indicated by the shaded, red, vertical bar.}
              \label{figforeground}%
    \end{figure}

Determining the foreground RM to W50 is challenging, as in combination with any possible errors on the distance to W50 itself, there are also additional uncertainties due to the derived electron model distances. A plot showing the RM of the pulsars and extragalactic sources, as a function of distance, is shown in the top panel of Fig.~\ref{figforeground}. The distance estimate and $1\sigma$-uncertainty for SS433 is indicated by the red-coloured vertical band. The pulsars in front of W50 have considerable scatter, with a median foreground RM of $+36\pm41$~rad~m$^{-2}$ -- thus consistent with zero foreground Faraday rotation or a low positive RM (on average across the field-of-view). The scatter in RM of pulsars behind W50 (and hence behind the region of the Milky Way in which W50 is situated) appears to be larger than the scatter in RM of pulsars in front of W50. As an overview of the general distribution of pulsars relative to W50, the pulsar with nearest angular distance to SS433 is located at a proximity of $1.09^{\circ}$, with an RM$=-127\pm7$~rad~m$^{-2}$. The DM distance to this pulsar is 2.95~kpc, also placing it in front of the radio nebula. The foreground RM as determined using the pulsars is evidently different from the median extragalactic RM towards W50 of $+209\pm80$~rad~m$^{-2}$ -- thus consistent with an integrated magnetic field pointing towards us in this region of the Galaxy.
As both the RM and the DM are \emph{integrated} quantities along the line-of-sight, it is also possible to plot one against the other to measure their relative variation as a function of DM-derived distance. Such RM versus DM techniques have been discussed before \citep[e.g.][]{1999MNRAS.306..371H,2004ApJS..150..317W,2006ApJ...642..868H}. Under reasonable assumptions for such a plot, the gradient is related to the average local magnetic field strength as $<B> = 1.232\times\textrm{RM}/\textrm{DM}$. The direction of the local field, at a given DM-distance, is determined by the slope of the RM/DM plot. Consequently, maxima and minima reveal the location of field reversals, and it has been argued these reversals correspond to the magnetic fields associated with different spiral arms \citep{2013IAUS..291..223H}. There is always a constraint on such a plot at $(\textrm{RM},\textrm{DM})=(0,0)$. This RM versus DM method, could feasibly be affected by magnetic fields in intervening HII regions and SNRs along the line-of-sight \citep[see e.g.][]{2003A&A...398..993M,2011ApJ...736...83H}. 

A plot showing the RM and DM of the pulsars is shown in the bottom panel of Fig.~\ref{figforeground}. The DM-distance estimate ($195.3$~pc~cm$^{-3}$) and $1\sigma$-uncertainty for SS433, as calculated using NE2001, is indicated by the red-coloured vertical band. To guide the eye, a dashed line shows a smooth interpolation of the data. The pulsars appear to show regular and consistent fluctuations of RM and DM. The data suggest that multiple magnetic field reversals are visible along the line-of-sight, with two field reversals occurring in the foreground of W50. At the distance corresponding to the location of W50, the local RM gradient appears to be positive in sign and therefore the parallel component of the magnetic field is pointing towards us. Our best estimate of the foreground Faraday rotation is $+15\pm30$~rad~m$^{-2}$ -- thus again consistent with zero foreground Faraday rotation.

However, due to the large angular size of W50, the DM varies across the source. These variations could change the DM-distance towards W50. The DM across the face of W50 is clearly expected to vary from East to West, i.e.\ with higher DMs closer to the Galactic plane. There is also possibly some, but considerably less, variation expected from North to South across the shell of W50. As shown in Table~\ref{table:2}, our data suggests a $\sim20$\% DM gradient from east to west, with no detectable DM gradient from north to south. The DMs measured at different locations towards the radio nebula, as calculated using NE2001, do not change sufficiently to change our foreground Faraday rotation estimate. Note that this does not rule out angular variations in the Faraday rotating foreground itself -- as it tells us nothing about the foreground magnetic fields, which could have both large- and small-scale fluctuations across the face of the source. We are therefore unable to constrain variations in foreground Faraday rotation across W50, and to do so would likely require a more finely-sampled pulsar grid.

\begin{table}
\caption{Variations in the Dispersion Measure for various lines-of-sight across the face of the W50/SS433 system, according to the NE2001 model and corresponding to a distance of $5.5\pm0.2$~kpc.}             
\label{table:2}      
\centering                          
\begin{tabular}{c c c c}        
\hline\hline                 
Location & RA (J2000) & Dec. (J2000) & DM (pc~cm$^{-3}$) \\    
\hline                        
   Near SS433       & 19h~11m~50s & 4d~59m~00s & $195\pm10$ \\   
   Eastern Ear      & 19h~16m~33s & 4d~47m~26s  & $175\pm8$ \\
   Western Ear       & 19h~08m~32s & 5d~07m~13s  & $218\pm13$ \\
   Top of Shell      & 19h~11m~47s & 5d~26m~30s & $198\pm11$ \\ 
   Bottom of Shell & 19h~11m~50s & 4d~30m~45s & $193\pm10$ \\
\hline                                   
\end{tabular}
\end{table}

\section{Discussion}
\label{discussion}

\subsection{Overall Magnetic Field Structure and Ionised Gas Distribution}
In Section~\ref{linpol}, we were able to see a `ring' of linearly polarized emission surrounding SS433 and corresponding to the shell of the central region of W50. This ring of ordered magnetic fields would naively appear to be consistent with field compression from an outgoing shock wave, and therefore consistent with a SNR hypothesis for the origin of the nebula. However, this does not take depolarization effects into account (see Section~\ref{depol}). 

In the presence of the diffuse H$\alpha$, the polarized emission from W50 itself, and also some of the Galactic polarized emission, both exhibit complete depolarization. This indicates that the warm ionised gas/thermal electrons that are traced by the diffuse H$\alpha$ emission must be acting as a Faraday screen that appears to be located in the foreground of W50/SS433. Similar anticorrelations between H$\alpha$ and polarization have been observed before in the Galactic plane \citep[e.g.][and references therein]{2014MNRAS.437.2936S}.

We interpret the Faraday thin nature of the depolarizing screen as being consistent with the thermal medium being located in the foreground. It is possible that the medium is Faraday thick, however the thickness must be at a level below the sensitivity of our observations. Even for reasonable estimates of magnetic field and electron density, e.g.\ $n_{e} = 0.1$~cm$^{-3}$, $B = 5$~$\muup$G, and $dl = 50$~pc, the total Faraday thick contribution through the nebula would be only $\approx20$~rad~m$^{-2}$ (which falls below our Faraday thickness sensitivity limits). A thickness of $\approx20$~rad~m$^{-2}$ therefore cannot be ruled out (also see Section~\ref{RMSynth}). However, a Faraday thin hypothesis is also consistent with the expected strong extinction in the Galactic plane which would preferentially allow us to observe only nearby ionised gas. This suggests that the diffuse Galactic emission to the East and West of the radio nebula also originates in the foreground, from a region nearer to us than the ionised gas.

One may argue that the source is unpolarized, rather than depolarized, however the W50 radio nebula has been shown to have steep optically-thin radio spectral indices -- see Section~\ref{backgroundw50}. This is typical of synchrotron radiation, which is intrinsically polarized. In combination with the 5~GHz observations of \citet{2011A&A...529A.159G} (shown in Fig.~\ref{fig7}), we interpret the observed strong depolarization as being due to spatially-varying Faraday rotation fluctuations across the face of the radio nebula. We can measure the magnitude of these fluctuations in the two most significantly depolarized regions: (i) the western ear, and (ii) the `funnel' leading towards the eastern ear. In the western ear, we find a polarized fraction of $\le2.5$\% at 5~GHz. From our 2.1~GHz Faraday cubes and band-averaged total intensity image, by measuring the noise, $\sigma$, in the $Q$ and $U$ Faraday depth cubes, we obtain $\sigma=0.0985$~mJy~beam$^{-1}$, and the brightest total intensity in the region reaches 51.54~mJy~beam$^{-1}$. This allows us to place a $2\sigma$ upper limit on the polarized fraction within the western ear of $\le0.38$\%. Similarly, within the eastern funnel, we find a polarized fraction of $\le20.4$\% at 5~GHz. From our 2.1~GHz Faraday cubes and band-averaged total intensity image, we obtain $\sigma=0.127$~mJy~beam$^{-1}$, and the brightest total intensity in the region reaches 25.93~mJy~beam$^{-1}$. This allows us to place a $2\sigma$ upper limit on the polarized fraction within the eastern funnel of $\le0.98$\%. Note that due to the limited spatial-scales sampled by the interferometer, there can be significant missing flux in the total intensity image at 2.1~GHz (see Section~\ref{limitations}). Our limits on the polarized fractions are therefore likely conservative.

We assume that the depolarizing Faraday screen traced by the diffuse H$\alpha$ consists of a magnetoionic region that has a negligible amount of relativistic particles and that exists somewhere along the line-of-sight between the observer and the source. If this Faraday screen contains a constant regular field and has a homogeneous distribution of free electrons, then the region causes Faraday rotation of the polarized emission from background sources, but does not cause any physical depolarization (remember that for depolarization, the magnetic field along the line-of-sight isn't necessarily relevant, but rather the magnetic field in the plane of the sky as seen in projection). Nevertheless, any deviation from a constant field within the synthesised beam, or any density fluctuations in the number of thermal electrons, will create a RM gradient and subsequently cause depolarization. In the case of magnetic fields, such anisotropy is caused by either turbulent, or systematically varying regular fields. We assume that the depolarization can be well-described by the model proposed by \citet{1966MNRAS.133...67B} (for a full review of external Faraday depolarization effects, see \citet{2014ApJS..212...15F}). In this case,
\begin{equation}
\Pi(\lambda) = \Pi_{0} \exp{\left( -2 \sigma_{\textrm{RM}}^2 \lambda^4 \right) } \\,
\end{equation}
where $\Pi$ is the measured fractional polarization, $\Pi_{0}$ is the intrinsic fractional polarization, $\lambda$ is the observing wavelength, and $\sigma_{\textrm{RM}}$ characterises the RM fluctuations within the observing beam, where we assume a Gaussian distribution of RMs within the beam. We can therefore derive an estimate of the RM fluctuations using,
\begin{equation}
\sigma_{\textrm{RM}}^2 =  \frac{1}{2 \left[ \lambda_{5.0~\textrm{GHz}}^4-\lambda_{2.1~\textrm{GHz}}^4 \right] } \ln \left( \frac{\Pi_{2.1~\textrm{GHz}}}{\Pi_{5.0~\textrm{GHz}}} \right) \\.
\end{equation}
In the western ear, this corresponds to Faraday rotation measure fluctuations of $\sigma_{\textrm{RM}}\ge$48~rad~m$^{-2}$ on scales smaller than 4.5 to 6~pc. In the eastern funnel, this corresponds to fluctuations of $\sigma_{\textrm{RM}}\ge$61~rad~m$^{-2}$ on scales smaller than 4.5 to 6~pc.

\begin{figure}
   \centering
   \includegraphics[trim=16cm 8cm 16cm 0cm,clip=true,angle=0,origin=c,totalheight=0.93\hsize]{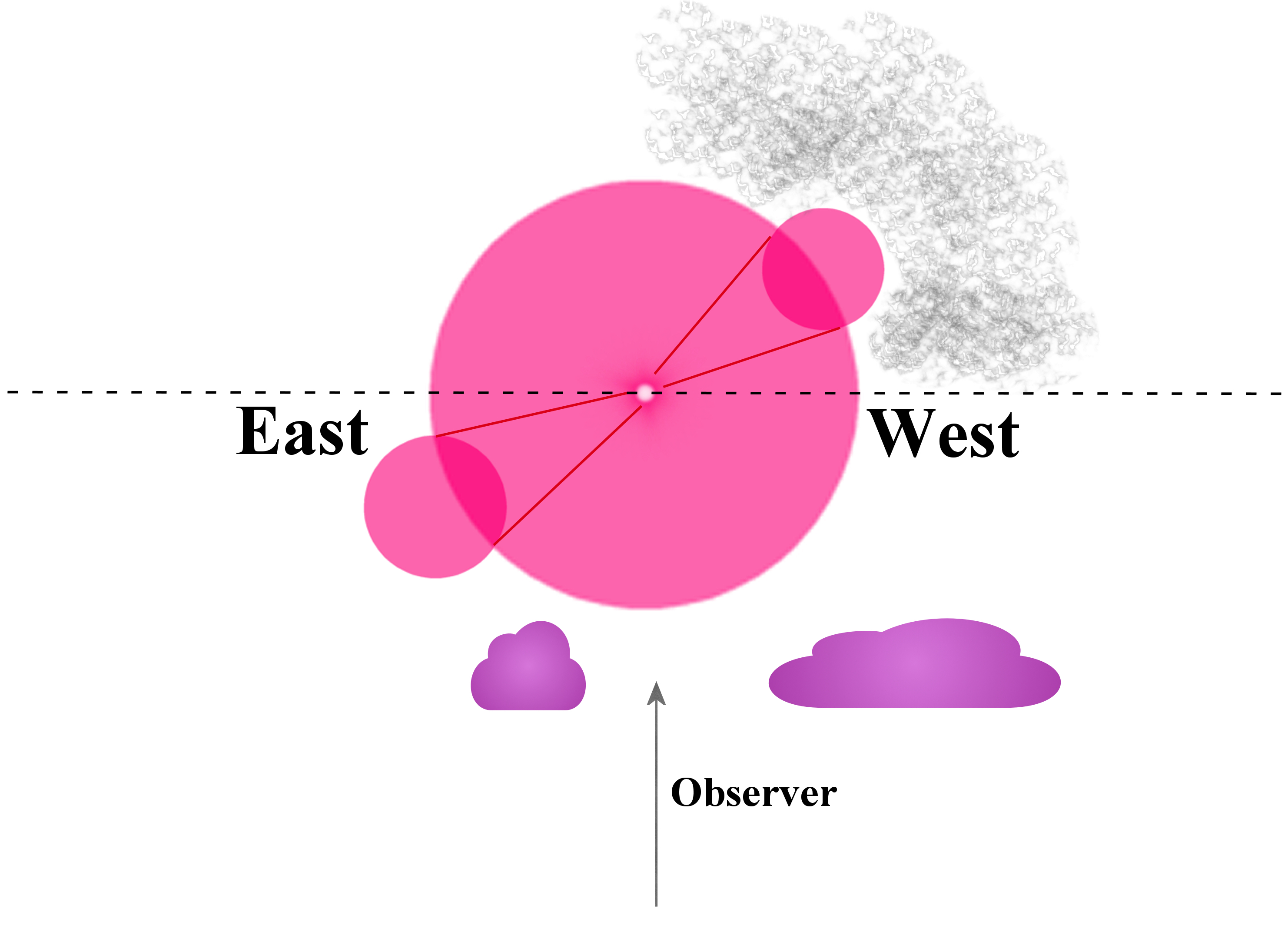}
   \caption{An explanatory cartoon of the line-of-sight towards W50. Note that the image is not to scale, and is merely an illustratory representation. W50/SS433 are shown in pink, alongside the known inclination of the system, with the Eastern ear nearer to us \citep{1979Natur.279..701A,1979IAUC.3358....1M}. The denser medium associated with the Galactic plane, located to the West of W50, is indicated as a gray patchy cloud. The foreground depolarizing H$\alpha$ clouds are shown in purple, and are located between W50 and the observer.}
              \label{figrgb3}%
    \end{figure}

For the case of the western ear, the depolarization could possibly be the result of both RM fluctuations due to variations in the number of intervening thermal electrons, and also possibly a quasi-Laing--Garrington effect \citep[e.g.][]{1988Natur.331..149L} as the ear lies further along the line-of-sight\footnote{The beams from SS433 are most likely inclined at $\approx$79$^{\circ}$ to the line-of-sight, or equivalently 11$^{\circ}$ to the plane of the sky \citep{1979Natur.279..701A,1979IAUC.3358....1M}, with the eastern ear pointing towards us.} (and is also dipping into the Galactic plane). This could mean that our estimate of the variations in the foreground screen could be overestimated, with a contribution to this depolarization also occurring in the local ISM immediately surrounding W50. Nevertheless, our estimate still reliably constrains the overall fluctuations in the plane of the sky towards the nebula -- which we have shown is substantially affected by the ionised gas distribution. We present a toy model interpretation of the location of the diffuse H$\alpha$ clouds relative to W50 itself in Fig.~\ref{figrgb3}.

It could be argued that the distribution of linear polarization from W50 also appears to show a correlation with the location of a diffuse molecular cloud, as inferred from CO observations \citep[e.g.][]{1983ApJ...272..609H,1984JRASC..78..210F}. However, there is no plausible mechanism that would allow such a cloud to preferentially allow the transmission of linearly polarized radio emission in contrast to the surrounding regions. More recent observations also suggest there is no convincing evidence for an interaction of the W50/SS433 system with this molecular cloud \citep{2007MNRAS.381..881L}, and so the cloud is most likely simply coincidentally aligned with W50.

\subsection{The Large-Scale Magnetic Fields as seen via Faraday Rotation}
\subsubsection{Interpretation of the Rotation Measures}
We here follow the results presented in Section~\ref{faradayrotation}, and consider the foreground Faraday rotation determined in Section~\ref{RMforeground}. The median foreground RM of the $5^{\circ}$ radius patch towards the W50/SS433 system is consistent with $0$~rad~m$^{-2}$. However, we are unable to constrain any spatial foreground variations at scales less than $5^{\circ}$. The location of W50 near to the Galactic plane, and situated in an area with rapidly-varying Galactic emission, and with a known depolarizing Faraday screen (see Section~\ref{depol}), suggests that the observed variation in RM across the face of W50 could be due to foreground variations. Indeed, the RMs of the diffuse polarized emission that surrounds W50 changes sign in both the East and the West, which shows that the foreground magnetic field structure is complicated. Nevertheless, the variation in the RMs of the polarized emission of W50 itself shows no obvious relation to the depolarizing screen, e.g.\ there are no enhancements or variations of the RM near to the diffuse H$\alpha$ clouds. This suggests either that the RM variations are intrinsic and related to the local environment, or that there are additional intervening screens of significant angular extent and of unknown origin.

The observed RMs towards the W50/SS433 system appear to show two key morphological features, as discussed in Section~\ref{atcaRMs}: (i) the central region of W50 appears to have an axis of symmetry that separates two large regions of oppositely-signed RM. This axis of symmetry appears to be located somewhere between $0^{\circ}$ to $45^{\circ}$ relative to the north--south axis that intersects SS433, and (ii) the eastern ear of W50 appears to have an axis of symmetry that lies parallel to the jet-axis, with positive RMs to the north and negative RMs to the south. As both of these features appear to be related to other morphological features of W50, and in combination with the average foreground RM being consistent with $0$~rad~m$^{-2}$, we interpret the observed RM variations to be intrinsic and directly related to the environment of W50. 

While such an interpretation is strongly dependent on our foreground estimate, we would require a very large error in our estimate to remove the RM-axis associated with the central region -- with a foreground offset of $\ge125$~rad~m$^{-2}$ required to completely remove any axis across the source. This further suggests that the foreground is likely not responsible for this feature. Similarly, the Galactic polarized emission that surrounds the eastern ear has a positive RM that is continuous across both the positive and negative regions of the emission associated with W50's ear. The sharp transition from the RM of $+230$~rad~m$^{-2}$ associated with the Galactic emission to the weaker $|\textrm{RM}|\approx40$~rad~m$^{-2}$ associated with the ear, also suggests that the Galactic foreground cannot be responsible for this feature. However, this transition from $+230$~rad~m$^{-2}$ to $|\textrm{RM}|\approx40$~rad~m$^{-2}$ suggests that the Galactic emission is located behind the ear. If the Galactic emission were originating in the foreground of the ear, one would expect the $+230$~rad~m$^{-2}$ to be imprinted on part of the eastern ear, which is clearly not the case. Nevertheless, the observed RM variations across some parts of the source could be due to Galactic foreground variations, which hinders enabling an accurate determination of the intrinsic properties of the magnetised plasma within the radio nebula -- especially at a Galactic latitude of $-2.0^\circ$. The RM of the foreground diffuse Galactic emission, visible in both the East and West of the ATCA mosaic, very clearly varies in both sign and magnitude on angular scales smaller than that of W50. It is likely that much more sophisticated techniques for modelling the foreground magnetic field structure in the Galaxy will need to be developed before a comprehensive understanding of the magnetic field structure in the object can be ascertained.

\subsubsection{Magnetic Fields in the Eastern Terminal Shock}
\label{earfield}
The Faraday rotation of the eastern ear can be interpreted as evidence that there is a toroidal- or helical-shaped magnetic field that loops about the axis from the jets, with the magnetic field facing away from us in the Southern part of the ear, and facing towards us in the Northern part of the ear. The magnetic field is not an exact loop, with positive RMs to the north, negative RMs to the south, and intermediately positive RMs in the conjoining intermediate region of $\sim25$~rad~m$^{-2}$. Interestingly, the axis of symmetry of the RMs would become much more strongly aligned with the jet-axis after subtracting our foreground RM estimate of $+36\pm41$~rad~m$^{-2}$. This RM-feature is also consistent with X-ray observations, which find that the eastern SS433 jet terminates in the eastern ear in a \emph{ring-like} `terminal shock' \citep{2007A&A...463..611B}. The spatial coincidence of X-ray and radio emission in this region suggests that the physical conditions of the terminal shock region are very similar to those found at the outer shocks of ordinary supernova remnants, and indicates the interaction between the terminal shock of the jet and the circumstellar medium. The X-ray emission is very strongly correlated with the radio-continuum emission and seems to form a ring-like structure confined by the outer boundary of the radio remnant, and indicative of the braking of a hollow-cone flow. \citet{2007A&A...463..611B} suggest this indicates that the final jet-flow may retain a hollow-cone morphology. The thermal X-ray component has a low-fitted temperature which implies that the terminal shock must not be very strong, and that the jet-flow must have been considerably slowed down from its initial $0.26c$. Where and how this deceleration happens, remains unclear \citep[e.g.][]{2004ASPRv..12....1F}.

\subsubsection{Magnetic Fields in the Central Shell}
The Faraday rotation of the central region can be interpreted as evidence that there is a shell-like magnetic field looping around the central region of the radio nebula. This is consistent with observations of SNRs \citep[e.g.][]{2010ApJ...712.1157H}. It has been shown by \citet{2009IAUS..259...75K} that the RM of a uniform ambient medium that is swept-up into a shell by an expanding SNR can leave a strong imprint on the RM variations of the resulting shell itself. Following \citet{2010ApJ...712.1157H}, the resulting variations in RM can therefore be used to infer the in-situ ambient magnetic field geometry of the gas into which the SNR has exploded. However, there is a known strong density gradient running E--W in W50, identified by both the spectral index and morphology of the Western ears (see Section~\ref{backgroundw50}), and yet this gradient is not identified in the RM structure. Similarly, as the shell of W50 is only clearly visible to the N and S, this provides an axis of symmetry for the radio nebula. If the SNR had swept up a uniform magnetic field it would be expected to leave identical RM gradients along each edge, with the gradient aligned with the objects axis of symmetry \citep{2009IAUS..259...75K}. Such a gradient is not observed. As an alternative explanation, a shell could have embedded an imprint in the RMs of the ambient magnetic field left behind by a progenitor wind. Again following \citet{2010ApJ...712.1157H}, at large distances from a stellar surface, stellar winds are expected to have largely toroidal fields -- if such a field geometry is preserved after material was swept up by an outgoing shock front, this could produce a RM pattern that was negative on one limb of the SNR and positive on the other. 

This does qualitatively appear to fit the observations of W50. Nevertheless, such an analysis is non-trivial for the W50/SS433 system, which could have a wind-blown bubble origin. In models of wind-blown bubbles in which a fast-wind (or supernova ejecta) overtakes and ``sweeps-up'' a slower magnetised wind from a prior state of stellar evolution, in the case of an azimuthal field, the resultant radio nebula displays two arc-like features with oppositely-signed RMs \citep[e.g.][]{2013ApJ...765...19I}. In particular, the negative RMs along the ridge to the SW of the central region below SS433 are also visible in the image of \citet{1986MNRAS.218..393D}, who find that unlike the rest of the object, this region does not have a tangential distribution of magnetic field relative to the shell but rather that the magnetic fields run directly along the ridge. This may imply that we are seeing polarization from the front layers of the shell in this region, with polarization further along the line-of-sight having already been depolarized by Faraday effects within the nebula. This would be consistent with the total intensity spectral index variations across W50, and with the diffuse depolarizing foreground screen near to this area -- which both imply a denser environment with more small-scale structure in the magnetic field towards the Galactic plane.

The analysis is complicated further due to the action of the SS433 jets that could have entirely remodified the magnetic field structure of W50. Indeed, one could argue that there is an axis of symmetry in the RMs of the central portion of W50 that are aligned perpendicularly to the axis of the SS433 jets: with negative RMs to the West and positive RMs to the East. However, this assessment does not incorporate the contradictory large positive RMs to the north/north west of the central region. Nevertheless, the northern portion of the shell has other features that suggest it should not be incorporated in an assessment of any symmetry axis and that this region could have an independently large RM. These anomalous features include its very high polarized fraction of $\sim50$\% (see Section~\ref{radiopolbackground}), which indicates that the northern rim has the most strongly-ordered magnetic fields across the whole object. Furthermore, the linear polarization of the northern rim is coincident with some of the optical filaments -- this is the only region in W50 in which this is the case. These features suggest that this region is dominated by classic compression of an interstellar field by a SN shock that is oriented approximately east--west. The most consistent interpretation therefore appears to be that there is a symmetry axis running approximately north--south (although the precise angle of this axis of symmetry is ambiguous), with a loop magnetic field that mostly runs east--west in the central region as seen in projection. This explanation also fits with the known orientation of W50 along the line-of-sight (with the Eastern lobe oriented slightly towards us, and the Western lobe directed away from us), and with other multiwavelength evidence that implies a preferential east--west axis in W50 (see Section~\ref{backgroundw50}). In combination, this therefore implies that the current magnetic field geometry is predominantly oriented parallel to the axis of the SS433 jets. Furthermore, as the symmetry axis of the RM-gradient in the circular region of W50 may be oriented perpendicularly to the jet-axis of SS433, this suggests that the central region of W50 is not a distinct object from the jets, and that the central region evolved in tandem with the jet-features. There must therefore be a strong degree of jet/remnant coupling in the W50/SS433 system, which includes coupling of the respective magnetic fields.

In combination, we interpret our data as evidence in favour of a loop magnetic field existing in the shell of W50, which is oriented almost exactly east--west. In this scenario, the polarization from the Northern rim probes a separate and distinct region of the shell with a large positive RM. This would naively appear to provide an intangible three-dimensional magnetic field geometry, however this is only as seen in projection and does not include e.g.\ depolarization effects along the line-of-sight. There is other evidence to suggest the northern rim could be distinct from the rest of the shell -- it is brighter in total intensity, and is considerably stronger in linear polarization compared to the rest of the shell. Furthermore, unlike the rest of the shell, the polarization angles in the Northern rim are typically noted as a classic case of compression of frozen-in magnetic fields (see Section~\ref{faradayrotation}). This is consistent with the northern rim being the only region in the W50/SS433 system that has been unmodified by the jets of SS433, with the northern rim therefore providing an archaeological glimpse into the initial SNR that compressed the ambient magnetic field of the surrounding ISM. 

Alternatively, we could invoke the presence of a pre-existing cavity that had been carved out prior to the supernova explosion, i.e.\ a stellar-wind bubble, and with the cavity being an ovoid that is preferentially larger along the east--west jet axis. It is known that an ambient ISM magnetic field oriented perpendicularly to the direction of the expansion of a bubble can lead to the formation of an ovoid, rather than spherical bubble. Strong magnetic fields, can even completely stop the expansion of the bubble in the direction perpendicular to the field, leading to the formation of a tube-like bubble \citep{2015A&A...584A..49V}. This effect would have strong consequences on the shape and evolution of nebulae such as W50, which may have been formed within a main wind-blown bubble. In this case, the supernova shock would then propagate undisturbed through the ovoid cavity, leaving behind a loop magnetic field structure. In the case of the Northern rim, the shock interacted strongly -- compressing the denser ambient ISM. Both interpretations clearly require a supernova progenitor, and therefore favour a \emph{SNR+Jets model}. In fact, this latter scenario suggests the possibility of a more complex scenario than has typically been considered in the literature, a \emph{Wind+SNR+Jets model}, with a stellar-wind bubble driven by the jets of SS433 expanding into a swept-up ISM that was itself created either entirely by the jets of SS433 or even by a continuous wind from the SS433 binary, followed by a supernova explosion -- with the SNR rapidly taking on the dimensions of the wind-blown bubble -- and finally an encounter between the jets from SS433 and the already-formed SNR, with the ram pressure from the jets `punching' through the shell to form the ears. Such a \emph{Wind+SNR+Jets model} has been partially suggested before \citep{1991MNRAS.251..318T}. This model could reconcile why the SNR has been previously interpreted as displaying features consistent with both a wind-blown bubble and a SNR \citep{1980ApJ...238..722B,1980Natur.287..806S,1983MNRAS.205..471K,1980MNRAS.192..731Z,1986MNRAS.218..393D,1996PASJ...48..819M,2000A&A...362..780V}. 

\subsection{Shocked Filaments and The Origin of W50}
Significant questions remain concerning the association of W50 and SS433 \citep[see e.g.][and references therein]{2004ASPRv..12....1F}. While a consensus has emerged that the jets of SS433 have punched holes in a preceding, expanding, supernova shell, the formation of the whole of W50's structure by SS433's jets alone has not been completely ruled out (see Section~\ref{intro}).

The lack of optical-emitting filaments in the shell of W50 has previously presented a challenge for a supernova origin of the shell \citep[e.g.][]{1987AJ.....94.1633E}. If the shell is an old SNR, then we would expect optical emission from some of the shell filaments, and for optical emission to be observed all along the shell and to trace out the region of radiative cooling of the shock wave in a relatively dense region of the ISM \citep[e.g.][]{1987ApJ...314..187H}. Although non-radiative shocks are also seen as faint filaments in the H$\alpha$ image of younger SNRs such as Kepler's SNR and RCW~86 \citep{2013MNRAS.435..910H,2015AAS...22514020S}, and ionised H$\alpha$ filaments have also been detected in systems such as the Orion-Eridanus superbubble \citep{2014MNRAS.441.1095P}. Previously proposed explanations for this discrepancy in W50 are that (i) there is a rift of higher extinction crossing the center of W50 \citep{1980ApJ...236L..23V,2007MNRAS.381..308B}, that could be obscuring any filaments, and (ii) alternatively, that the radio emission from a large SNR, like W50, of $\approx50$~pc radius would come from the crushing of low-density clouds which would produce only weak optical emission (as predicted by the model of \citet{1982ApJ...260..625B}). We have resolved this dilemma by identifying weak optical filaments distributed throughout the shell. We have identified counterparts in the radio morphology for nearly all of the newly-identified optical filaments. An alignment is known for the brightest optical features at the breakout regions of the shell \citep{1987AJ.....94.1633E}, while an alignment between optical and radio filaments in the shell itself would be expected for an old SNR. While it could be argued that the newly discovered optical filaments are suggestive of a SN shock, this is by no means certain, and it is not possible -- using the current data -- to directly rule out an origin of the H$\alpha$ filaments due to a wind-blown bubble. Nevertheless, the combined weight of the multiwavelength evidence does typically suggest a SNR origin \citep{2004ASPRv..12....1F}. Either way, future studies will be able to use these new filaments to obtain in-depth information on the kinematics of the shell of W50 -- potentially allowing a definitive experiment to determine how W50 was formed.

In W50, the lack of a detection of an additional, more extensive, network of filamentary nebulosity throughout the shell is consistent with previous deep-images and high-resolution spectroscopy of the northern radio ridge of W50 that suggested that patchy foreground dust along the 5.5~kpc line-of-sight is inhibiting the detection of all of the optical nebulosity associated with W50 \citep{2007MNRAS.381..308B}. Our detection of the faintest filaments, in parts of the shell, lend support to a very patchy foreground. The discovery of a more extensive network of optical filaments lends additional support to the \emph{SNR+Jets model} for the formation of the W50/SS433 system. However, it does not rule out a mixed-type origin for the nebula, and could also imply a \emph{Wind+SNR+Jets model} with a SNR expanding in a pre-formed wind-blown bubble \citep[see][]{1991MNRAS.251..318T,1993MNRAS.261..674R,1995ApJ...442..679K,1999ApJ...527..866L}. There is also the possibility of most of the expansion of the shell being driven by a wind that follows, rather than precedes, the SNR -- a \emph{SNR+Wind+Jets model} (or perhaps a \emph{Wind+SNR+Wind+Jets model} as it is not mutually exclusive from there being a pre-existing cavity before the supernova explosion). In this scenario, a preceding SNR left a shell, which was then further inflated by a wind from the binary system, and with the eastern and western regions later being strongly distorted by outflows from the jets. This potentially better explains the morphology of the filled-centre almost plerion-like central component which has a steep-spectrum interior, however this scenario would appear to be inconsistent with the observed magnetic field structure in the northern rim that differs from the rest of the central region. What is clear is that the magnetic contributions of the shell and the jets are coupled, and that either (i) the field in the shell has helped to form SS433's axis and to collimate the jets in this direction, (ii) the axis of the jets has modified the magnetic fields in the surrounding medium, (iii) there is a strong ambient field that has influenced both the intrinsic axes of the shell and the jets, or (iv) some combination of the above.

Numerical simulations of explosions inside pre-existing wind-driven bubbles, particularly of non-spherical SNRs, and predictions of the effect on the observed RM geometry will help substantially in unpicking these various interpretations \citep[e.g.][]{1990MNRAS.244..563T,1995RvMP...67..661B,2012JSARA...7...23P,2015A&A...578A..24G}. While beyond the scope of this paper, demonstrating the application of $QU$-fitting \citep{2012MNRAS.421.3300O,2015AJ....149...60S} to an extended object such as W50 would form a useful future study. We would require the total intensity short-spacing data in order to perform reliable $QU$-fitting on our ATCA data, as we are currently unable to retrieve reliable polarized fractions due to the limited and changing $uv$-coverage across the very broad band. Future observations would therefore ideally include broadband data from a single-dish in order to recover all of the spatial scales in the object. Nevertheless, $QU$-fitting should be able to identify separate signatures from the Northern rim and the east--west aligned ring-like magnetic field. This would allow one to test the various scenarios for W50's formation, although it appears a supernova explosion is likely a necessary feature in the object's evolution. 

\subsection{A Unique Class of Zombie Supernova Remnants?}
W50 has implications for how SNRs which contain compact objects fit into the overall picture of SNR evolution. The combined evidence of our own and previous observations is consistent with W50 being the reanimated corpse of a shell-like SNR, driven back to life by the powerful outflows from its central engine. This places W50 into a unique class of Zombie supernova remnants. The study of such outflow-dominated objects may provide unique methods in which to identify remnants with massive star progenitors, and may also assist in revealing why the fraction of SNRs in which a compact source is identified is much less than that expected by supernovae type alone \citep[e.g.][]{1991ARA&A..29..363V}. This is doubtlessly partially due to observational selection effects. Nevertheless, future studies of W50, and other possible Zombie SNRs, may allow us to begin recognising more subtle signatures that allow the presence of a central source to be inferred rather than directly detected. An extensive sample of similar Zombie remnants are already known, and a number of potential candidates are already visible in the MOST SNR catalogue \citep{1996A&AS..118..329W,1998MNRAS.299..812G}. Future observations of W50 and these other candidates may therefore be able to help us understand the larger picture of where SNRs with compact objects fit on the evolutionary ladder, and to create an even more comprehensive catalogue of similar objects for future study. Observations of similar ununsual large-angular size nebulae are already underway \citep[e.g.][]{2015ApJ...804...22P}, and the use of facilities such as the ATCA and the VLA will continue to improve upon Galactic SNR imaging techniques \citep[e.g.][]{2011ApJ...739L..20B}. It is clear that the W50/SS433 system is in a special and unique class of outflow-driven SNRs. 


\section{Conclusions}
\label{conclusions}

Using the ATCA, we have obtained a 198 pointing mosaic of a $3 \times 2$~degree region of the sky surrounding W50, with continuous radio frequency coverage from 1.4 to 3.1~GHz, and in full-Stokes. Following RM Synthesis, we have obtained a noise level in our Faraday cubes of 0.125~mJy~beam$^{-1}$~rmsf$^{-1}$. This is the most sensitive mosaic of W50 that has been created in linear polarization, and provides an unrivalled view of the large-scale magnetic fields in the radio nebula. To our knowledge, these data are nominally also the most sensitive images of W50 in the radio continuum to date, although we are limited in total intensity by our $uv$-coverage. We have found the limitation that the frequency-dependent $uv$-coverage inherent to ultra broadband radio observations tends to provide an overly steep spectral index of extended sources.

We have complemented our radio data by also creating a large mosaic of IPHAS continuum-corrected H$\alpha$ data, which maps out the ionised gas distribution surrounding the W50/SS433 system -- thereby providing a unique multi-wavelength perspective of the object.

Our conclusions are summarised as follows:
\begin{enumerate}
\item We have identified new, faint, optical, filamentary emission across the nebula. Conversely to previous studies, we find optical emission coincident with both the Northern shell and the bright radio filaments in the Western ear. We also find additional new filaments within the filled-centre of the W50 nebula. Unlike previously discovered filaments that are most likely associated with breakout regions from the microquasar jets, the newly discovered faint filaments are most consistent with being associated with the initial SNR shell itself. All of these optical filaments are either directly traced out, or have a nearby filamentary counterpart, in the radio continuum image of W50.

\item We have mapped the linearly polarized radio emission across the nebula. Unlike the radio continuum data, the linear radio polarization appears to be largely independent of the brightest optical filaments. Nevertheless, the polarization is coincident with some of the fainter filaments in the Northern ridge -- also consistent with classic field compression by a supernova shock. The linear radio polarization also shows a strong anti-correlation with the surrounding diffuse H$\alpha$ emission, demonstrating that the foreground ionised gas traced by the H$\alpha$ emission is depolarizing the observed emission due to spatially-varying Faraday rotation fluctuations across the face of the radio nebula. The simultaneous depolarization of both W50 and the Galactic polarized emission indicates that these clouds of thermal electrons must be located in the foreground of W50/SS433.

\item The structure of the Faraday foreground towards W50 is complicated and features many large-scale changes of sign of the magnetic field -- both in the plane of the sky (as traced by the RM from linearly polarized emission) and along the line-of-sight (as traced by the RM and DM of nearby pulsars). As the median foreground RM towards the nebula is consistent with $0$~rad~m$^{-2}$, the pattern of RMs across W50 appears to be best described by a loop magnetic field surrounding the central region itself. The symmetry axis of this loop-field is oriented perpendicularly to, and appears to be connected to, the symmetry axis of the jets/ears -- suggesting that both the shell and the ears are related and dependent phenomena. The evolution of magnetic fields in both the jet and remnant components of the W50/SS433 system must be intertwined.

\item The Faraday rotation data suggest that there is a ring-like magnetic field that is threaded through the termination shock associated with the eastern ear, at the location where the shock of the jet interacts with the circumstellar medium. This is consistent with the ring-like structure previously inferred by X-ray observations.
\end{enumerate}

Finally, it should be noted that the enhanced radio and X-ray emitting Eastern ear structure is possibly similar to a hotspot in FRII radio galaxies that are also the physical manifestation of a jet-termination shock. Thus W50, and similar objects of its Zombie class, present a unique opportunity to study structures similar to hotspots using Galactic jets. Curiously, a similar magnetic field structure to that seen in W50's eastern ear has also been implicated in the magnetic field structure of jets from active galactic nuclei \citep[e.g.][]{2015MNRAS.450.2441G}. This further suggests that investigations of the magnetic field structure using tools such as $QU$-fitting, will be important in unravelling the full Faraday structure along the line-of-sight towards Galactic supernova remnants. In particular, future radio observations with for example the Square Kilometre Array (SKA) will allow us to learn about magnetic field structures in SNRs through detection of the emission itself, and through a background grid of polarized sources, due to the vast increase in the available sample size \citep{2015aska.confE..96H}. Such studies will help us further understand the ISM itself and features such as SNRs, and to therefore refine and remove our estimates for the Galactic foreground for a large number of extragalactic studies.

\section*{Acknowledgments}
We thank Martin Bell, Ilana Feain, and Craig Anderson for useful discussions during the preparation of this project, and Janet Drew and Nick Wright for discussions on mosaicing data from the IPHAS Survey. We are also grateful to Naomi McClure-Griffiths for helpful advice on observing with the ATCA. J.S.F.~ is very grateful to Filippo Mannucci and the staff of the Arcetri Astrophysical Observatory for hosting him as a visitor during the writing of this paper. J.S.F. \& B.M.G.~acknowledge the support of the Australian Research Council through grant DP0986386. Parts of this research were conducted by the Australian Research Council Centre of Excellence for All-sky Astrophysics (CAASTRO), through project number CE110001020. S.P.O.~acknowledges support from UNAM through the PAPIIT project IA103416. The Australia Telescope Compact Array is part of the Australia Telescope National Facility which is funded by the Commonwealth of Australia for operation as a National Facility managed by CSIRO. This paper makes use of data obtained as part of the INT Photometric H$\alpha$ Survey of the Northern Galactic Plane (IPHAS, www.iphas.org) carried out at the Isaac Newton Telescope (INT). The INT is operated on the island of La Palma by the Isaac Newton Group in the Spanish Observatorio del Roque de los Muchachos of the Instituto de Astrofisica de Canarias. All IPHAS data are processed by the Cambridge Astronomical Survey Unit, at the Institute of Astronomy in Cambridge. The bandmerged DR2 catalogue was assembled at the Centre for Astrophysics Research, University of Hertfordshire, supported by STFC grant ST/J001333/1. This research made use of Montage, which is funded by the National Science Foundation under Grant Number ACI-1440620, and was previously funded by the National Aeronautics and Space Administration's Earth Science Technology Office, Computation Technologies Project, under Cooperative Agreement Number NCC5-626 between NASA and the California Institute of Technology.


\appendix

\section{Morphological Descriptions of the Filamentary H$\alpha$ Emission Regions}
\label{appendix}

\subsection{The Eastern Breakout Region (Region 1)}
\label{1stregion}
Region 1 shows the eastern breakout region, where the eastern ear punches through the eastern boundary of the circular region of W50, and is shown in Fig.~\ref{figreg1}. The images show the pseudo-colour IPHAS continuum-corrected H$\alpha$ image, overlaid with the total intensity radio continuum contours from the VLA image at 1.4~GHz \citep{1998AJ....116.1842D}, and with the linearly polarized intensity contours from the ATCA images at both 3.1~GHz and 2.2~GHz. For the regions discussed in Appendices~\ref{2ndregion} to \ref{7thregion}, the representation of various emission by different colours and contours is the same as in Fig.~\ref{figreg1}. The exact position of all the regions relative to the full extent of the W50/SS433 system is shown in Fig.~\ref{figregions}.

The most striking features are the very brightest optical filaments, that have been identified before \citep{1980ApJ...236L..23V,1980MNRAS.192..731Z}, although the quality of the IPHAS data allows us to observe them with unprecedented resolution. The network of filaments is very extensive, and includes numerous previously-unidentified smaller filamentary structures adjacent to the main optical arcs. As seen in projection, the entanglement of the filaments is highly pronounced.

From 19$^{\textrm{h}}$15$^{\textrm{m}}$00$^{\textrm{s}}$, +5$^{\circ}$4$^{\textrm{m}}$ to 19$^{\textrm{h}}$14$^{\textrm{m}}$20$^{\textrm{s}}$, +5$^{\circ}$4$^{\textrm{m}}$ the edge of the radio continuum emission is traced out by an H$\alpha$ filament. Just southwest of this edge, near 19$^{\textrm{h}}$14$^{\textrm{m}}$15$^{\textrm{s}}$, +5$^{\circ}$2$^{\textrm{m}}$ is a bright compact radio source which is likely extragalactic. Surrounding the compact source is diffuse radio emission associated with the very brightest H$\alpha$ filaments. The emission appears continuous and extended, and is brighter at radio wavelengths in the areas with brighter optical emission. It is likely that improved angular resolution observations would be able to identify individual radio counterparts to each optical filament within the observing beam. To the east, at 19$^{\textrm{h}}$14$^{\textrm{m}}$40$^{\textrm{s}}$, +5$^{\circ}$0$^{\textrm{m}}$ another region of extended radio emission is associated with another group of optical filaments. The edge of this extended emission is traced out by several filaments, notably in the southwest of the region, where the boundary of the radio emission drops off rapidly at the location of the optical filament.

There is another bright optical filament oriented almost exactly along north--south and centred at 19$^{\textrm{h}}$13$^{\textrm{m}}$50$^{\textrm{s}}$, +4$^{\circ}$52$^{\textrm{m}}$. The filament approximately traces out the missing portion of the circular central part of W50, where the jet punches through the periphery. The filament has a counterpart in the radio continuum image, also aligned north--south, with a circular compact `knot' located near the centre of the vertical feature. Further southeast, near 19$^{\textrm{h}}$14$^{\textrm{m}}$35$^{\textrm{s}}$, +4$^{\circ}$48$^{\textrm{m}}$ there is another group of optical filaments -- fainter than the main filaments -- that again appear to be preferentially oriented north--south. These features have faint diffuse radio emission associated with them, that again approximately traces out the boundary of the optical filaments. To the west of this region, at 19$^{\textrm{h}}$14$^{\textrm{m}}$25$^{\textrm{s}}$, +4$^{\circ}$46$^{\textrm{m}}$ are four small optical filaments $\approx1$~arcmin in length and again oriented north--south. These features have no radio counterpart, but are not related to any specific imaging artefact -- they are most likely real features associated with W50 itself.

There are no apparent counterparts to any of the optical filaments in the linearly polarized radio emission, either at 3.1~GHz or 2.2~GHz. This is due to the foreground of diffuse H$\alpha$ emission that is depolarizing the radio nebula, as discussed in Section~\ref{depol}.

\subsection{The Eastern Ear (Region 2)}
\label{2ndregion}
Region 2 shows the eastern ear, and where the radio emission from the eastern ear terminates, and is shown in Fig.~\ref{figreg2}.

A significant number of the features in the west of this image have already been described in Section~\ref{1stregion}, although this image provides a zoomed-out perspective of the same region -- while also including views further south and east. To the south of the brightest optical nebulosity, even more faint optical filaments are identified. The filaments described in Section~\ref{1stregion} at 19$^{\textrm{h}}$14$^{\textrm{m}}$35$^{\textrm{s}}$, +4$^{\circ}$48$^{\textrm{m}}$ extend even further south than was previously seen, with the southern extension curving steadily towards the east at lower declinations. At the most southern end of these filaments, at 19$^{\textrm{h}}$14$^{\textrm{m}}$50$^{\textrm{s}}$, +4$^{\circ}$40$^{\textrm{m}}$, and where the optical filaments are almost below the detection threshold, there is diffuse radio emission that is slightly extended along the length of the filaments and which we therefore interpret as a radio counterpart.

Further west, at 19$^{\textrm{h}}$14$^{\textrm{m}}$10$^{\textrm{s}}$, +4$^{\circ}$40$^{\textrm{m}}$ is another optical filament that is again largely oriented north--south albeit with several curves -- such that the filament appears to have an `S'-shape. There is nearby diffuse radio emission, that is brightest towards the southwest of the `S', although there are no obvious morphological similarities between the optical and radio continuum emission. The radio emission may be unrelated to the `S' feature, which may therefore lack a counterpart at radio wavelengths.

Further south from the `S' feature, are two more small filaments. The brightest of the two is located at 19$^{\textrm{h}}$14$^{\textrm{m}}$10$^{\textrm{s}}$, +4$^{\circ}$36$^{\textrm{m}}$ and is $\approx1$~arcmin in length and oriented at $\approx45^{\circ}$ towards the east. A fainter filament is located at 19$^{\textrm{h}}$14$^{\textrm{m}}$20$^{\textrm{s}}$, +4$^{\circ}$34$^{\textrm{m}}$. Neither of these two filaments appear to have a radio counterpart. There are an additional two optical filaments nearby to where the radio continuum emission begins to increase rapidly in brightness towards the eastern ear, and located at 19$^{\textrm{h}}$15$^{\textrm{m}}$15$^{\textrm{s}}$, +4$^{\circ}$53$^{\textrm{m}}$ and 19$^{\textrm{h}}$15$^{\textrm{m}}$10$^{\textrm{s}}$, +4$^{\circ}$47$^{\textrm{m}}$. These features also have no obvious radio continuum counterpart. 

There are no optical filaments associated with the end of the eastern ear itself. The radio continuum chimney feature described by \citet{1998AJ....116.1842D} is visible extending to the north from the outskirts of the ear, and has no optical counterparts. The foreground of diffuse H$\alpha$ emission described in Section~\ref{depol} is visible running through the image. The diffuse H$\alpha$ appears to decline in brightness towards the eastern ear, with the boundary approximately aligned with the north--south radio filament at 19$^{\textrm{h}}$15$^{\textrm{m}}$30$^{\textrm{s}}$, +4$^{\circ}$54$^{\textrm{m}}$. In combination with the brightest radio filament further towards the east, and also with the mentioned optical filaments, there appears to be a significant number of features that are approximately aligned north--south. We interpret these features as being related to the jet axis of SS433 -- which is aligned approximately perpendicularly along east--west. The optical and radio filaments are then all consistent with shocked material due to ram pressure from the jets.

There are no apparent counterparts to any of the southern optical filaments in the linearly polarized radio emission, either at 3.1~GHz or 2.2~GHz. Similarly to the filaments described in Section~\ref{1stregion}, this is due to the foreground of diffuse H$\alpha$ emission that is depolarizing the radio nebula. However there is very strong linearly polarized emission from the bright radio filaments in the eastern ear. The linear polarization is almost entirely associated with the termination shock of the eastern jet, indicating a well-ordered magnetic field and a weakly depolarizing foreground. This polarized emission appears to terminate relatively abruptly along the edge of the diffuse H$\alpha$ emission, as discussed in Section~\ref{depol}. Due to spectral index effects, there is a region to the north of the ear where the polarized intensity is brighter at lower radio frequencies, although this places no constraint on whether the polarized fraction is decreasing. There are other regions clearly showing polarized intensity, particularly located along the eastern and southern edges of the image. This is most likely diffuse Galactic polarized emission, as described in Section~\ref{linpol}.

\subsection{The Western Ear \& Breakout Region (Region 3)}
\label{3rdregion}
Region 3 shows the western ear, the western breakout region where the jet punches through the western boundary of the circular region of W50, and where the radio emission from the western ear terminates, and is shown in Fig.~\ref{figreg3}. The most striking features are the very brightest optical filaments, that have been identified before \citep{1980ApJ...236L..23V,1980MNRAS.192..731Z}, although the quality of the IPHAS data allows us to observe them with unprecedented resolution. The pre-identified filaments are those that trace out the apparent circular edge of the central region of W50, extending from 19$^{\textrm{h}}$10$^{\textrm{m}}$15$^{\textrm{s}}$, +5$^{\circ}$10$^{\textrm{m}}$ down to 19$^{\textrm{h}}$9$^{\textrm{m}}$30$^{\textrm{s}}$, +4$^{\circ}$55$^{\textrm{m}}$. We also identify an additional filament that traces out the periphery of the Southern central region, which is located at 19$^{\textrm{h}}$9$^{\textrm{m}}$25$^{\textrm{s}}$, +4$^{\circ}$44$^{\textrm{m}}$. All of these filaments are located along the approximately circular boundary of W50. There are radio continuum contours associated with almost all of these features, with another north--south feature seen tracing out the brightest optical filament. This is consistent with shocked material due to ram pressure from the jets of SS433, and is similar to the north--south features also seen in the eastern ear (see Section~\ref{2ndregion}).

A small H$\alpha$ filament of $\approx1$~arcmin in length is located at 19$^{\textrm{h}}$10$^{\textrm{m}}$40$^{\textrm{s}}$, +4$^{\circ}$43$^{\textrm{m}}$, and is both coincident with, and oriented in the same direction as the bright radio filament that extends from near to SS433 and curves down towards the southwest. The full extent of this radio filament lies outside of Fig.~\ref{figreg3}, but is more clearly visible in Fig.~\ref{figrgb}. We find no other optical features associated with this radio filament.

Along the northern edge of the western radio ear, a bright H$\alpha$ filament traces out the entire northern boundary that is visible in the radio continuum images. A network of filaments extend further south into the interior of the western ear itself, with several bright knots of emission. The northeastern part of the ear, located at 19$^{\textrm{h}}$9$^{\textrm{m}}$45$^{\textrm{s}}$, +5$^{\circ}$15$^{\textrm{m}}$ also contains fainter filaments that connect the outer ear to the central region of W50. The filaments very closely follow the boundary of the radio structure. There are no visible filaments associated exclusively with the circular portion of W50 -- the location is most consistent with jet activity. Interestingly there is a further filament located to the far west, approximately $7$~pc further than the western edge of the radio ear. The alignment of the filament with the general shape of the western ear suggests that it is related to the W50/SS433 system and implies that SS433 is influencing the surrounding ISM at greater distances than is typically deduced from the radio morphology.

There is also considerable radio linear polarization associated with the central portion of W50. However, any intrinsic linear polarization associated with the ear is depolarized by the foreground diffuse H$\alpha$ (see Section~\ref{depol}). There is a clear difference between the 3.1~GHz and 2.2~GHz distributions of polarized emission, with the 3.1~GHz distribution reaching out to a larger radius from SS433 and consistent with W50 being a magnetised spherical structure. At lower frequencies, the depolarization effects are more considerable, such that only polarization on the near side of W50 can reach us. In projection, this appears as a shrinking radius of the polarized emission at lower radio frequencies. This suggests that the depolarization happens inside W50 itself. This is further justified as the inner part of the 2.1~GHz data is more rapidly beginning to be depolarized than the outer ring, as would be expected due to the longer line-of-sight through the turbulent inner gas in the central region of W50. This trend becomes more apparent at the even lower frequencies down to 1.4~GHz, as shown in Fig.~\ref{fig2}, with the ring both shrinking in radius, and also becoming more `ring-like', as the central portion becomes a polarized hole. 

There are largely no connections between the linear polarization and the optical filaments, with the notable exception of the $\approx1$~arcmin in length filament located at 19$^{\textrm{h}}$10$^{\textrm{m}}$40$^{\textrm{s}}$, +4$^{\circ}$43$^{\textrm{m}}$. The full polarized distribution in this region can be seen from RM Synthesis in Fig.~\ref{figrgb}. The optical filament coincides with the edge of a bright peak in the linearly polarized emission. The shocking of material in this filament has therefore possibly given rise to a compressed and well-ordered magnetic field, although it may also be a coincidental alignment.

\subsection{The Northeastern Interior of W50 (Region 4)}
\label{4thregion}
Region 4 shows the northeastern interior of W50, including SS433 within the field of view, and is shown in Fig.~\ref{figreg4}. In this region, there is no optical emission visible in the area in closest proximity to SS433. This is consistent with the previously inferred patchy absorption of optical emission across W50 (see Section~\ref{discussion} and \citet{2007MNRAS.381..308B}). To the far southeast of the region are the bright optical filaments associated with the eastern breakout region and which were discussed in Section~\ref{1stregion}. From the east, there are several fainter H$\alpha$ filaments that appear to stretch inwards towards the interior of the radio nebula. The largest two of these filaments are oriented approximately along the east--west axis, and become increasingly wispy towards the east (possibly implying that they are associated with the jets of SS433). Of these two larger filaments, the brighter one to the north will be referred to as filament `A', and the fainter one to the south will be referred to as filament `B'.

Filament A has a direct radio continuum counterpart. This radio counterpart is a linear filamentary feature and appears to be offset very slightly so that the emission originates along the southern edge of the H$\alpha$ emission. While filament B has no obvious radio counterpart, to the west there is an associated linear radio feature leading directly between the end of the H$\alpha$ filament and that terminates at SS433 itself. The H$\alpha$ filament and the associated radio filament both appear to be part of one, aligned, physical structure. By interpreting this feature as a single filament, this further suggests that filament B is associated with the jets of SS433. The implication is that this filamentary feature traces the magnetic field lines. The lack of optical emission near to SS433 and the abruptness of the transition is consistent with the inferred patchy absorption \citep{2007MNRAS.381..308B}. For only the western edge of the filament to be bright in radio continuum suggests that the charged particles in the eastern edge are: either being swept-away so as not being constantly replenished (although ram-pressure stripping would also leave an imprint on the H$\alpha$ filament), or that the magnetic field is increasingly ordered further from SS433 such that incident electrons are expelled before they are able to significantly contribute to the synchrotron emission, in a manner similar to that theorised for linear radio filaments near to the Galactic centre \citep{2011ApJ...741...95L}. This would imply that either the filament rapidly transitions to a region that contains a very strong magnetic field, on the order of 100~$\muup$G, or that the filaments are possibly aligned as seen in projection.

In linear radio polarization, both filament A and B appear to be depolarizing the diffuse polarized emission from the radio nebula, with even the small extension to the north-west from filament A having a possibly offset counterpart in the radio polarization. Note the very different intrinsic angular resolution of both the radio polarization and H$\alpha$ data. This implies that the filaments depolarize the radio shell, and so the filaments must be located in the foreground of the magnetoionic layer in the shell from which the polarization is being emitted. This possibly suggests that the polarization in this region emanates from the rear-side of the shell at higher-frequencies. The brightest radio polarization throughout the region is dominated by the emission in the northern and northwestern rim as discussed in Sections~\ref{5thregion} and \ref{6thregion}.

\subsection{The Northern Rim (Region 5)}
\label{5thregion}
Region 5 shows a small section from the northern rim of W50 and is shown in Fig.~\ref{figreg5}. In this small region, diffuse H$\alpha$ appears to fill the field of view. There are again several H$\alpha$ filaments, near to the sensitivity limit of the observations. To the very top of the region are three parallel filaments located at 19$^{\textrm{h}}$12$^{\textrm{m}}$15$^{\textrm{s}}$, +5$^{\circ}$27$^{\textrm{m}}$, each running tangentially to the shell of the circular region of W50. At the bottom of the region is another thicker filament, located at 19$^{\textrm{h}}$12$^{\textrm{m}}$10$^{\textrm{s}}$, +5$^{\circ}$21$^{\textrm{m}}$, with some indication of a bifurcation at the western edge of the filament.

In the radio continuum, the three parallel filaments are coincident with a peak in the total intensity and must therefore also contain a peak in the density of charged particles and/or a localised strong magnetic field. These filaments are located at the northern edge of the radio nebula, and trace out the periphery of W50. The filament at the bottom of the region very approximately traces out a boundary where the radio continuum rapidly increases in brightness, however the filament is offset at an angle unrelated to the radio emission, and we interpret the approximate alignment as likely being coincidental. At the noisiest levels of the image, there is a very faint tentative indication of a possible bridge filament that links the parallel filaments with the filament further to the south.

In radio polarization, there are numerous linearly polarized features throughout the region, with a strong variation from 3.1~GHz to 2.2~GHz that indicates significant Faraday depolarization effects. There is no obvious connection between the optical and the linearly polarized radio features, suggesting that while the polarized emission is originating from within the radio nebula, it is from nearer to the observer along the line-of-sight.

\subsection{The Northwestern Rim (Region 6)}
\label{6thregion}
Region 6 shows a small section from the northwestern rim of W50 and is shown in Fig.~\ref{figreg6}. Throughout the H$\alpha$ image, the field appears to be full with diffuse and patchy emission. The region indicated by the white box in Fig.~\ref{figreg6} shows the approximate area in which H$\alpha$+NII emission has been previously reported, albeit without continuum-subtraction \citep{2007MNRAS.381..308B}. There are no identified H$\alpha$ filaments within this box. Due to the location of the diffuse nebulosity, it is not clear whether the diffuse emission measured by \citet{2007MNRAS.381..308B} is associated with W50 itself or with the surrounding ISM.

The optical image again shows numerous optical filaments. Most appear to have radio continuum counterparts, although the field is essentially full of emission. Interestingly, an H$\alpha$ filament located at 19$^{\textrm{h}}$11$^{\textrm{m}}$10$^{\textrm{s}}$, +5$^{\circ}$13$^{\textrm{m}}$, appears to trace out a boundary of the 3.1~GHz polarized radio emission, and which has depolarized at 2.2~GHz.

\subsection{G38.7--1.4 (Region 7)}
\label{7thregion}
Region 7 shows the SNR candidate G38.7--1.4 and is shown in Fig.~\ref{figreg7}. At this outer edge of the large optical mosaic, there are several imaging artefacts as shown by the `blocky' features and bright horizontal/vertical stripes. This supernova remnant candidate was first identified in the ROSAT All-Sky Survey \citep{2002ASPC..271..391S}. The presence of optical emission indicates that this object is indeed most likely an old supernova remnant, and this has previously been reported using IPHAS data in \citet{2013MNRAS.431..279S} (which reports the SNR as G038.7-1.3). The work of \citet{2014ApJ...785..118H} also agrees that it is an old SNR.

Optical filamentary emission clearly traces out the rim between the edge of the radio continuum emission and the surrounding ambient medium, and additional diffuse optical emission is coincident along the entire Eastern part of the radio shell. No linearly polarized radio emission is associated with G38.7--1.4 at these sensitivity levels. As the supernova remnant is completely depolarized across the entire ATCA band, the depolarization could be associated with the H$\alpha$ emission itself, or alternatively, G38.7--1.4 could be depolarized due to bandwidth depolarization -- which would imply a very high RM$\ge$1,997~rad~m$^{-2}$ (see Section~\ref{imaging}). It is most likely that depolarization effects are strong due to the same foreground variations that are affecting detection of linear polarization from W50 (see Section~\ref{depol}).

Referring back to Fig.~\ref{figreg7}, there is also clearly H$\alpha$ emission associated with S74 (Sh2-74), located in the Galactic plane to the northwest of W50 \citep[e.g.][]{1998AJ....116.1842D}. Sh2-74 is a partially-obscured HII region that is situated at a distance of more than 3.1~kpc, on the outer edge of the Sagittarius Arm. It is therefore in the foreground of W50. The diffuse H$\alpha$ (see Section~\ref{depol}) that depolarizes the western ear is likely associated with this complex, and is therefore feasibly in front of G38.7--1.4 (hence the SNR candidate's complete depolarization). This places a weak distance estimate of $>3.1$~kpc on the distance to G38.7--1.4.

\bsp
\label{lastpage}

\end{document}